\documentclass[aps,prd,reprint,showpacs,nofootinbib,superscript address]{revtex4-2}
\usepackage[utf8]{inputenc}
\usepackage{scalerel,stackengine}
\usepackage{physics}
\usepackage{tocbasic}
\usepackage{amssymb}
\usepackage{upgreek}
\usepackage{hhline}
\usepackage{amsmath}
\usepackage{mathtools}
\usepackage[dvipsnames]{xcolor}
\usepackage{multirow,tabularx}
\usepackage{siunitx}
\usepackage{multirow}
\usepackage{graphicx}
\usepackage{etoolbox}
\usepackage{notoccite}
\usepackage{natbib}
\usepackage{mathrsfs}
\usepackage{tensor}
\usepackage{accents}

\usepackage{xspace}

\newrobustcmd{\PSALTer}{\textit{PSALTer}\xspace}
\newrobustcmd{\xAct}{\textit{xAct}\xspace}
\newrobustcmd{\xTensor}{\textit{xTensor}\xspace}
\newrobustcmd{\xCoba}{\textit{xCoba}\xspace}
\newrobustcmd{\xPerm}{\textit{xPerm}\xspace}
\newrobustcmd{\xCore}{\textit{xCore}\xspace}
\newrobustcmd{\xTras}{\textit{xTras}\xspace}
\newrobustcmd{\SymManipulator}{\textit{SymManipulator}\xspace}
\newrobustcmd{\RectanglePacking}{\textit{RectanglePacking}\xspace}
\newrobustcmd{\Inkscape}{\textit{Inkscape}\xspace}
\newrobustcmd{\Mathematica}{\textit{Mathematica}\xspace}
\newrobustcmd{\xPert}{\textit{xPert}\xspace}
\newrobustcmd{\MathGR}{\textit{MathGR}\xspace}
\newrobustcmd{\HiGGS}{\textit{HiGGS}\xspace}
\newrobustcmd{\Windows}{\textit{Microsoft Windows}\xspace}
\newrobustcmd{\Mac}{\textit{macOS}\xspace}
\newrobustcmd{\Linux}{\textit{Linux}\xspace}
\newrobustcmd{\GitHub}{\textit{GitHub}\xspace}
\newrobustcmd{\Bash}{\textit{bash}\xspace}
\newrobustcmd{\WolframLanguage}{\textit{Wolfram Language}\xspace}
\newrobustcmd{\CPP}{\textit{C++}\xspace}

\usepackage{xspace}
\usepackage{xstring}
\usepackage{titlesec}

\usepackage{stix}
\allowdisplaybreaks

\newrobustcmd{\pea}[1]{%
	\emph{#1}\textbf{\ \ \ ---}
}
\titleformat{\paragraph}[runin]{\normalfont\normalsize\bfseries}{\emph\theparagraph}{1em}{\pea}

\usepackage{listings}
\usepackage{sourcecodepro}
\usepackage{upquote}
\newrobustcmd{\dollar}{\mbox{\color{gray}\textbf\textdollar}}

 \lstdefinestyle{ascii-tree}{ 
    literate=
    {├}{{%
	\hphantom{\raisebox{0.5ex}{\rule{0.5ex}{1pt}}}%
	\smash{\raisebox{-1ex}{\rule{1pt}{1.1\baselineskip}}}%
	\raisebox{0.5ex}{\rule{0.5ex}{1pt}}%
	}}1 
    {│}{{%
	\hphantom{\raisebox{0.5ex}{\rule{0.5ex}{1pt}}}%
	\smash{\raisebox{-1ex}{\rule{1pt}{1.1\baselineskip}}}%
	\hphantom{\raisebox{0.5ex}{\rule{0.5ex}{1pt}}}%
	}}1 
    {─}{{%
	\smash{\raisebox{0.5ex}{\rule{1ex}{1pt}}}%
	}}1 
    {└}{{%
	\hphantom{\raisebox{0.5ex}{\rule{0.5ex}{1pt}}}%
	\smash{\raisebox{0.5ex}{\rule{1pt}{0.5\baselineskip}}}%
	\raisebox{0.5ex}{\rule{0.5ex}{1pt}}%
	}}1 
    { }{ }1 
 }

\lstdefinelanguage[PSALTer]{Mathematica}[]{Mathematica}{
	morekeywords=[2]{brown,PacletInstall,DrazinInverse,mx,wl,Association,KernelID,OptionsPattern,OptionValue,Private,Head,DistributeDefinitions},
	morekeywords=[3]{blue,SymmetryOf,xAct,xAct`,xTensor,xCore,xPerm,xTras,SymManipulator,AllowUpperDerivatives,Antisymmetric,ChangeCovD,Christoffel,ConstantSymbolQ,ContractMetric,ContractMetrics,DefConstantSymbol,DefScalarFunction,DefTensor,delta,DependenciesOfTensor,IndicesOfVBundle,Labels,LI,LieDToCovD,MakeRule,NoScalar,OverDerivatives,ParamD,PD,PrintAs,Projected,ScalarFunctionQ,ScreenDollarIndices,SeparateMetric,SlotsOfTensor,Symmetric,SymmetryGroupOfTensor,ToCanonical,CommuteCovDs,xTensorQ,Zero,Tensors,ConstantSymbols,DefManifold,IndexRange,DefMetric,FlatMetric,SymCovDQ,DefCovD},
	morekeywords=[4]{green,a,Action,b,c,cartesian,CD,ChristoffelCD,ChristoffelPDcartesian,d,Def,DefField,DetG,e,EinsteinCCCD,EinsteinCCPDcartesian,EinsteinCD,En,Eps,epsilonG,etaDowncartesian,etaUpcartesian,f,g,G,h,i,j,k,KretschmannCD,l,m,M4,MaxLaurentDepth,Mo,n,o,p,P,ParticleSpectrum,PDcartesian,PoleResidue,PrintSourceAs,q,r,RicciCD,RicciPDcartesian,RicciScalarCD,RicciScalarPDcartesian,RiemannCD,RiemannPDcartesian,s,SchoutenCCCD,SchoutenCCPDcartesian,SchoutenCD,SchoutenPDcartesian,SymRiemannCD,SymRiemannPDcartesian,t,TangentM4,TetraG,TFRicciCD,TheoryName,TorsionCD,TorsionPDcartesian,u,v,V,w,WeylCD,x,y,z},
	morekeywords=[5]{red,Coupling1,Coupling2,Coupling3,Coupling4,Coupling5,Coupling6,ScalarField,VectorField,TwoFormField,MetricPerturbation,TetradPerturbation,SpinConnection,HigherSpinField,Connection,Alp0,Alp1,Alp2,Alp3,Alp4,Alp5,Alp6,Bet1,Bet2,Bet3,kT1,A0,A1,A2,A3,A4,A5,A6,A7,A8,A9,A10,A11,C0,C1,C2,C3,C4,C5,C6,C7,C8,C9,C10,C11,C12,C13,C14,C15,C16,MetricPerturbation}}

\lstdefinelanguage{Special}{%
morekeywords=[1]{%
PSALTer.m,ToNewCanonical.m,NewFramed.m,NewParallelSubmit.m,Colours.m,DefGeometry.m,ReMagnify.m,ParallelGrid.m,RaggedBlock.m,MonitorParallel.m,Diagnostic.m,Vectorize.m,ReloadPackage.m,ConstructSourceConstraints.m,ConstructUnitarityConditions.m,ConstructMasslessAnalysis.m,ConjectureNullSpace.m,NonTrivialDot.m,ToCovariantForm.m,CreateList.m,CleanNullVector.m,IsNullVectorOfSpace.m,EnsureLinearInCouplings.m,RemoveReferencesToMomentum.m,CommonNullVector.m,SummariseResults.m,UpdateTheoryAssociation.m,ValidateMethod.m,ConstructWaveOperator.m,ValidateTheoryName.m,ConstructMassiveAnalysis.m,PrintSecularEquation.m,PrintMasslessSpectrum.m,PrintMassiveSpectrum.m,PrintParticle.m,GetDiagram.m,StripFactors.m,CLIPrint.m,SummariseTheory.m,Status.m,ShowIfSmall.m,ResultsMosaic.m,MakeLabel.m,ShrinkPackRectangles.m,GraphicsMosaic.m,PrintSourceConstraints.m,SplitWignerGrid.m,DetailCell.m,PrintUnitarityConditions.m,WignerGrid.m,PrintSpectrum.m,ConstructSaturatedPropagator.m,ConstructLinearAction.m,IntermediateRules.m,ParameterisedNullVectorQ.m,CarefullyOrthogonalise.m,ManualPseudoInverse.m,ConsolidateFinalElement.m,GradualExpand.m,BatchExpanded.m,GradualExpandSubTask.m,SimplifyIfSmall.m,GrabExpression.m,InitialExpand.m,ConsolidateUnmakeSymbolic.m,UnmakeSymbolic.m,MakeSymbolic.m,ConjectureInverse.m,DistributeConjugate.m,MatrixFromSymbolic.m,FullyCanonicalise.m,ConstrainInLightcone.m,ExaminePoleOrder.m,MatrixToSymbolic.m,FullyExpandSources.m,MakeSaturatedMatrix.m,ReparameteriseSources.m,ExtractSecularEquation.m,MasslessAnalysisOfTotal.m,NullResidue.m,ExtractDenominator.m,ExtractReparameterisationMatrix.m,ExtractPart.m,Repartition.m,ExpressInLightcone.m,ConstructLightcone.m,MakeConstraintComponentList.m,DefFreeSourceVariables.m,AllIndependentComponents.m,RescaleNullVector.m,MakeFreeSourceVariables.m,ConstraintComponentToLightcone.m,IndependentComponents.m,ConvertLightcone.m,ValidateMaxLaurentDepth.m,ValidateLagrangian.m,NormaliseRescalings.m,CatalogueInvariant.m,CacheContexts.m,GenerateAnsatz.m,DefPlaceholderSpins.m,SimplifyMasses.m,MassiveGhost.m,MassiveAnalysisOfSector.m,IsolatePoles.m,PoleToSquareMass.m,CombineAssociations.m,GetHermitianPart.m,ConstructOperator.m,FourierLagrangian.m,DefField.m,ValidateSO3Irreps.m,DefAllComponentValues.m,DefSummary.m,ValidateInverseMode.m,ValidateSymmetryField.m,ValidateSpatial.m,ValidateInverseField.m,ValidateTraceless.m,ValidateSymmetryMode.m,RegisterFieldRank3Symmetric13.m,SummariseField.m,MakeAutomaticallyTraceless.m,DefSymbol.m,MakeUniqueQuadratic.m,MakeAutomaticallyNotAntisymmetric.m,RegisterFieldRank3Antisymmetric23.m,RegisterFieldRank1.m,RegisterFieldRank3Symmetric12.m,DefSO3Irrep.m,RegisterFieldRank3TotallySymmetric.m,RegisterFieldRank3Antisymmetric13.m,CombineRules.m,RegisterFieldRank3Symmetric23.m,AppendToField.m,RegisterFieldRank3TotallyAntisymmetric.m,DefFiducialField.m,RegisterFieldRank2.m,RegisterFieldRank2Antisymmetric.m,RegisterFieldRank3.m,RegisterFieldRank0.m,RegisterFieldRank3Antisymmetric12.m,ExpansionTable.m,FieldMosaic.m,DecompositionTable.m,RegisterFieldRank2Symmetric.m,ParticleSpectrum.m},morekeywords=[2]{%
init.wl},morekeywords=[3]{%
Documentation.nb},morekeywords=[4]{%
Documentation.pdf,FeynmanDiagramSpin1ParityEven.pdf,FeynmanDiagramHexic.pdf,FeynmanDiagramSpin0ParityEven.pdf,FeynmanDiagramQuadratic.pdf,FeynmanDiagramQuartic.pdf,FeynmanDiagramSpin0ParityOdd.pdf,FeynmanDiagramSpin2ParityOdd.pdf,FeynmanDiagramSpin3ParityOdd.pdf,FeynmanDiagramSpin2ParityEven.pdf,FeynmanDiagramSpin1ParityOdd.pdf,FeynmanDiagramSpin3ParityEven.pdf,GitHubLogo.pdf,FieldKinematics.pdf,ParticleSpectrograph.pdf,GitLabLogo.pdf},morekeywords=[5]{%
FieldKinematics.png,GitHubLogo.png,GitLabLogo.png,ParticleSpectrograph.png},morekeywords=[6]{%
Preamble.tex,FeynmanDiagramSpin1ParityOdd.tex,FeynmanDiagramSpin2ParityOdd.tex,FeynmanDiagramSpin0ParityOdd.tex,FeynmanDiagramSpin3ParityOdd.tex,FeynmanDiagramSpin3ParityEven.tex,FeynmanDiagramSpin1ParityEven.tex,FeynmanDiagramQuadratic.tex,FeynmanDiagramQuartic.tex,FeynmanDiagramHexic.tex,FeynmanDiagramSpin0ParityEven.tex,FeynmanDiagramSpin2ParityEven.tex},morekeywords=[7]{%
MakeDiagrams.sh,GitHubLogo.sh,GitLabLogo.sh},morekeywords=[8]{%
ASCIILogo.txt},morekeywords=[9]{%
GitHubLogo.svgz,GitLabLogo.svgz},morekeywords=[10]{%
LICENSE.md,README.md},morekeywords=[11]{cd,git,clone,r,cp,tree,mathematica,sudo,pacman,S},
keywordstyle=[1]\color{Green}\textbf,keywordstyle=[2]\color{Brown}\textbf,keywordstyle=[3]\color{RoyalBlue}\textbf,keywordstyle=[4]\color{Cyan}\textbf,keywordstyle=[5]\color{SkyBlue}\textbf,keywordstyle=[6]\color{Blue}\textbf,keywordstyle=[7]\color{Purple}\textbf,keywordstyle=[8]\color{Black}\textbf,keywordstyle=[9]\color{teal}\textbf,keywordstyle=[10]\color{teal}\textbf,keywordstyle=[11]\color{black}\textbf,alsoletter={./},basicstyle=\color{gray}\ttfamily,moredelim=[s][\color{gray}\textbf]{[user@system }{]}}

\definecolor{backing}{rgb}{0.88,1,0.88}

\newrobustcmd{\MinimumPower}[1]{%
	{n_{#1}} 
}
\newrobustcmd{\ComponentTotal}{%
	{\mathsf{j}_{\text{T}}} 
}
\newrobustcmd{\ConstraintTotal}{%
	{\mathsf{v}_{\text{T}}} 
}
\newrobustcmd{\MPl}{%
	{M_{\mathrm{Pl}}} 
}
\newrobustcmd{\IdentityJ}[1]{%
	{\mathsf{1}_{}}%
}
\newrobustcmd{\IdentityJP}[2]{%
	{\mathsf{1}_{{#1^#2}}}%
}
\newrobustcmd{\Jp}{%
	{J{'}}%
}
\newrobustcmd{\Pp}{%
	{P{'}}%
}
\newrobustcmd{\Sp}{%
	{s{'}}%
}
\newrobustcmd{\Ap}{%
	{a{'}}%
}
\newrobustcmd{\Bp}{%
	{b{'}}%
}
\newrobustcmd{\Cp}{%
	{c{'}}%
}
\newrobustcmd{\States}[4]{%
	{\tensor*{{#4}}{^{#3}_{{{#1}^{#2}}}}}%
}
\newrobustcmd{\FieldIndices}[2]{%
	{#2_#1}%
}
\newrobustcmd{\ParallelFieldIndices}[3]{%
	{{\overline{#3}}_{{#1^#2}}}%
}
\newrobustcmd{\FieldUp}[2]{%
	{\tensor{\zeta}{^{{\FieldIndices{#1}{#2}}}}}%
}
\newrobustcmd{\FieldDown}[2]{%
	{\tensor{\zeta}{_{\FieldIndices{#1}{#2}}}}%
}
\newrobustcmd{\SourceUp}[2]{%
	{\tensor{j}{^{{\FieldIndices{#1}{#2}}}}}%
}
\newrobustcmd{\SourceDown}[2]{%
	{\tensor{j}{_{\FieldIndices{#1}{#2}}}}%
}
\newrobustcmd{\FieldUpState}[5]{%
	{\tensor{\zeta\big(\States{#1}{#2}{#3}{#4}\big)}{^{{\ParallelFieldIndices{#1}{#2}{#5}}}}}%
}
\newrobustcmd{\SourceUpFullState}[5]{%
	{\tensor{j\big(\States{#1}{#2}{#3}{#4}\big)}{^{{\FieldIndices{#3}{#5}}}}}%
}
\newrobustcmd{\FieldDownState}[5]{%
	{\tensor{\zeta\big(\States{#1}{#2}{#3}{#4}\big)}{_{\ParallelFieldIndices{#1}{#2}{#5}}}}%
}
\newrobustcmd{\FieldDownFullState}[5]{%
	{\tensor{\zeta\big(\States{#1}{#2}{#3}{#4}\big)}{_{\FieldIndices{#3}{#5}}}}%
}
\newrobustcmd{\WaveOperatorTensorUpDown}[4]{%
	{\tensor{\mathcal{O}}{^{{\FieldIndices{#1}{#3}}}_{\FieldIndices{#2}{#4}}}}%
}
\newrobustcmd{\Normalisation}[4]{%
	{c\big(\States{#1}{#2}{#3}{#4}\big)}%
}
\newrobustcmd{\SPODownUp}[8]{%
	{\tensor{%
	\mathcal{P}%
	\big(\States{#1}{#2}{#3}{#5},\States{#1}{#2}{#4}{#6}\big)%
	}{_{\FieldIndices{#3}{#7}}^{{\FieldIndices{#4}{#8}}}}}%
}
\newrobustcmd{\SPOUpDown}[8]{%
	{\tensor{%
	\mathcal{P}%
	\big(\States{#1}{#2}{#3}{#5},\States{#1}{#2}{#4}{#6}\big)%
	}{^{{\FieldIndices{#3}{#7}}}_{\FieldIndices{#4}{#8}}}}%
}
\newrobustcmd{\ReducedSPODownUp}[6]{%
	{\tensor{%
	\mathcal{P}%
	\big(\States{#1}{#2}{#3}{#4}\big)%
	}{_{\ParallelFieldIndices{#1}{#2}{#5}}^{{\FieldIndices{#3}{#6}}}}}%
}
\newrobustcmd{\ReducedSPOUpDown}[6]{%
	{\tensor{%
	{\mathcal{P}%
	\big({\States{#1}{#2}{#3}{#4}}\big)}%
	}{^{{\ParallelFieldIndices{#1}{#2}{#5}}}_{{\FieldIndices{#3}{#6}}}}}%
}
\newrobustcmd{\GaugeVarying}[3]{%
	{\tensor*{g}{_{#3_{{#1^#2}}}}}%
}
\newrobustcmd{\GaugeVaryingConj}[3]{%
	{\tensor*{g}{^*_{#3_{{#1^#2}}}}}%
}
\newrobustcmd{\GaugeFixing}[3]{%
	{\tensor*{h}{_{#3_{{#1^#2}}}}}%
}
\newrobustcmd{\Mass}[3]{%
	{\tensor*{M}{_{#3_{{#1^#2}}}}}%
}
\newrobustcmd{\SquareMass}[3]{%
	{\tensor*{M}{^2_{#3_{{#1^#2}}}}}%
}
\newrobustcmd{\NullVectors}[3]{%
	{#3_{{#1^#2}}}%
}
\newrobustcmd{\NullVector}[3]{%
	{\tensor*{\mathsf{v}}{_{#3_{{#1^#2}}}}}%
}
\newrobustcmd{\NullVectorComponents}[5]{%
	{\tensor{{\left[{\NullVector{#1}{#2}{#3}}\right]}}{_{{\States{#1}{#2}{#4}{#5}}}}}%
}
\newrobustcmd{\ConstraintNullVector}[1]{%
	{\tensor*{\mathsf{v}}{_{#1}}}%
}
\newrobustcmd{\ConstraintNullVectorConj}[1]{%
	{\tensor*{\mathsf{v}}{^\dagger_{#1}}}%
}
\newrobustcmd{\ReducedEigenVector}[1]{%
	{\tensor*{\mathsf{u}}{_{#1}}}%
}
\newrobustcmd{\ReducedEigenVectorConj}[1]{%
	{\tensor*{\mathsf{u}}{^\dagger_{#1}}}%
}
\newrobustcmd{\ReducedEigenValue}[1]{%
	{\tensor*{\lambda}{_{#1}}}%
}
\newrobustcmd{\ReducedSource}{%
	{\tensor*{\mathsf{K}}{}}%
}
\newrobustcmd{\ReducedSourceConj}{%
	{\tensor*{\mathsf{K}}{^\dagger}}%
}
\newrobustcmd{\ReducedPropagator}{%
	{\mathsf{\Pi}}%
}
\newrobustcmd{\SpinMin}{%
	{J_{\mathrm{F}}}%
}
\newrobustcmd{\SpinMax}{%
	{J_{\mathrm{L}}}%
}
\newrobustcmd{\ParityMin}{%
	{P_{\mathrm{F}}}%
}
\newrobustcmd{\ParityMax}{%
	{P_{\mathrm{L}}}%
}
\newrobustcmd{\FieldMin}{%
	{X_{\mathrm{F}}}%
}
\newrobustcmd{\FieldMax}{%
	{X_{\mathrm{L}}}%
}
\newrobustcmd{\VectorMin}{%
	{a_{\mathrm{F}}}%
}
\newrobustcmd{\VectorMax}{%
	{a_{\mathrm{L}}}%
}
\newrobustcmd{\ZeroVector}{%
	{\mathsf{0}}%
}
\newrobustcmd{\ConstraintMatrix}{%
	{\mathsf{V}}%
}
\newrobustcmd{\SourceVector}{%
	{\mathsf{J}}%
}
\newrobustcmd{\NullVectorConj}[3]{%
	{\tensor*{\mathsf{v}}{^\dagger_{#3_{{#1^#2}}}}}%
}
\newrobustcmd{\WaveOperator}{%
	{\tensor{\mathsf{O}}{}}%
}
\newrobustcmd{\WaveOperatorConj}{%
	{\tensor{\mathsf{O}}{^\dagger}}%
}
\newrobustcmd{\MoorePenrose}{%
	{\tensor{\mathsf{O}}{^+}}%
}
\newrobustcmd{\MoorePenroseJP}[2]{%
	{\tensor*{\mathsf{O}}{^+_{{#1^#2}}}}%
}
\newrobustcmd{\WaveOperatorJP}[2]{%
	{\tensor*{\mathsf{O}}{_{{#1^#2}}}}%
}
\newrobustcmd{\WaveOperatorJPComponents}[6]{%
	{\tensor{\left[\WaveOperatorJP{#1}{#2}\right]}{_{\States{#1}{#2}{#3}{#5}\States{#1}{#2}{#4}{#6}}}}%
}
\newrobustcmd{\Propagator}{%
	{\tensor{\mathsf{P}}{}}%
}
\newrobustcmd{\Similarity}{%
	{\tensor{\mathsf{U}}{}}%
}
\newrobustcmd{\Source}{%
	{\tensor{\mathsf{j}}{}}%
}
\newrobustcmd{\SourceConj}{%
	{\tensor{\mathsf{j}}{^\dagger}}%
}
\newrobustcmd{\Field}{%
	{\tensor{\upzeta}{}}%
}
\newrobustcmd{\FieldConj}{%
	{\tensor{\upzeta}{^\dagger}}%
}
\newrobustcmd{\Powers}[3]{%
	{#3_{{#1^#2}}}%
}
\newrobustcmd{\Masses}[3]{%
	{#3_{{#1^#2}}}%
}
\newrobustcmd{\NewRes}[2]{%
	{\ \mathop{\mathrm{Res}}_{#1\mapsto#2}}%
}
\newrobustcmd{\AMat}[1]{%
	{\mathsf{A}}%
}
\newrobustcmd{\AMatT}[1]{%
	{\mathsf{A}}^{\text{T}}%
}
\newrobustcmd{\AMatU}[1]{%
	\tensor[^{\text{s}}]{\mathsf{A}}{}%
}
\newrobustcmd{\AMatUt}[1]{%
	\tensor[^{\text{s}}]{\tilde{\mathsf{A}}}{}%
}
\newrobustcmd{\AMatO}[1]{%
	\tensor[^{\text{a}}]{\mathsf{A}}{}%
}
\newrobustcmd{\BMat}[1]{%
	{\mathsf{B}}%
}
\newrobustcmd{\BMatT}[1]{%
	{\mathsf{B}}^{\text{T}}%
}
\newrobustcmd{\BMatU}[1]{%
	\tensor[^{\text{s}}]{\mathsf{B}}{}%
}
\newrobustcmd{\BMatO}[1]{%
	\tensor[^{\text{a}}]{\mathsf{B}}{}%
}
\newrobustcmd{\BMatOt}[1]{%
	\tensor[^{\text{a}}]{\tilde{\mathsf{B}}}{}%
}
\newrobustcmd{\CMat}[1]{%
	{\mathsf{C}}%
}
\newrobustcmd{\CMatT}[1]{%
	{\mathsf{C}}^{\text{T}}%
}
\newrobustcmd{\CMatU}[1]{%
	\tensor[^{\text{s}}]{\mathsf{C}}{}%
}
\newrobustcmd{\CMatUt}[1]{%
	\tensor[^{\text{s}}]{\tilde{\mathsf{C}}}{}%
}
\newrobustcmd{\HigherSpinField}[1]{%
  \tensor{h}{#1}
}
\newrobustcmd{\HigherSpinSource}[1]{%
  \tensor{F}{#1}
}
\newrobustcmd{\MAGA}[1]{%
  \tensor{\Gamma}{#1}
}
\newrobustcmd{\MAGF}[1]{%
	\tensor{\mathcal{R}}{#1}
}
\newrobustcmd{\MAGFa}[1]{%
	\tensor{\mathcal{R}}{^{(14)}#1}
}
\newrobustcmd{\MAGFb}[1]{%
	\tensor{\mathcal{R}}{^{(13)}#1}
}
\newrobustcmd{\MAGT}[1]{%
	\tensor{\mathcal{T}}{#1}
}
\newrobustcmd{\MAGQ}[1]{%
	\tensor{\mathcal{Q}}{#1}
}
\newrobustcmd{\MAGQt}[1]{%
	\tensor{\tilde{\mathcal{Q}}}{#1}
}
\newrobustcmd{\B}[1]{%
	\tensor{\mathcal{B}}{#1}%
}
\newrobustcmd{\PD}[1]{%
	\tensor{\partial}{#1}%
}
\newrobustcmd{\BConj}[1]{%
	\tensor{{B^\dagger}}{#1}%
}
\newrobustcmd{\N}[1]{%
	\tensor{n}{#1}%
}
\newrobustcmd{\J}[1]{%
	\tensor{\mathcal{J}}{#1}%
}
\newrobustcmd{\En}[1]{%
	\mathcal{E}%
}
\newrobustcmd{\Mo}[1]{%
	p%
}
\newrobustcmd{\JConj}[1]{%
	\tensor{{J^\dagger}}{#1}%
}
\newrobustcmd{\K}[1]{%
	\tensor{K}{_#1}%
}
\newrobustcmd{\KConj}[1]{%
	\tensor*{{K^{\dagger}}}{_{#1}}%
}
\newrobustcmd{\glfourr}{%
  {\mathrm{GL}(4,\mathbb{R})}%
}
\newrobustcmd{\sltwoc}{%
  {\mathrm{SL}(2,\mathbb{C})}%
}
\newrobustcmd{\poincare}{%
  {\mathbb{R}^{1,3}\rtimes\mathrm{SO}^+(1,3)}%
}
\newrobustcmd{\poincaref}{%
  {\mathrm{P}(1,3)}%
}
\newrobustcmd{\weyl}{%
  {\mathrm{W}(1,3)}%
}
\newrobustcmd{\conformal}{%
  {\mathrm{C}(1,3)}%
}
\newrobustcmd{\diffeomorphism}{%
  {\mathbb{R}^{1,3}}%
}
\newrobustcmd{\soonethree}{%
  {\mathrm{SO}^+(1,3)}%
}
\newrobustcmd{\othree}{%
  {\mathrm{SO}(3)}%
}
\newrobustcmd{\sothree}{%
  {\mathrm{SO}(3)}%
}
\newrobustcmd{\sotwo}{%
  {\mathrm{SO}(2)}%
}
\newrobustcmd{\suthreec}{%
  {\mathrm{SU}(3)_{\text{c}}}%
}
\newrobustcmd{\sutwol}{%
  {\mathrm{SU}(2)_{\text{L}}}%
}
\newrobustcmd{\uoney}{%
  {\mathrm{U}(1)_{\text{Y}}}%
}
\newrobustcmd{\uone}{%
  {\mathrm{U}(1)}%
}
\newrobustcmd{\uoneem}{%
  {\mathrm{U}(1)_{\text{em}}}%
}
\newrobustcmd{\sutwo}{%
  {\mathrm{SU}(2)}%
}
\newrobustcmd{\Planck}{%
	{M_{\text{Pl}}}%
}
\newrobustcmd{\ovl}[1]{%
	\overline{#1}%
}
\newrobustcmd{\FieldG}[1]{%
	\tensor{g}{#1}
}

\DeclareTOCStyleEntry[numwidth=20pt,linefill=\bfseries\TOCLineLeaderFill]{tocline}{section}
\DeclareTOCStyleEntry[entryformat=\textit,numwidth=10pt,linefill=\TOCLineLeaderFill]{tocline}{subsection}
\DeclareTOCStyleEntry[entryformat=\textit,numwidth=10pt,linefill=\TOCLineLeaderFill]{tocline}{subsubsection}

\usepackage{hyperref}
\hypersetup{%
     colorlinks = true,%
     linkcolor = Blue,%
     citecolor = Blue,%
     filecolor = Blue,%
     urlcolor = Blue%
     }%
\usepackage[capitalize,nameinlink]{cleveref}
\crefname{paragraph}{paragraph}{paragraphs}
\Crefname{paragraph}{Paragraph}{Paragraphs}

\begin{document}

\title{\PSALTer{}: Particle Spectrum for Any Tensor Lagrangian}

\author{W. Barker}
\email{wb263@cam.ac.uk}
\affiliation{Astrophysics Group, Cavendish Laboratory, JJ Thomson Avenue, Cambridge CB3 0HE, UK}
\affiliation{Kavli Institute for Cosmology, Madingley Road, Cambridge CB3 0HA, UK}

\author{C. Marzo}
\email{carlo.marzo@kbfi.ee}
\affiliation{Laboratory for High Energy and Computational Physics, NICPB, R\"{a}vala 10, Tallinn 10143, Estonia}

\author{C. Rigouzzo}
\email{claire.rigouzzo@kcl.ac.uk}
\affiliation{Laboratory for Theoretical Particle Physics and Cosmology, King's College London, London WC2R 2LS, UK}

\begin{abstract}
	We present the \PSALTer{} software for efficiently computing the mass and energy of the particle spectrum for any (e.g. higher-rank) tensor field theory in the \WolframLanguage{}. The user must provide a Lagrangian density which is expanded quadratically in the fields around a Minkowski vacuum, is linear in the coupling coefficients, and otherwise built from the partial derivative and Minkowski metric. \PSALTer{} automatically computes the spin-projection operators, saturated propagator, bare masses, residues of massive and massless poles and overall unitarity conditions in terms of the coupling coefficients. The constraints on the source currents and total number of gauge symmetries are produced as a by-product. We provide examples from scalar, vector, tensor and gauge theories of gravity. Each example, including spectra of higher-spin modified gravity theories, may be obtained on a personal computer in a matter of minutes. The software is also parallelised for use on high-performance computing resources. The initial release allows for parity-preserving operators constructed from fields of up to rank three: this functionality will be extended in future versions. \PSALTer{} is a contribution to the \xAct{} project.
\end{abstract}

\maketitle

\tableofcontents

\section{Introduction}\label{Introduction}
\paragraph*{Particle spectra} In theoretical physics it is often necessary to extract the propagating degrees of freedom (d.o.f) from a specified model. For higher-rank tensor fields, unlike for the simple vector bosons of the standard model, this can quickly become very serious undertaking. If the model admits a perturbative, Lorentz-covariant quantum field theory (QFT) near Minkowski spacetime, then the free (non-interacting) QFT is completely specified by its propagator structure, or \emph{particle spectrum}: a statement of the polarisations of propagating particles and their bare masses (where nonzero). Perturbatively, the particle spectrum is encoded by the leading-order Lagrangian, which is quadratic in the fields. The cubic and higher-order corrections constitute the interacting part of the QFT, i.e. the Feynman vertices and all non-linear aspects of the model. Without these vertices, the free QFT makes only tree-level predictions which can just as well be obtained by classical means. Indeed, by varying the \emph{quadratic} part of the Lagrangian we obtain the \emph{linear} part of the field equations, and these will also yield the particle spectrum. Stability of these field equations corresponds to the elimination of ghostly and tachyonic d.o.f from the particle spectrum, which is a fundamental requirement for most physical vacua.

\paragraph*{Spin and parity} Massive particles in Minkowski spacetime are associated with quantum states of definite \emph{spin}~$J$ and \emph{parity}~$P$ (denoted \emph{spin-parity}~$J^P$). In the massless limit,~$J^P$ is superceded by the \emph{helicity} quantum number. Higher-rank tensor fields typically decompose under~$\othree$ into a large number of~$J^P$ states, including higher-spin states with~$J\geq 1$. A simple-looking Lagrangian may contain a complicated collection of bilinear operators in these~$J^P$ states. 

\paragraph*{The algorithm} In such cases, particle physics provides a sophisticated algorithm for extracting the particle spectrum~\cite{Aurilia:1969bg,Buoninfante:2016iuf}. The algorithm uses \emph{spin-projection operators} (SPOs) to express the wave operator in a canonical matrix form. The null space of the wave operator encodes all the gauge symmetries of the free QFT, and the associated constraints on any external source currents to which the theory is coupled. The saturated (physical) propagator is the (pseudo)inverse of this matrix. The roots of the (pseudo)determinant encode the positions of the propagator poles, i.e. the bare masses whose reality ensures the absence of tachyons. The positivity of the residues over these poles meanwhile ensures the absence of ghosts. In this way, the SPO algorithm reduces the particle spectrum of any given theory to a unique problem in linear algebra.

\paragraph*{Demand in gravity} The SPO algorithm is especially useful for studying theories of gravity. Even in the most conservative case, Einstein's general relativity (GR)~\cite{Einstein:1915} already propagates a spin-two particle from within a rank-two metric tensor field. Due to Lovelock's theorem~\cite{Einstein:1915,Lovelock:1972vz}, many alternative models couple additional tensorial fields of higher or lower rank to the metric~\cite{Clifton:2011jh}. Extra fields can contribute dozens of new, independent~$J^P$ sectors, necessitating a systematic approach to the propagator structure. 

\paragraph*{Previous advances} The SPO algorithm becomes challenging, but also rewarding, when applied to field theories of rank three or higher. Several such applications can be found in the literature, which are particularly impressive because they appear to have been computed manually~\cite{VanNieuwenhuizen:1973fi,Neville:1978bk,Neville:1979rb,Sezgin:1981xs,Sezgin:1979zf,Kuhfuss:1986rb,Karananas:2014pxa,Karananas:2016ltn,Mendonca:2019gco,Percacci:2019hxn,Marzo:2021esg,Marzo:2021iok,Mikura:2023ruz,Mikura:2024mji} --- see also more general works~\cite{Aurilia:1969bg,Buoninfante:2016iuf}. In recent years, Lin, Hobson and Lasenby used the \MathGR{}~\cite{Wang:2013mea} package for \Mathematica{}~\cite{Mathematica:2024} for computer-assisted analyses in~\cite{Lin:2018awc,Lin:2019ugq,Lin:2020phk,Lin:2020fuo}. Hitherto, however, no computer algebra implementation of the algorithm was made available to the community.

\paragraph*{Aims of this paper} We present a robust \WolframLanguage{} framework for the SPO algorithm. Particle Spectrum for Any Tensor Lagrangian (\PSALTer{}) is a package for the \Mathematica{} software system. It is not a derivative of the previous implementation in~\cite{Lin:2018awc,Lin:2019ugq,Lin:2020phk,Lin:2020fuo}, but a fresh, open-source contribution to the \xAct{} tensor computer algebra project~\cite{Martin-Garcia:2007bqa,Martin-Garcia:2008ysv,Martin-Garcia:2008yei,Nutma:2013zea,xCore:2018,xCoba:2020,SymManipulator:2021}. Previous iterations of \PSALTer{} were used in~\cite{Barker:2023bmr,Barker:2024ydb,Barker:2024dhb}. \PSALTer{} is available for download at the public \GitHub{} repository \href{https://github.com/wevbarker/PSALTer}{\texttt{github.com/wevbarker/PSALTer}}. Current functionality is limited to parity-preserving actions depending on fields which are tensors of up to rank three. The aim of this paper is to provide an illustration of (and user manual for) this current functionality.

\paragraph*{Structure of this paper} In~\cref{SymbolicImplementation}, we present \PSALTer{}. We use the software to obtain particle spectrographs of theories formed from a scalar field, a vector field, and rank-two and rank-three tensor fields with and without index symmetries. The examples provided are mostly weak-field approximations to motivated physical theories, facilitating comparisons with the literature. Having illustrated the software, in~\cref{TheoreticalDevelopment} we provide a theoretical statement of the underlying SPO algorithm. As an illustration, we provide worked examples from vector field theory. Conclusions follow in~\cref{Conclusions}, with technical appendices.

\paragraph*{When to use spectral analysis} We are concerned with tensor theories which can be expanded to quadratic order around the background of Minkowski spacetime. Such theories are associated with a \emph{free} (i.e. quadratic, non-interacting) action 
\begin{align}\label{BasicPositionAction}
	S_{\text{F}}=\int\mathrm{d}^4x\ 
	\sum_X
\FieldDown{X}{\mu}
	\Big[&
		\sum_Y\WaveOperatorTensorUpDown{X}{Y}{\mu}{\nu}\FieldUp{Y}{\nu}
	-\SourceUp{X}{\mu}\Big],
\end{align}
where~\cref{BasicPositionAction} contains the following ingredients:
\begin{enumerate}
	\item The fields~$\FieldDown{X}{\mu}$ are real tensors. Distinct fields carry the index~$X$, each field has some collection of spacetime indices~$\FieldIndices{X}{\mu}$, perhaps with some symmetry. Implemented index symmetries for~$X$ are the scalar~$\tensor{\zeta}{}$, the vector~$\tensor{\zeta}{_{\mu}}$, the asymmetric tensor~$\tensor{\zeta}{_{\mu\nu}}$ or its symmetrizations~$\tensor{\zeta}{_{\mu\nu}}\equiv\tensor{\zeta}{_{[\mu\nu]}}$ or~$\tensor{\zeta}{_{(\mu\nu)}}$, and the asymmetric rank-three field~$\tensor{\zeta}{_{\mu\nu\sigma}}$ or its symmetrizations~$\tensor{\zeta}{_{\mu\nu\sigma}}\equiv\tensor{\zeta}{_{[\mu\nu\sigma]}}$,~$\tensor{\zeta}{_{(\mu\nu\sigma)}}$,~$\tensor{\zeta}{_{[\mu\nu]\sigma}}$,~$\tensor{\zeta}{_{[\mu|\nu|\sigma]}}$,~$\tensor{\zeta}{_{\mu[\nu\sigma]}}$,~$\tensor{\zeta}{_{(\mu\nu)\sigma}}$,~$\tensor{\zeta}{_{(\mu|\nu|\sigma)}}$ or~$\tensor{\zeta}{_{\mu(\nu\sigma)}}$.
	\item The wave operator~$\WaveOperatorTensorUpDown{X}{Y}{\mu}{\nu}$ is a real differential operator constructed from~$\tensor{\eta}{_{\mu\nu}}$ and~$\tensor{\partial}{_{\mu}}$ and even powers of~$\tensor{\epsilon}{_{\mu\nu\sigma\lambda}}$ (i.e. it is parity-preserving), linearly parameterised by a collection of coupling coefficients.
	\item The real source currents~$\SourceUp{X}{\mu}$ are conjugate to the fields~$\FieldDown{X}{\mu}$. They encode all external interactions to second order in fields, whilst keeping the external dynamics completely anonymous.
\end{enumerate}
For theories of the form in~\cref{BasicPositionAction}, the SPO algorithm detailed in~\cref{TheoreticalDevelopment} applies and the \PSALTer{} package illustrated in~\cref{SymbolicImplementation} may be used. Of course, spectra can also be obtained for more exotic theories (e.g. parity-odd, complex fields, etc.), but these require the algorithm to be modified beyond its minimal form --- we will discuss such extensions in~\cref{Conclusions}.

\paragraph*{Conventions} The \PSALTer{} software assumes a flat, four-dimensional spacetime and flat metric tensor~$\tensor{\eta}{_{\mu\nu}}\equiv\mathrm{diag}\left(1,-1,-1,-1\right)$. In some inertial frame, Greek indices are associated with Cartesian coordinates. The totally antisymmetric tensor is~$\tensor{\epsilon}{_{\mu\nu\sigma\lambda}}$, with~$\tensor{\epsilon}{_{0123}}\equiv 1$. We also borrow the \emph{weak equality} notation `$\approx$' from canonical analysis: this means an equality which is true \emph{on-shell} with respect to the field equations. Further conventions are introduced as needed.

\section{Examples with code}\label{SymbolicImplementation}
\paragraph*{Syntax highlighting} This section and~\cref{TheoreticalDevelopment} contain code listings with syntax highlighting. Symbols defined in \WolframLanguage{} (i.e. \Mathematica{}) appear in \lstinline!brown!, those defined in \xAct{} appear in \lstinline!blue!, those defined in \PSALTer{} appear in \lstinline!green!, and those to be defined in an example user session appear in \lstinline!red!. The symbol `\lstinline!In[#]:=!' denotes the start of a new input cell in a \Mathematica{} notebook, and `\lstinline!Out[#]:=!' denotes the start of an output cell. Comments are shown in \lstinline!(*gray*)! and strings in \lstinline!"orange"!. With the exception of error messages, all of the outputs from the two functions provided by \PSALTer{} (namely \lstinline!DefField! and \lstinline!ParticleSpectrum!) take the form of \texttt{.pdf} figures. These are prepared to a high-enough standard that they can be used directly in scientific publications, eliminating the need for \LaTeX.

\paragraph*{Loading the software} Instructions for obtaining and installing the software may be found in~\cref{Install}. The package is designed to be used in the \Mathematica{} Front End, or graphical user interface (GUI). In a fresh notebook, the package will typically be loaded with the \lstinline!Get! command:
\begin{lstlisting}
In[#]:= Get["xAct`PSALTer`"]; 
\end{lstlisting}
This will first load various other \Mathematica{} dependencies, specifically the five \xAct{} packages \xTensor{}~\cite{Martin-Garcia:2007bqa,Martin-Garcia:2008yei}, \SymManipulator{}~\cite{SymManipulator:2021}, \xPerm{}~\cite{Martin-Garcia:2008ysv}, \xCore{}~\cite{xCore:2018}, \xTras{}~\cite{Nutma:2013zea} and \xCoba{}~\cite{xCoba:2020}, and also the package \RectanglePacking{}~\cite{RectanglePacking:2024}. Having loaded the dependencies, the \PSALTer{} package will load its own sources and automatically pre-define the geometric environment. This environment is made up from a four-dimensional Lorentzian manifold \lstinline!M4!, flat-space metric \lstinline!G!, antisymmetric tensor \lstinline!epsilonG! and covariant derivative \lstinline!CD!. All the lower-case Latin letters \lstinline!a!, \lstinline!b!, etc., through to \lstinline!z! are reserved for holonomic indices derived from Cartesian coordinates on \lstinline!M4!. These indices format as their Greek counterparts~$\alpha$,~$\beta$, etc., through to~$\zeta$ in the output. Thus, \lstinline!G[-m,-n]! corresponds to~$\tensor{\eta}{_{\mu\nu}}$ and \lstinline!CD[-m]@! corresponds to~$\PD{_\mu}$, whilst \lstinline!epsilonG[-m,-n,-s,-l]! corresponds to~$\tensor{\epsilon}{_{\mu\nu\sigma\lambda}}$. For those familiar with \xTensor, the \PSALTer{} geometric environment corresponds to that which would result from the following use of the \lstinline!DefManifold! and \lstinline!DefMetric! commands:
\begin{lstlisting}
In[#]:=DefManifold[M4, 4, IndexRange[{a,z}]];DefMetric[-1, G[-a,-c], CD, {",","\[PartialD]"}, PrintAs->"\[Eta]", FlatMetric->True, SymCovDQ->True];
\end{lstlisting}
Because they are executed automatically at startup, these commands do not need to be run at any time by the user. No attempt should be made to tamper with this geometry, at the risk of compromising the \PSALTer{} package.

\begin{figure*}[t!]
	\includegraphics[width=\linewidth]{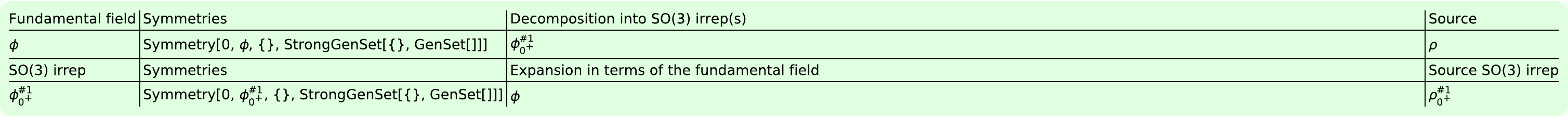}
	\caption{The declaration of the single scalar field \lstinline!ScalarField! which contains only a single~$\sothree{}$ irrep in the form of the~$J^P=0^+$ mode. This mode is formatted as~$\tensor*{\phi}{_{0^+}^{\#1}}$. Further~$0^+$ irreps contained within the same tensor field would carry the labels~$\#2$,~$\#3$ etc. Higher-rank tensors may contain many irreps, and in such cases figures such as this start to become useful (we will see this in~\cref{FieldKinematicsVectorField,FieldKinematicsTwoFormField,FieldKinematicsMetricPerturbation,FieldKinematicsTetradPerturbation,FieldKinematicsSpinConnection,FieldKinematicsHigherSpinField,FieldKinematicsConnection}). There are no index symmetries. In general, the contents of the second columns are the outputs of calls to \lstinline!SymmetryOf! from \xTensor{}. These definitions are used in~\cref{ParticleSpectrographMasslessScalarTheory,ParticleSpectrographMassiveScalarTheory}.}
\label{FieldKinematicsScalarField}
\end{figure*}

\subsection{Scalar theory}\label{ScalarTheory}
\paragraph*{Kinematics} The simplest theory to consider is that of a single, real scalar field~$\phi$. The most general parity-preserving scalar action is 
\begin{equation}\label{ScalarTheoryAction}
	S_{\text{F}}=\int\mathrm{d}^4x\left[\alpha\PD{_\mu}\phi\PD{^\mu}\phi-\beta\phi^2+\phi\rho\right],
\end{equation}
where~$\alpha$ and~$\beta$ are parameters and~$\rho$ is the source conjugate to~$\phi$. It is clear that~\cref{ScalarTheoryAction} conforms to the format specified in~\cref{BasicPositionAction}. We will denote~$\phi$ with the new symbol \lstinline!ScalarField!, which must be defined using the \lstinline!DefField! command:
\begin{lstlisting}
In[#]:= DefField[ScalarField[], PrintAs -> "\[Phi]", PrintSourceAs -> "\[Rho]"]; 
\end{lstlisting}
Notice how the syntax is very similar to \lstinline!DefTensor! from \xTensor{}, including the use of the \lstinline!PrintAs! option for formatting the symbol as a Greek letter. The conjugate source~$\rho$ is defined internally by \lstinline!DefField! and should never be needed by the user, however since it appears in the output its formatting should be specified with the additional option \lstinline!PrintSourceAs!. Unlike in the case of \lstinline!DefTensor!, there is no need to specify an underlying manifold: all fields defined using \lstinline!DefField! are assumed to be associated with \lstinline!M4!. The output of \lstinline!DefField! is shown in~\cref{FieldKinematicsScalarField}. This vectorised figure is automatically exported to \texttt{FieldKinematicsScalarField.pdf}. If the notebook is saved, then the export path should be set to the output of the \lstinline!NotebookDirectory! command. If the notebook is not saved, then \texttt{.pdf} files will save to the path specified by the output from \lstinline!Directory!. To build theories such as in~\cref{ScalarTheoryAction}, the coupling coefficients~$\alpha$ and~$\beta$ are defined as \lstinline!Coupling1! and \lstinline!Coupling2! using the \lstinline!DefConstantSymbol! command from \xTensor{}:
\begin{lstlisting}
In[#]:= DefConstantSymbol[Coupling1, PrintAs -> "\[Alpha]"]; DefConstantSymbol[Coupling2, PrintAs -> "\[Beta]"]; DefConstantSymbol[Coupling3, PrintAs -> "\[Gamma]"]; 
\end{lstlisting}
Here, we have also taken the opportunity to define the extra coupling \lstinline!Coupling3! which will format as~$\gamma$. As a quick aside: when entering commands into the notebook Front End, some users may prefer to use drop-down menus to directly select formatted symbols. Since \PSALTer{} was developed from the command line, the examples in this paper use more cumbersome string-based representations such as \lstinline!"\[Alpha]"!.

\begin{figure}[t!]
	\includegraphics[width=\linewidth]{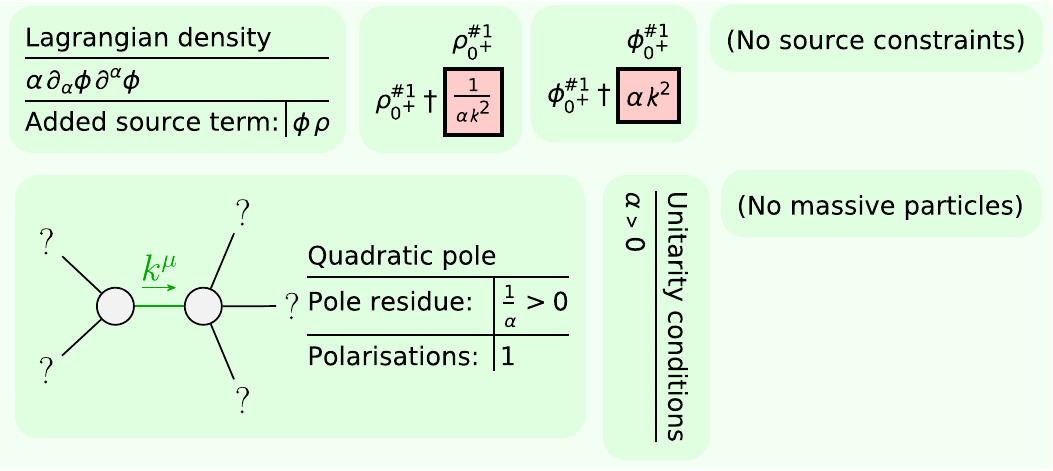}
	\caption{The particle spectrum of a massless scalar field obtained by setting~$\beta=0$ in~\cref{ScalarTheoryAction}. The unitarity condition~$\alpha>0$ is the expected no-ghost condition. The theory contains no gauge symmetries, and therefore no source constraints. There are no massive particles, because there are no explicit mass scales in the theory. The wave operator and saturated propagator appear as matrices --- in this case of dimension~$1\times 1$ because there is only a single~$\sothree{}$ irrep. The \emph{red} background signifies that the matrix element is an interaction between \emph{positive} parity states. Note also that some information boxes are automatically rotated, to save space on the page. All quantities are defined in~\cref{FieldKinematicsScalarField}.}
\label{ParticleSpectrographMasslessScalarTheory}
\end{figure}

\paragraph*{Massless scalar} To begin with, the simplest version of~\cref{ScalarTheoryAction} is the massless scalar, i.e.~$\beta=0$\footnote{In the conventions of this paper, it would make sense to further set~$\alpha=1/2$, however we will leave~$\alpha$ free because \PSALTer{} requires that all terms in the action be linearly parameterised by symbolic coupling coefficients.}. We use the \lstinline!ParticleSpectrum! command to perform the calculation:
\begin{lstlisting}
In[#]:= ParticleSpectrum[Coupling1 * CD[-a][ScalarField[]] * CD[a][ScalarField[]], TheoryName -> "MasslessScalarTheory", Method -> "Hard", MaxLaurentDepth -> 3]; 
\end{lstlisting}
The first argument is the Lagrangian density, without the source coupling (which is always included automatically). The option \lstinline!TheoryName! allocates a string which will be used to label the output file. The option \lstinline!Method! (which is already a built-in symbol in \Mathematica{}) can be either of the strings \lstinline!"Easy"! or \lstinline!"Hard"!, where the latter is recommended for more complicated theories. The \lstinline!"Easy"! mode relies internally on the built-in symbol \lstinline!PseudoInverse! to invert the (possibly singular) wave operator matrices. However, the \Mathematica{} implementation of \lstinline!PseudoInverse! is very limited, and unsuitable for large symbolic matrices\footnote{This is also true of the alternative algorithm \lstinline!DrazinInverse!.}. The \lstinline!"Hard"! mode relies internally on a custom algorithm for the Moore--Penrose pseudoinverse~\cite{Moore:1920,Penrose:1955}, which is parallelised to take advantage of available CPU cores. The \lstinline!Method->"Hard"! mode is slower for simple theories, and the \lstinline!Method->"Easy"! mode may be insufficient for complicated theories\footnote{We will see in~\cref{HigherSpinBosons} that \lstinline!Method->"Easy"! also permits the user to bypass the restriction that Lagrangia be linearly parameterised by coupling coefficients.}. The option \lstinline!MaxLaurentDepth! specifies the maximum order of null poles to be investigated in the propagator. Setting \lstinline!MaxLaurentDepth->1! is sufficient for determining the \emph{quadratic} poles, corresponding to potentially healthy (depending on the residue) propagating massless d.o.f. Setting \lstinline!MaxLaurentDepth->2! or \lstinline!MaxLaurentDepth->3! (the maximum) triggers the investigation of any \emph{quartic} or \emph{hexic} poles: these poles are always sick, but the process of finding them may become expensive for more complicated models. For further details about the massless spectrum, see~\cref{MasslessSpectrum}. The output is shown in~\cref{ParticleSpectrographMasslessScalarTheory}. This vectorised figure is automatically exported to \texttt{ParticleSpectrographMasslessScalarTheory.pdf} in the export path. It is worth understanding the anatomy of the so-called `particle spectrograph' produced by \PSALTer{}. The graphic is composed of seven kinds of boxes containing the following information:
\begin{enumerate}
	\item The input Lagrangian density.
	\item The wave operator (one box per~$J$-sector).
	\item The saturated propagator (one box per~$J$-sector).
	\item The constraints on the sources.
	\item The massive particles (if any, one box per particle).
	\item The massless particles (if any, one box per particle).
	\item The overall unitarity conditions.
\end{enumerate}
As will be explained more thoroughly in~\cref{TheoreticalDevelopment}, these are the essential elements for understanding the particle spectrum of any given theory. The question of how to optimally display all this information in a publishable format poses a surprisingly difficult problem. This is because both the numbers and sizes of various expressions can vary wildly from model to model. The current format shown in this paper uses two-dimensional rectangle bin packing, implemented in \CPP{}, to compactly arrange the boxes\footnote{This functionality is provided by \RectanglePacking{}~\cite{RectanglePacking:2024}, which is based on algorithms presented in~\cite{Jylanki:2010}.}. This is a heuristic solution: economy of space on the page is prioritised at the cost rotating some boxes by~$90^{\circ}$.

\begin{figure}[t!]
	\includegraphics[width=\linewidth]{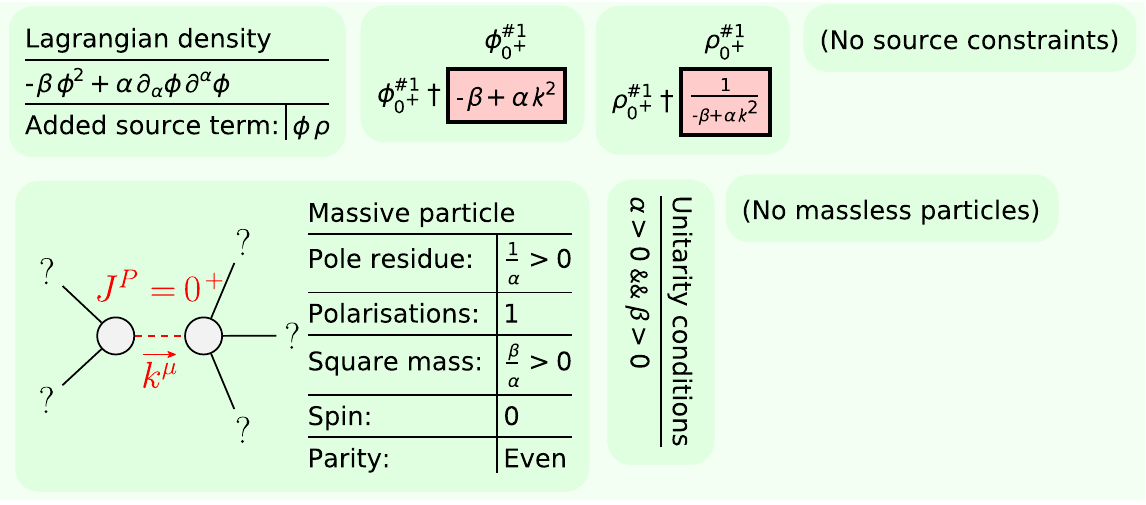}
	\caption{The particle spectrum of the massive scalar field in~\cref{ScalarTheoryAction}, to be compared with~\cref{ParticleSpectrographMasslessScalarTheory}. The no-ghost condition~$\alpha>0$ remains, and is accompanied by the no-tachyon condition~$\beta>0$. All quantities are defined in~\cref{FieldKinematicsScalarField}.}
\label{ParticleSpectrographMassiveScalarTheory}
\end{figure}

\paragraph*{Klein--Gordon theory} Now add a mass term~\cite{Klein:1926tv,Gordon:1926emj}:
\begin{lstlisting}
In[#]:= ParticleSpectrum[-(Coupling2 * ScalarField[]^2) + Coupling1 * CD[-a][ScalarField[]] * CD[a][ScalarField[]], TheoryName -> "MassiveScalarTheory", Method -> "Hard", MaxLaurentDepth -> 3]; 
\end{lstlisting}
The massive output is shown in~\cref{ParticleSpectrographMassiveScalarTheory}. Note that the spin of the particle only becomes meaningful in the massive case.

\begin{figure*}[t!]
	\includegraphics[width=\linewidth]{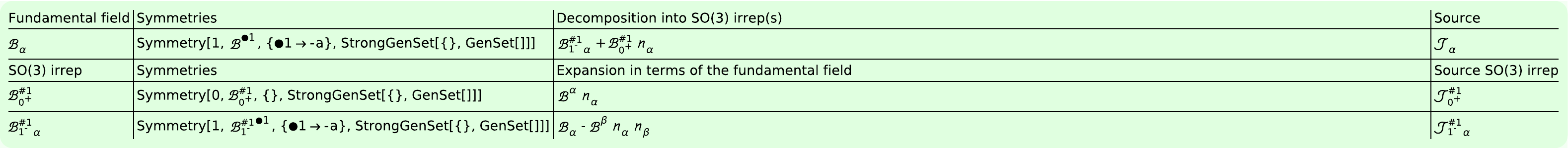}
	\caption{The declaration of a vector field \lstinline!VectorField!, which contains both~$0^+$ and~$1^-$ modes. As with~\cref{FieldKinematicsScalarField} there are no index symmetries. These definitions are used in~\cref{ParticleSpectrographMaxwellTheory,ParticleSpectrographProcaTheory,ParticleSpectrographSickMaxwellTheory,ParticleSpectrographSickProcaTheory,ParticleSpectrographLongitudinalMassless,ParticleSpectrographLongitudinalMassive}.}
\label{FieldKinematicsVectorField}
\end{figure*}

\subsection{Vector theory}\label{SuitabilityPractice}
\paragraph*{Kinematics} As another example of~\cref{BasicPositionAction}, consider the general vector theory in~$\B{_\mu}$, conjugate to the source~$\J{^\mu}$, which we can write as
\begin{align}\label{GeneralVectorLagrangian}
	S_{\text{F}}=\int\mathrm{d}^4x\ 
	\Big[&
		\alpha\PD{_\mu}\B{_\nu}\PD{^\mu}\B{^\nu}
		+\beta\PD{_\mu}\B{^\mu}\PD{_\nu}\B{^\nu}
		\nonumber\\
		&
		+\gamma\B{_\mu}\B{^\mu}
		+\B{_\mu}\J{^\mu}
	\Big],
\end{align}
where~$\alpha$,~$\beta$ are dimensionless coupling coefficients which completely parameterise the permissible kinetic operators up to total derivatives, and~$\gamma$ is a coupling with mass dimension two --- we have defined it already as \lstinline!Coupling3!. We denote the vector with the symbol \lstinline!VectorField! again by using the \lstinline!DefField! command:
\begin{lstlisting}
In[#]:= DefField[VectorField[-a], PrintAs -> "\[ScriptCapitalB]", PrintSourceAs -> "\[ScriptCapitalJ]"]; 
\end{lstlisting}
The output is shown in~\cref{FieldKinematicsVectorField}. This time we start to see the utility of the field kinematics graphics. There is a four-momentum~$\tensor{k}{^\mu}$ which is associated with massive or massless particle states. In the massive case, there is a finite~$k>0$ such that~$k^2\equiv\tensor{k}{^\mu}\tensor{k}{_\mu}$ and one may define the unit-timelike vector~$\N{^\mu}\equiv\tensor{k}{^\mu}/k$ as the four-velocity of the rest-frame of the particle. The vector~$\N{^\mu}$ may then be used to define the various~$\sothree$ irreps contained within a Lorentz-covariant tensor field through various projections. In the vector case, the extraction of the~$0^+$ and~$1^-$ states is very simple (see~\cref{SpinProjection}). For higher-rank fields, the process becomes more involved and requires the use of Young tableaux: \PSALTer{} performs this decomposition automatically.

\begin{figure}[t!]
	\includegraphics[width=\linewidth]{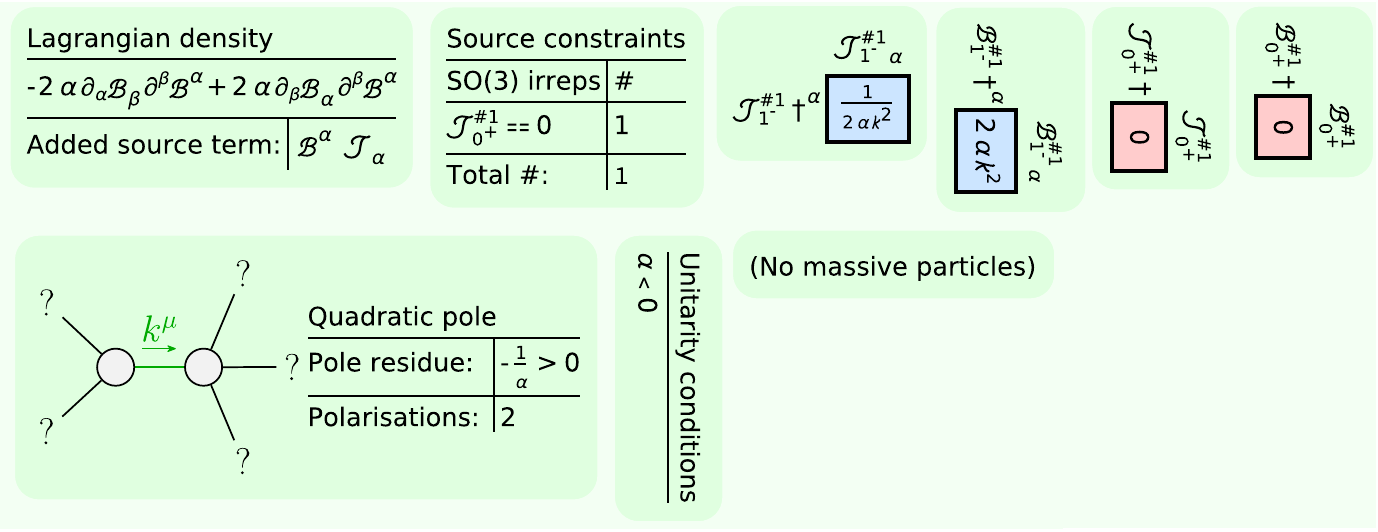}
	\caption{The particle spectrum of Maxwell theory, reached by setting~$\beta=-\alpha$ and~$\gamma=0$ in~\cref{GeneralVectorLagrangian}. Note the first appearance of source constraints~$\PD{_\mu}\J{^\mu}\approx 0$. These constraints are induced by the the singularity of the~$0^+$ wave operator and saturated propagator. Note also the appearance of a second set of matrices for the~$1^-$ sector. The \emph{blue} backgrounds signify interactions between \emph{negative} parity states. The two massless polarisations are those of the photon. In Maxwell's electrodynamics it is conventional to take~$\alpha=-1/2$, which agrees with the~$\alpha<0$ no-ghost condition. All quantities are defined in~\cref{FieldKinematicsVectorField}.}
\label{ParticleSpectrographMaxwellTheory}
\end{figure}

\paragraph*{Maxwell theory} The special case of Maxwell's electromagnetism~\cite{Maxwell:1865zz} is reached by setting~$\beta=-\alpha$ and~$\gamma=0$:
\begin{lstlisting}
In[#]:= ParticleSpectrum[-2 * Coupling1 * CD[-a][VectorField[-b]] * CD[b][VectorField[a]] + 2 * Coupling1 * CD[-b][VectorField[-a]] * CD[b][VectorField[a]], TheoryName -> "MaxwellTheory", Method -> "Easy", MaxLaurentDepth -> 3]; 
\end{lstlisting}
Once again, standard conventions would further imply that~$\alpha=-1/2$, but to avoid errors we should leave~$\alpha$ as a free coupling coefficient to linearly parameterise all the terms in the Lagrangian. The output is shown in~\cref{ParticleSpectrographMaxwellTheory}. In this case~$\J{^\mu}$ is identified as the conserved~$\uone$ current: the flow of electrical charge. The details of the spinor (or complex scalar) fields which comprise~$\J{^\mu}$ are not needed if we are just trying to understand the physics of~$\B{_\mu}$, so they can remain suppressed. The physics of~$\B{_\mu}$ does however allow us to draw conclusions about~$\J{^\mu}$, such as the conservation law~$\tensor{\partial}{_\mu}\J{^\mu}\approx 0$. We will return to the specifics of the Maxwell theory, and the broader question of source constraints and conservation laws in~\cref{MassiveSpectrum}.

\begin{figure}[t!]
	\includegraphics[width=\linewidth]{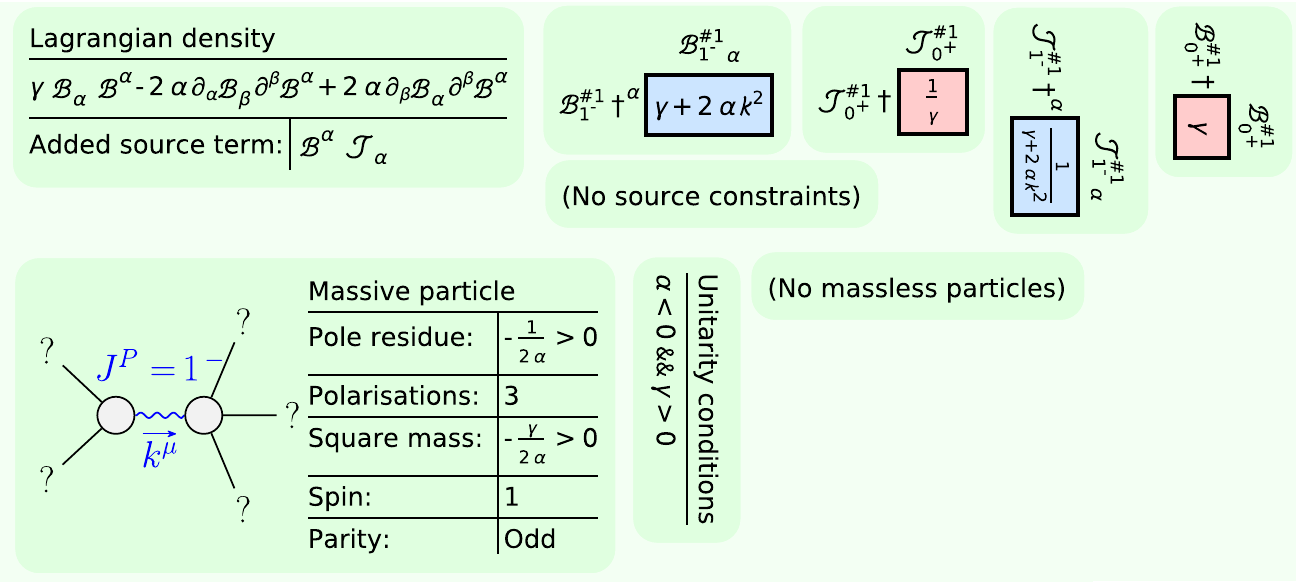}
	\caption{The particle spectrum of a massive vector, reached by setting~$\beta=-\alpha$ in~\cref{GeneralVectorLagrangian}. The result should be compared with~\cref{ParticleSpectrographMaxwellTheory}, and the~$\gamma>0$ condition prevents a tachyon. All quantities are defined in~\cref{FieldKinematicsVectorField}.}
\label{ParticleSpectrographProcaTheory}
\end{figure}

\paragraph*{Proca theory} We consider next the Proca theory~\cite{Proca:1936fbw}:
\begin{lstlisting}
In[#]:= ParticleSpectrum[Coupling3 * VectorField[-a] * VectorField[a] - 2 * Coupling1 * CD[-a][VectorField[-b]] * CD[b][VectorField[a]] + 2 * Coupling1 * CD[-b][VectorField[-a]] * CD[b][VectorField[a]], TheoryName -> "ProcaTheory", Method -> "Easy", MaxLaurentDepth -> 3]; 
\end{lstlisting}
The output is shown in~\cref{ParticleSpectrographProcaTheory}. Whilst Proca theory is healthy, we will next show how it is possible to construct \emph{inconsistent} vector models.

\begin{figure}[t!]
	\includegraphics[width=\linewidth]{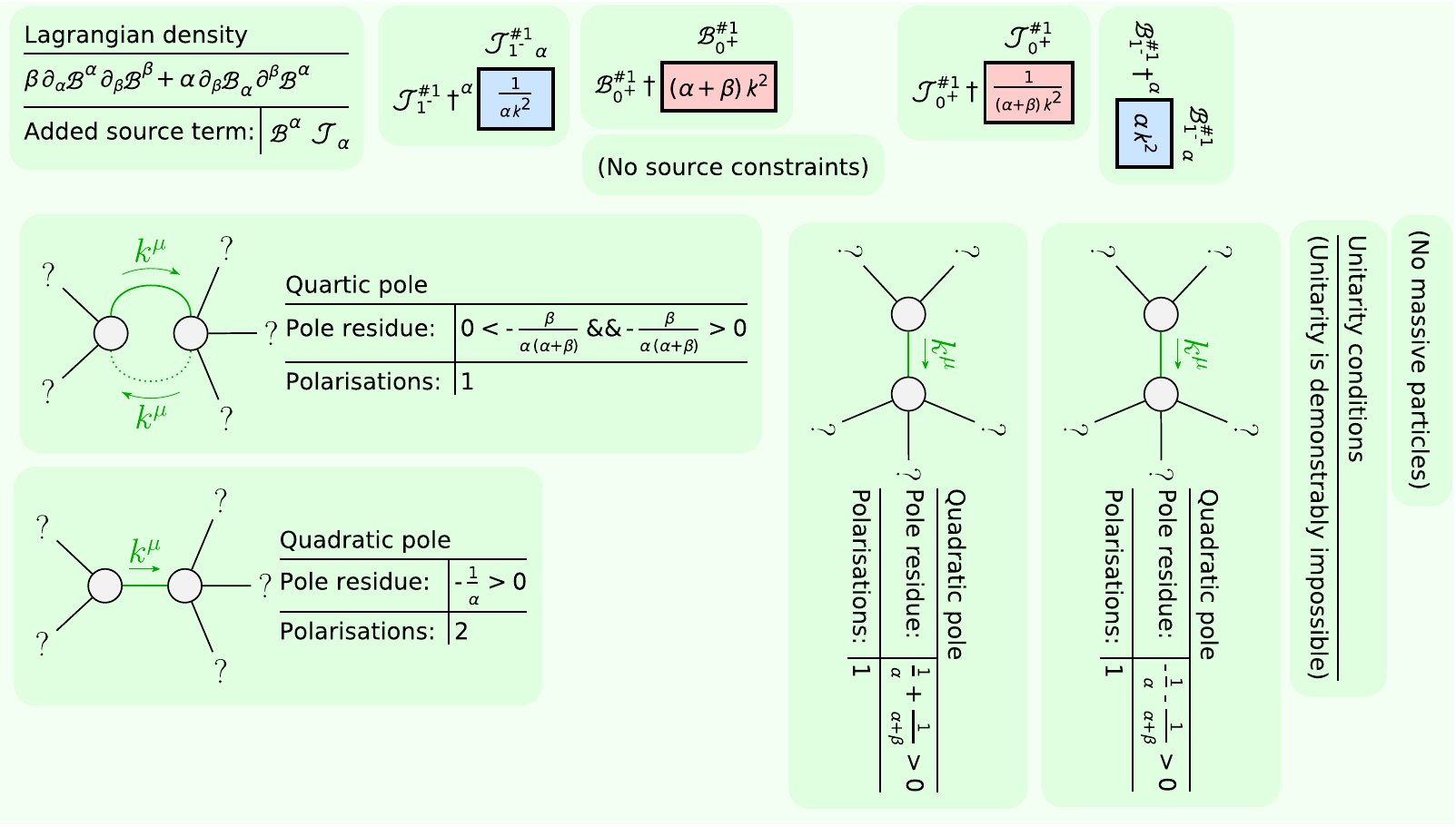}
	\caption{The particle spectrum of a massless vector without tuning of the kinetic terms, reached by setting~$\gamma=0$ in~\cref{GeneralVectorLagrangian}. The result should be compared with~\cref{ParticleSpectrographMaxwellTheory}. The gauge symmetry is destroyed, and the appearance of a quartic pole renders the theory non-unitary. The quartic pole is visible with the settings \lstinline!MaxLaurentDepth->2! or \lstinline!MaxLaurentDepth->3!. All quantities are defined in~\cref{FieldKinematicsVectorField}.}
\label{ParticleSpectrographSickMaxwellTheory}
\end{figure}

\paragraph*{Sick Maxwell theory} Consider the general massless vector theory:
\begin{lstlisting}
In[#]:= ParticleSpectrum[Coupling2 * CD[-a][VectorField[a]] * CD[-b][VectorField[b]] + Coupling1 * CD[-b][VectorField[-a]] * CD[b][VectorField[a]], TheoryName -> "SickMaxwellTheory", Method -> "Easy", MaxLaurentDepth -> 3]; 
\end{lstlisting}
The output is shown in~\cref{ParticleSpectrographSickMaxwellTheory}. When~$\alpha$ and~$\beta$ are not tuned, the vector theory generically propagates a total of four d.o.f. Two of these d.o.f result from a quartic propagator pole, one of which always constitutes a ghost state. This matter will be discussed further in~\cref{MassiveSpectrum}. In cases such as this, where unitarity cannot be guaranteed by imposing \emph{inequalities} on the coupling coefficients~$\alpha$ and~$\beta$, \PSALTer{} will simply declare the theory to be non-unitary. It is up to the user to identify \emph{equalities} such as the Maxwell condition~$\beta=-\alpha$ which qualitatively change the spectrum. A natural and straightforward extension of the software will be to automatically identify such special cases, in a recursive manner. It has been shown already in~\cite{Lin:2018awc,Lin:2019ugq} that such automated analyses are practical, even for very complicated theories such as Poincar\'e gauge theory (PGT). We will consider PGT further in~\cref{PoincareGaugeTheory}.

\begin{figure}[t!]
	\includegraphics[width=\linewidth]{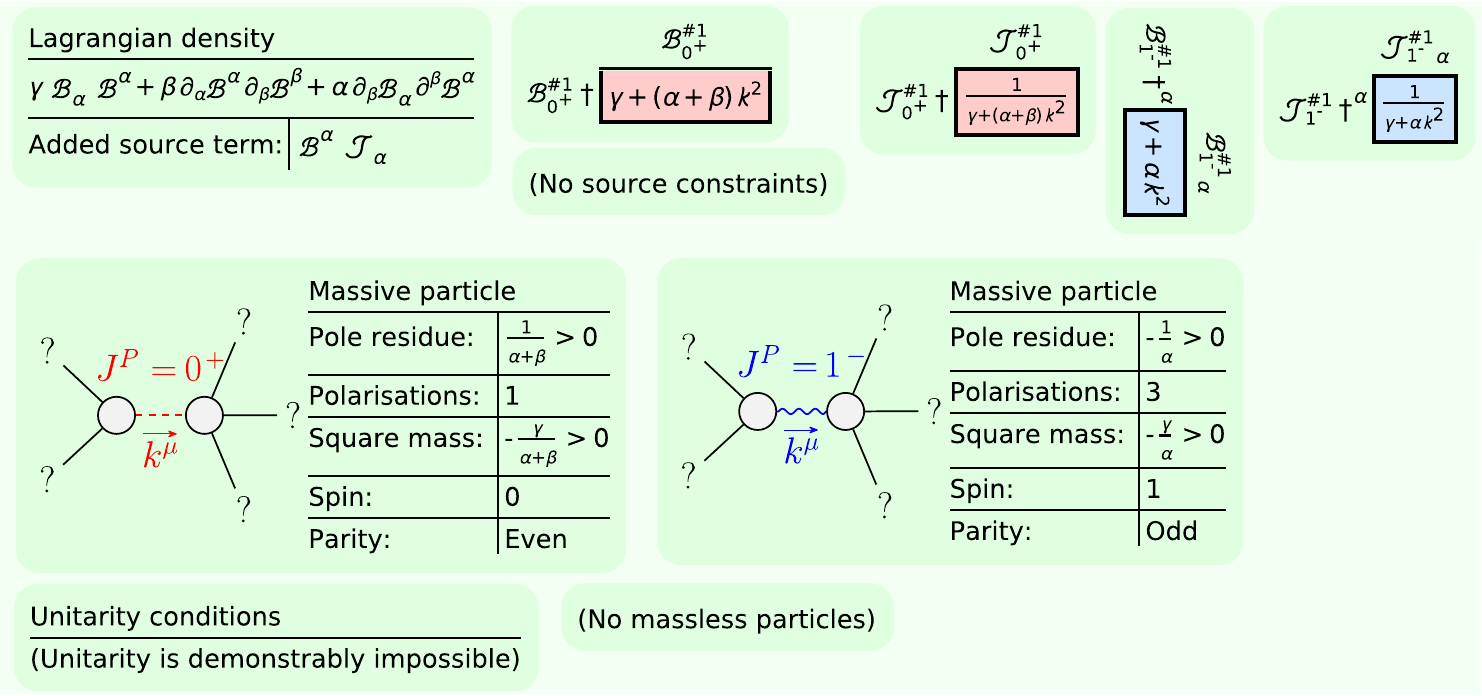}
	\caption{The particle spectrum of a sick massive vector, reached without constraints on~\cref{GeneralVectorLagrangian}. The result should be compared with~\cref{ParticleSpectrographProcaTheory}. All quantities are defined in~\cref{FieldKinematicsVectorField}.}
\label{ParticleSpectrographSickProcaTheory}
\end{figure}

\paragraph*{Sick Proca theory} We also consider the sick version of Proca theory:
\begin{lstlisting}
In[#]:= ParticleSpectrum[Coupling3 * VectorField[-a] * VectorField[a] + Coupling2 * CD[-a][VectorField[a]] * CD[-b][VectorField[b]] + Coupling1 * CD[-b][VectorField[-a]] * CD[b][VectorField[a]], TheoryName -> "SickProcaTheory", Method -> "Easy", MaxLaurentDepth -> 3]; 
\end{lstlisting}
The output is shown in~\cref{ParticleSpectrographSickProcaTheory}. The theory is again sick, as can be seen by trying to reconcile the simultaneous no-ghost and no-tachyon conditions for the massive~$0^+$ and~$1^-$ modes.

\begin{figure}[t!]
	\includegraphics[width=\linewidth]{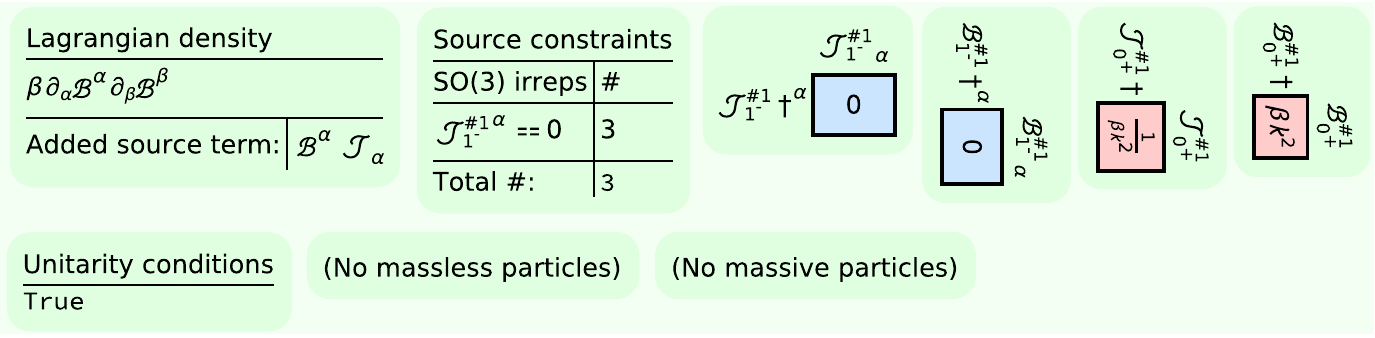}
	\caption{The particle spectrum of the longitudinal vector theory, reached by imposing~$\alpha=\gamma=0$ on~\cref{GeneralVectorLagrangian}. The spectrum is entirely empty. All quantities are defined in~\cref{FieldKinematicsVectorField}.}
\label{ParticleSpectrographLongitudinalMassless}
\end{figure}

\paragraph*{Longitudinal massless vector} As another alternative to the Maxwell model, we consider~$\alpha=\gamma=0$:
\begin{lstlisting}
In[#]:= ParticleSpectrum[Coupling2 * CD[-a][VectorField[a]] * CD[-b][VectorField[b]], TheoryName -> "LongitudinalMassless", Method -> "Easy", MaxLaurentDepth -> 3]; 
\end{lstlisting}
The output is shown in~\cref{ParticleSpectrographLongitudinalMassless}. In this case the spectrum is entirely empty, so the ghost in~\cref{ParticleSpectrographSickMaxwellTheory} is avoided.

\begin{figure}[t!]
	\includegraphics[width=\linewidth]{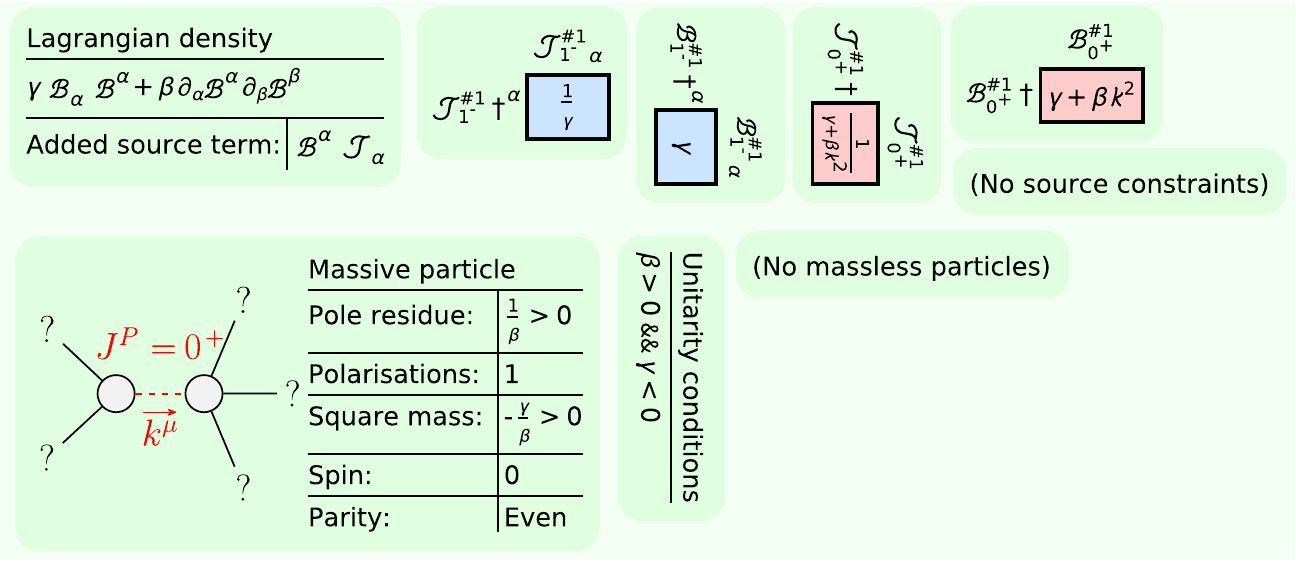}
	\caption{The particle spectrum of the massive longitudinal vector theory, reached by imposing~$\alpha=0$ on~\cref{GeneralVectorLagrangian}. The result is a healthy theory of a massive scalar. All quantities are defined in~\cref{FieldKinematicsVectorField}.}
\label{ParticleSpectrographLongitudinalMassive}
\end{figure}

\paragraph*{Longitudinal massive vector} Whilst the massless longitudinal model turns out to be trivial, it is possible to populate the spectrum by readmitting the mass. This is done by restricting to the single condition~$\alpha=0$:
\begin{lstlisting}
In[#]:= ParticleSpectrum[Coupling3 * VectorField[-a] * VectorField[a] + Coupling2 * CD[-a][VectorField[a]] * CD[-b][VectorField[b]], TheoryName -> "LongitudinalMassive", Method -> "Easy", MaxLaurentDepth -> 3]; 
\end{lstlisting}
The output is shown in~\cref{ParticleSpectrographLongitudinalMassive}. In this case the theory propagates a massive scalar, again without any gauge symmetries. This often-overlooked model of a vector is unitary, and therefore viable.

\begin{figure*}[t!]
	\includegraphics[width=\linewidth]{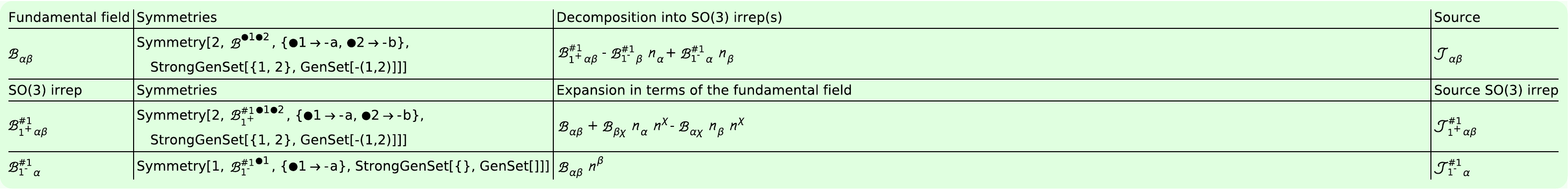}
	\caption{The declaration of an antisymmetric, rank-two tensor \lstinline!TwoFormField!, which contains both~$1^+$ and~$1^-$ modes. For the first time we see the appearance of index symmetries. These definitions are used in~\cref{ParticleSpectrographTwoFormEectrodynamics,ParticleSpectrographTwoFormEectrodynamicsMassive}.}
\label{FieldKinematicsTwoFormField}
\end{figure*}

\subsection{Two-form theory}
\paragraph*{Kinematics} We now move on to fields of second rank. The simplest such field is a two-form~$\B{_{\mu\nu}}\equiv\B{_{[\mu\nu]}}$ conjugate to the source~$\J{^{\mu\nu}}\equiv\J{^{[\mu\nu]}}$. A suitably general two-form action for our purposes is given by~\cite{Freedman:1980us,Aurilia:1981xg}
\begin{align}
	S_{\text{F}}\equiv\int\mathrm{d}^4x\Big[&\alpha\PD{_{[\mu}}\B{_{\nu\sigma]}}\PD{^{[\mu}}\B{^{\nu\sigma]}}
	\nonumber
	\\
	&
	+\beta\B{_{\mu\nu}}\B{^{\mu\nu}}+\B{_{\mu\nu}}\J{^{\mu\nu}}\Big].\label{TwoFormLagrangian}
\end{align}
We denote the two-form with the symbol \lstinline!TwoFormField!:
\begin{lstlisting}
In[#]:= DefField[TwoFormField[-a, -b], Antisymmetric[{-a, -b}], PrintAs -> "\[ScriptCapitalB]", PrintSourceAs -> "\[ScriptCapitalJ]"]; 
\end{lstlisting}
Note that we need to specify the index symmetry, and for this we use the \lstinline!Antisymmetric! command just as we would do within \lstinline!DefTensor! from \xTensor{}. The two-form field contains one~$1^+$ and one~$1^-$ state. The output is shown in~\cref{FieldKinematicsTwoFormField}.

\begin{figure}[t!]
	\includegraphics[width=\linewidth]{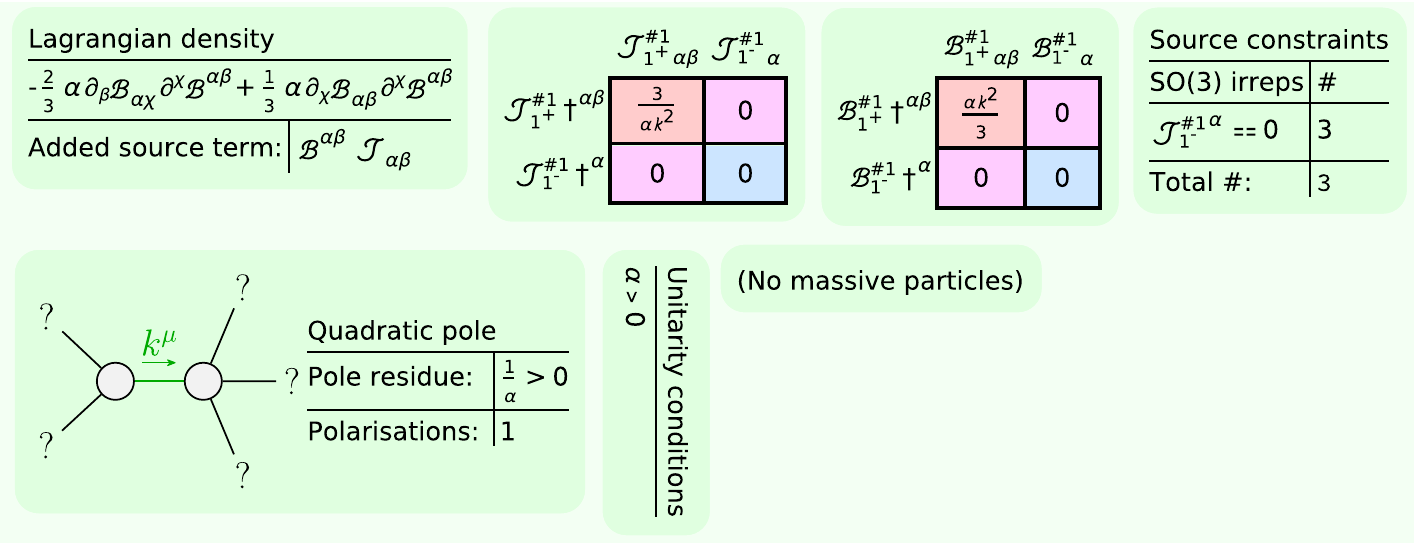}
	\caption{The particle spectrum of Kalb--Ramond theory, obtained by setting~$\beta=0$ in~\cref{TwoFormLagrangian}. The condition~$\alpha>0$ prevents the Kalb--Ramond scalar from being a ghost. In the output, there is a new set of matrices for each spin, but not for each parity. For the~$1^+$ and~$1^-$ modes of the two-form, this means that the spin-one matrix is divided into red and blue diagonal blocks of even and odd parity respectively. Off-diagonal blocks, which \emph{mix} parities, are \emph{purple}. These become populated in Lagrangia with odd powers of~$\tensor{\epsilon}{_{\mu\nu\sigma\lambda}}$. According to the suitability criteria set out in~\cref{Introduction}, \PSALTer{} can presently only consider theories where purple elements vanish. All quantities are defined in~\cref{FieldKinematicsTwoFormField}.}
\label{ParticleSpectrographTwoFormEectrodynamics}
\end{figure}

\paragraph*{Kalb--Ramond theory} The~$\beta=0$ limit of~\cref{TwoFormLagrangian} is a viable action for a parity-preserving, massless two-form:
\begin{lstlisting}
In[#]:= ParticleSpectrum[(-2 * Coupling1 * CD[-b][TwoFormField[-a, -c]] * CD[c][TwoFormField[a, b]])/3 + (Coupling1 * CD[-c][TwoFormField[-a, -b]] * CD[c][TwoFormField[a, b]])/3, TheoryName -> "TwoFormEectrodynamics", Method -> "Easy", MaxLaurentDepth -> 1]; 
\end{lstlisting}
This corresponds to a single Kalb--Ramond d.o.f, with three gauge generators~\cite{Kalb:1974yc}. The output is shown in~\cref{ParticleSpectrographTwoFormEectrodynamics}.

\begin{figure}[t!]
	\includegraphics[width=\linewidth]{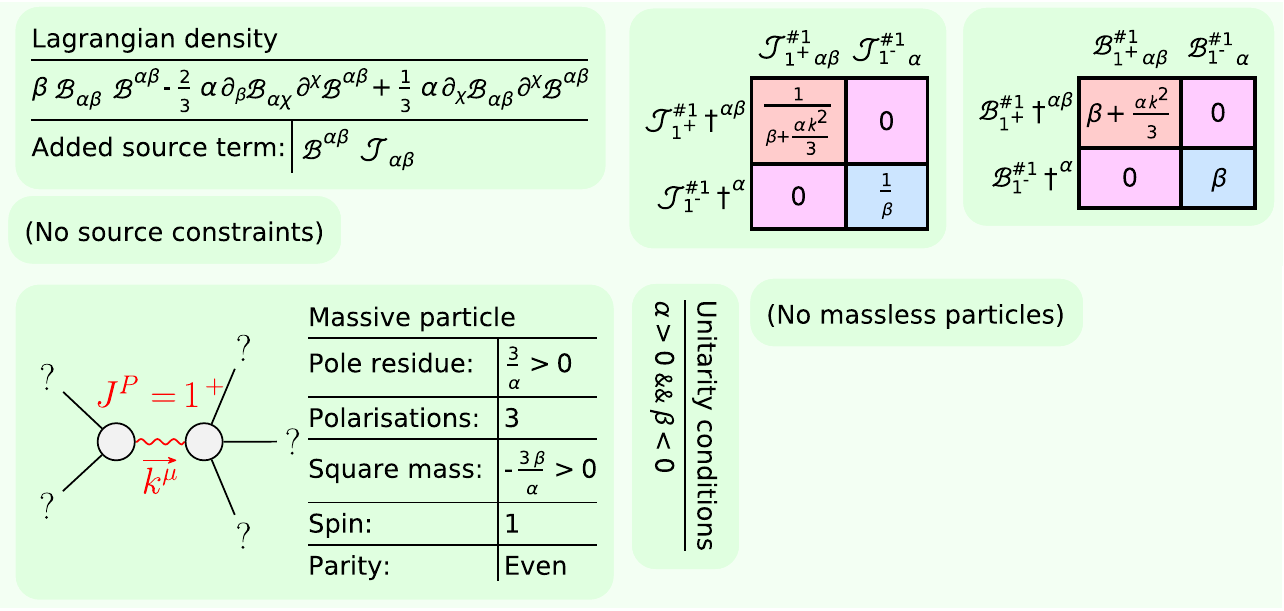}
	\caption{The particle spectrum of the massive two-form in~\cref{TwoFormLagrangian}, to be compared with the result in~\cref{ParticleSpectrographTwoFormEectrodynamics}. The gauge symmetry is destroyed and the condition~$\beta<0$ prevents a tachyon. All quantities are defined in~\cref{FieldKinematicsTwoFormField}.}
\label{ParticleSpectrographTwoFormEectrodynamicsMassive}
\end{figure}

\paragraph*{Massive two-form} Once again, a mass can straightforwardly be added:
\begin{lstlisting}
In[#]:= ParticleSpectrum[Coupling2 * TwoFormField[-a, -b] * TwoFormField[a, b] - (2 * Coupling1 * CD[-b][TwoFormField[-a, -c]] * CD[c][TwoFormField[a, b]])/3 + (Coupling1 * CD[-c][TwoFormField[-a, -b]] * CD[c][TwoFormField[a, b]])/3, TheoryName -> "TwoFormEectrodynamicsMassive", Method -> "Easy", MaxLaurentDepth -> 1]; 
\end{lstlisting}
The effect is to break the gauge symmetry and excite a massive~$1^+$ mode. The output is shown in~\cref{ParticleSpectrographTwoFormEectrodynamicsMassive}. It should be noted that there are, in general, two Lorentz-invariant kinetic operators for a two-form, which generalise the operator used in~\cref{TwoFormLagrangian}. However, as with the model in~\cref{ParticleSpectrographSickMaxwellTheory}, these lead to ghosts without the appropriate tuning. Such effects can readily be explored with \PSALTer{}, though we will not do so here.

\begin{figure*}[t!]
	\includegraphics[width=\linewidth]{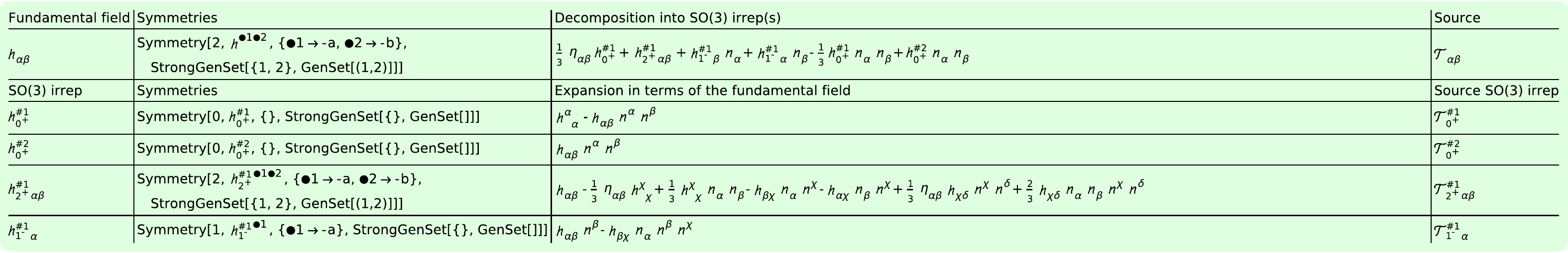}
	\caption{The declaration of a symmetric, rank-two tensor field \lstinline!MetricPerturbation!. There are more~$\sothree{}$ irreps than in the antisymmetric case in~\cref{FieldKinematicsTwoFormField}. Note for the first time the appearance of multiple~$\othree$ irreps with the same~$J^P$, respectively denoted~$\tensor*{h}{_{0^+}^{\#1}}$ and~$\tensor*{h}{_{0^+}^{\#2}}$. These definitions are used in~\cref{ParticleSpectrographFierzPauliTheory,ParticleSpectrographMassiveGravity,ParticleSpectrographSickMassiveGravity}.}
\label{FieldKinematicsMetricPerturbation}
\end{figure*}

\subsection{Tensor theory}\label{TensorTheory}
\paragraph*{Kinematics} Having dealt with the antisymmetric tensor, we will consider the symmetric case. The obvious motivation is the metric perturbation~$\tensor*{h}{_{\mu\nu}}\equiv\tensor*{h}{_{(\mu\nu)}}$ which we define using the scheme 
\begin{equation}\label{DefMetricPerturbation}
	\tensor{g}{_{\mu\nu}}\equiv\tensor{\eta}{_{\mu\nu}}+\tensor*{h}{_{\mu\nu}},\quad
	\tensor{g}{^{\mu\nu}}\equiv\tensor{\eta}{^{\mu\nu}}-\tensor*{h}{^{\mu\nu}}+\mathcal{O}\big(h^2\big),
\end{equation}
where~$\tensor{g}{_{\mu\nu}}$ is the metric tensor of \emph{curved} manifold to which the flat \lstinline!M4! is a good approximation. The conjugate source is the perturbative stress-energy tensor~$\tensor{T}{^{\mu\nu}}\equiv\tensor{T}{^{(\mu\nu)}}$. Accordingly we define \lstinline!MetricPerturbation! using:
\begin{lstlisting}
In[#]:= DefField[MetricPerturbation[-a, -b], Symmetric[{-a, -b}], PrintAs -> "\[ScriptH]", PrintSourceAs -> "\[ScriptCapitalT]"]; 
\end{lstlisting}
Note the use of the \lstinline!Symmetric! command from \xTensor{}. There is now one~$2^+$, one~$1^-$ and two~$0^+$ modes. The output is shown in~\cref{FieldKinematicsMetricPerturbation}. A suitably general action will be
\begin{align}
	S_{\text{F}}=\int\mathrm{d}^4x\Big[&\frac{1}{2}\alpha\PD{_\mu}\tensor*{h}{}\PD{^\mu}h+\alpha\PD{_\mu}\tensor*{h}{^\mu_\nu}\PD{_\sigma}\tensor*{h}{^{\sigma\nu}}-\alpha\PD{^\mu}\tensor*{h}{}\PD{_\nu}\tensor*{h}{_\mu^\nu}
	\nonumber\\
	&\hspace{-30pt}-\frac{1}{2}\alpha\PD{_\mu}\tensor*{h}{_{\nu\sigma}}\PD{^\mu}\tensor*{h}{^{\nu\sigma}}+\beta\tensor*{h}{_{\mu\nu}}\tensor*{h}{^{\mu\nu}}-\gamma\tensor*{h}{}^2+\tensor*{h}{_{\mu\nu}}\tensor{T}{^{\mu\nu}}\Big],\label{TensorLagrangian}
\end{align}
where~$\tensor*{h}{}\equiv\tensor*{h}{^\mu_\mu}$ is the trace of the metric perturbation.

\begin{figure}[t!]
	\includegraphics[width=\linewidth]{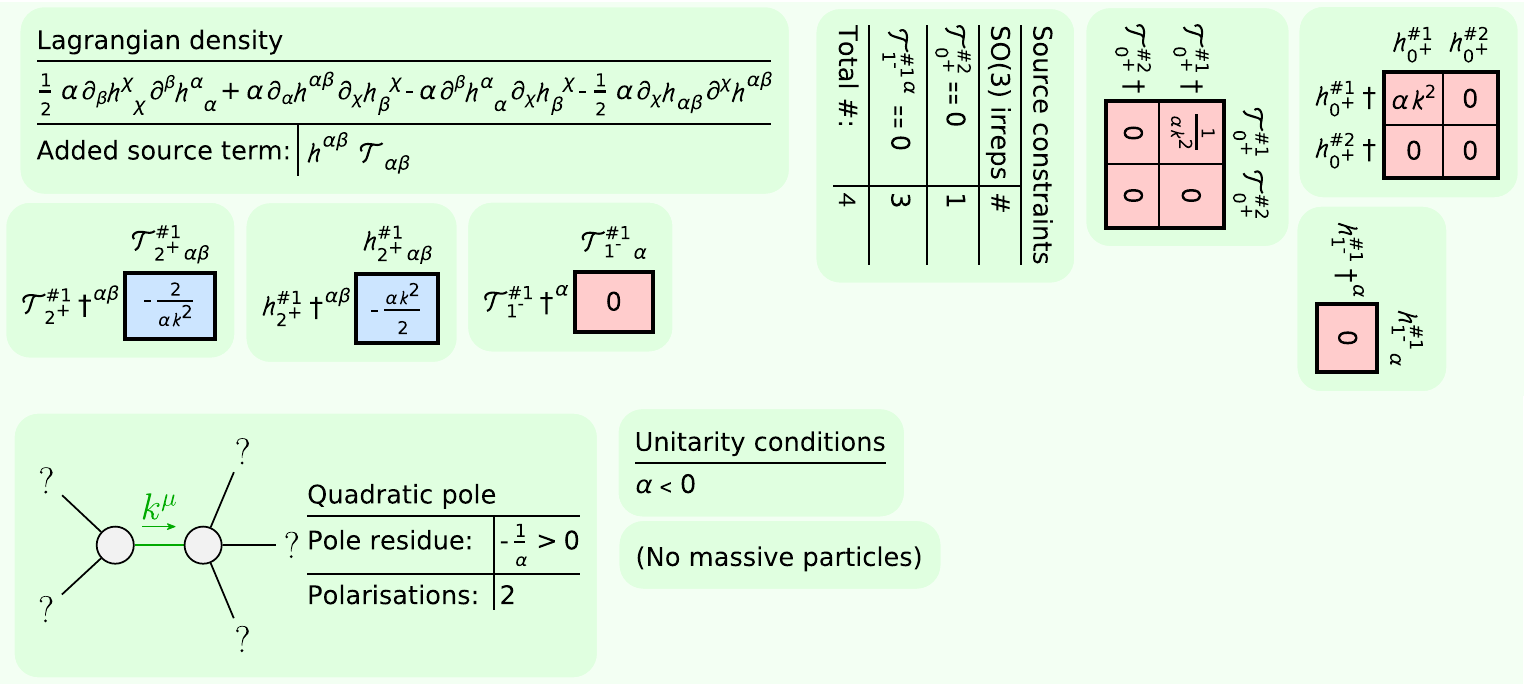}
	\caption{The particle spectrum of massless Fierz--Pauli theory. The theory propagates a graviton, with the no-ghost condition~$\alpha<0$ (note that~$\alpha=-\MPl{}^2$ when the theory is viewed as the linearisation of GR). Linearised diffeomorphism invariance gives rise to the source constraint~$\PD{_\mu}\tensor{T}{^{\mu\nu}}\approx 0$, i.e. the conservation of matter stress-energy. All quantities are defined in~\cref{FieldKinematicsMetricPerturbation}.}
\label{ParticleSpectrographFierzPauliTheory}
\end{figure}

\paragraph*{Fierz--Pauli theory} The most relevant example is that of Fierz--Pauli theory~\cite{Fierz:1939ix}, setting~$\beta=\gamma=0$ in~\cref{TensorLagrangian}:
\begin{lstlisting}
In[#]:= ParticleSpectrum[(Coupling1 * CD[-b][MetricPerturbation[c, -c]] * CD[b][MetricPerturbation[a, -a]])/2 + Coupling1 * CD[-a][MetricPerturbation[a, b]] * CD[-c][MetricPerturbation[-b, c]] - Coupling1 * CD[b][MetricPerturbation[a, -a]] * CD[-c][MetricPerturbation[-b, c]] - (Coupling1 * CD[-c][MetricPerturbation[-a, -b]] * CD[c][MetricPerturbation[a, b]])/2, TheoryName -> "FierzPauliTheory", Method -> "Hard", MaxLaurentDepth -> 3]; 
\end{lstlisting}
This propagates the two massless polarisations of the Einstein graviton. There are four gauge generators, corresponding to diffeomorphism gauge symmetry. The conjugate source constraint is~$\PD{_\mu}\tensor{T}{^{\mu\nu}}\approx 0$, i.e. the conservation of matter stress-energy. The output is shown in~\cref{ParticleSpectrographFierzPauliTheory}. The results are consistent with the fact that the Fierz--Pauli model is the weak-field limit of GR, with the Einstein--Hilbert action. The no-ghost condition~$\alpha<0$ is consistent with the convention that~$\alpha=-\MPl{}^2$, where~$\MPl{}$ is the Planck mass.

\begin{figure}[t!]
	\includegraphics[width=\linewidth]{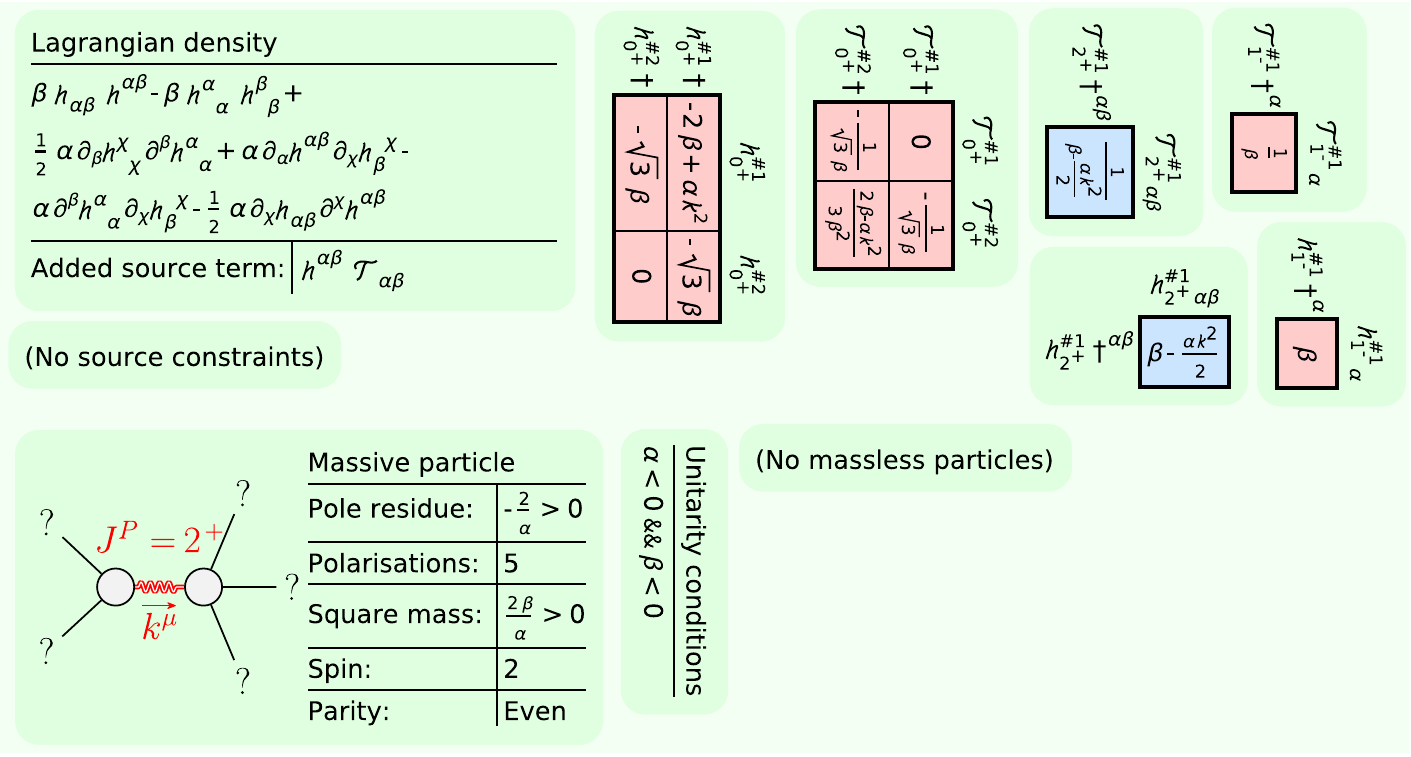}
	\caption{The particle spectrum of massive gravity, to be compared with~\cref{ParticleSpectrographFierzPauliTheory}. This time, the gauge diffeomorphism invariance is destroyed and the graviton acquires a mass. The condition~$\beta<0$ prevents a tachyon. All quantities are defined in~\cref{FieldKinematicsMetricPerturbation}.}
\label{ParticleSpectrographMassiveGravity}
\end{figure}

\paragraph*{Massive gravity} By setting~$\gamma=\beta$ in~\cref{TensorLagrangian}, we arrive at the tuned Fierz--Pauli mass term and the linearisation of massive gravity:
\begin{lstlisting}
In[#]:= ParticleSpectrum[Coupling2 * MetricPerturbation[-a, -b] * MetricPerturbation[a, b] - Coupling2 * MetricPerturbation[a, -a] * MetricPerturbation[b, -b] + (Coupling1 * CD[-b][MetricPerturbation[c, -c]] * CD[b][MetricPerturbation[a, -a]])/2 + Coupling1 * CD[-a][MetricPerturbation[a, b]] * CD[-c][MetricPerturbation[-b, c]] - Coupling1 * CD[b][MetricPerturbation[a, -a]] * CD[-c][MetricPerturbation[-b, c]] - (Coupling1 * CD[-c][MetricPerturbation[-a, -b]] * CD[c][MetricPerturbation[a, b]])/2, TheoryName -> "MassiveGravity", Method -> "Easy", MaxLaurentDepth -> 3]; 
\end{lstlisting}
This propagates a massive~$2^+$ mode. The output is shown in~\cref{ParticleSpectrographMassiveGravity}. 

\begin{figure}[t!]
	\includegraphics[width=\linewidth]{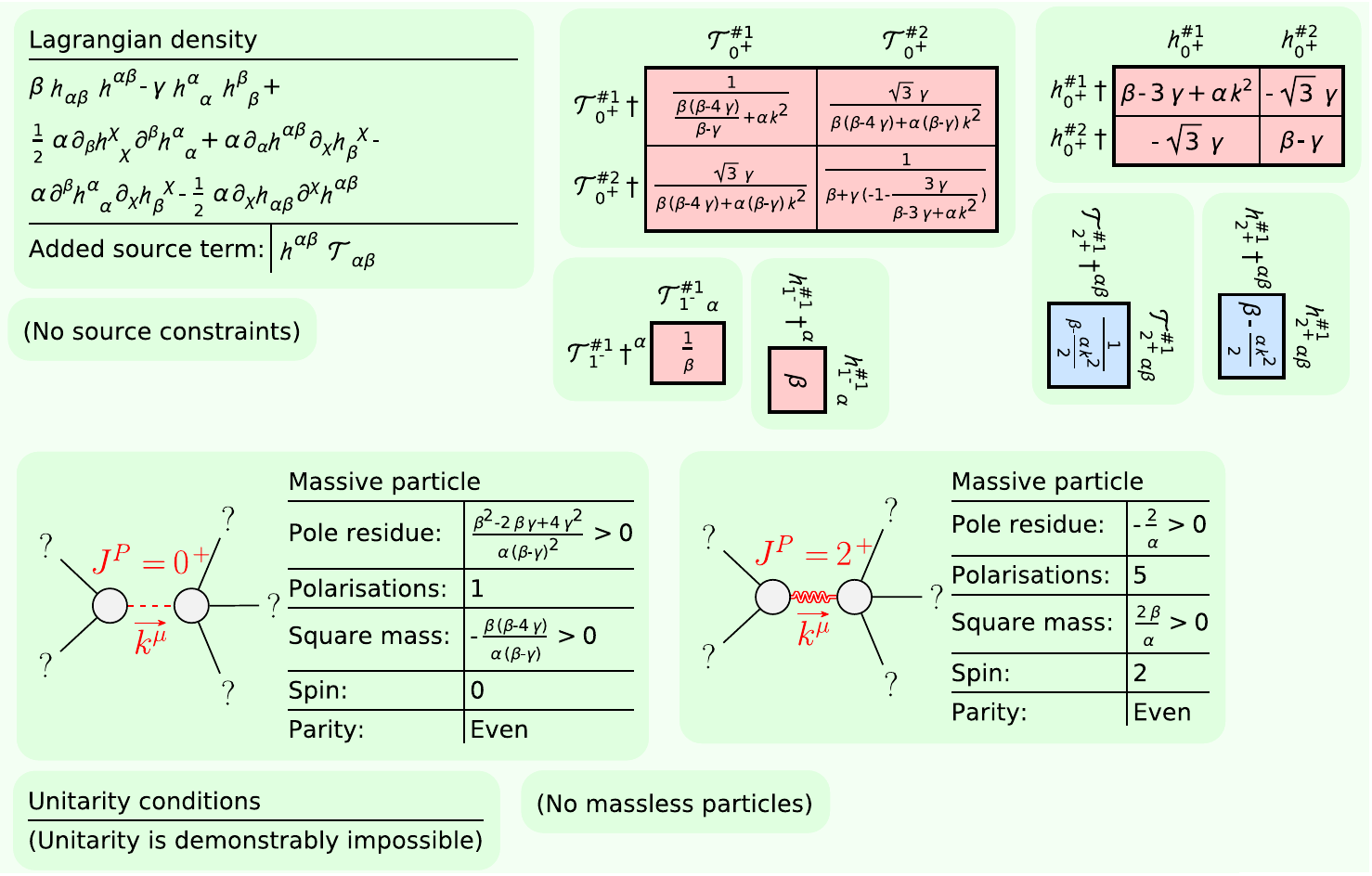}
	\caption{The particle spectrum of sick massive gravity, to be compared with~\cref{ParticleSpectrographMassiveGravity}. The extra scalar is the Boulware--Deser ghost. All quantities are defined in~\cref{FieldKinematicsMetricPerturbation}.}
\label{ParticleSpectrographSickMassiveGravity}
\end{figure}

\paragraph*{Sick massive gravity} The tuning of the Fierz--Pauli mass term can be broken by imposing no constraints on~\cref{TensorLagrangian}:
\begin{lstlisting}
In[#]:= ParticleSpectrum[Coupling2 * MetricPerturbation[-a, -b] * MetricPerturbation[a, b] - Coupling3 * MetricPerturbation[a, -a] * MetricPerturbation[b, -b] + (Coupling1 * CD[-b][MetricPerturbation[c, -c]] * CD[b][MetricPerturbation[a, -a]])/2 + Coupling1 * CD[-a][MetricPerturbation[a, b]] * CD[-c][MetricPerturbation[-b, c]] - Coupling1 * CD[b][MetricPerturbation[a, -a]] * CD[-c][MetricPerturbation[-b, c]] - (Coupling1 * CD[-c][MetricPerturbation[-a, -b]] * CD[c][MetricPerturbation[a, b]])/2, TheoryName -> "SickMassiveGravity", Method -> "Easy", MaxLaurentDepth -> 3]; 
\end{lstlisting}
This does not produce higher-order poles, but it does activate a massive~$0^+$ particle known as the Boulware--Deser ghost~\cite{Boulware:1972yco}. The output is shown in~\cref{ParticleSpectrographSickMassiveGravity}.

\begin{figure*}[t!]
	\includegraphics[width=\linewidth]{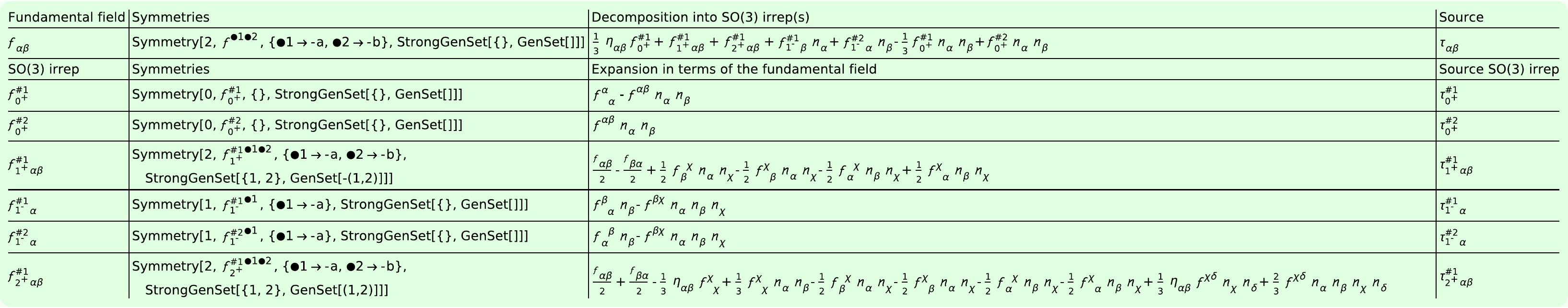}
	\caption{The declaration of an asymmetric, rank-two tensor field \lstinline!TetradPerturbation!. The kinematic content is the union of that found in~\cref{FieldKinematicsTwoFormField,FieldKinematicsMetricPerturbation}. These definitions are used in~\cref{ParticleSpectrographEinsteinCartanTheory,ParticleSpectrographGeneralPGT} and~\crefrange{ParticleSpectrographCase1}{ParticleSpectrographCase19}.}
\label{FieldKinematicsTetradPerturbation}
\end{figure*}
\begin{figure*}[t!]
	\includegraphics[width=\linewidth]{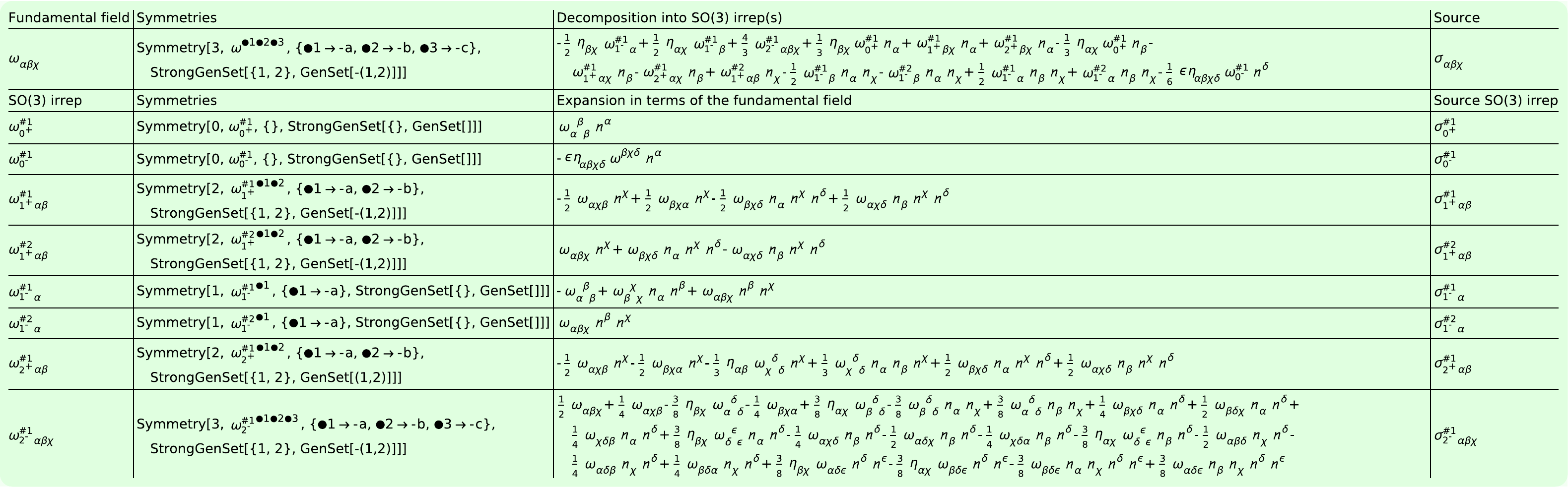}
	\caption{The declaration of a pair-antisymmetric, rank-three tensor field \lstinline!SpinConnection!. These definitions are used in~\cref{ParticleSpectrographEinsteinCartanTheory,ParticleSpectrographGeneralPGT} and~\crefrange{ParticleSpectrographCase1}{ParticleSpectrographCase19}.}
\label{FieldKinematicsSpinConnection}
\end{figure*}

\subsection{Poincar\'e gauge theory}\label{PoincareGaugeTheory}
\paragraph*{Kinematics} A useful feature of \PSALTer{} is that \lstinline!ParticleSpectrum! accepts theories depending simultaneously on multiple different kinds of fields. For example, one can reach scalar-tensor theories by combining \lstinline!ScalarField! with \lstinline!MetricPerturbation!, or the Einstein--Maxwell and Einstein--Proca theories by combining \lstinline!MetricPerturbation! with \lstinline!VectorField!. By defining a second symmetric tensor field, bimetric gravity could also be reached. We will consider the more ambitious example of Poincar\'e gauge theory (PGT), i.e. the formulation of gravity which uses a tetrad and spin-connection. We will use conventions close to~\cite{Barker:2020elg,Barker:2020gcp,Barker:2021oez,VandepeerBarker:2022xnp,Barker:2022jsh,Barker:2023bmr,Barker:2023bmr} when discussing PGT. Like scalar-tensor theory, PGT is not a specific model, but rather a kinematic framework for developing specific models. None of the fields defined so far using \lstinline!DefField! encompass these kinematics. We introduce~$\tensor{e}{^i_\mu}$ and~$\tensor{e}{_i^\mu}$ as the co-tetrad and tetrad components, which are associated with Roman Lorentz (i.e. anholonomic) indices, so that the curved-space metric introduced in~\cref{TensorTheory} is~$\tensor{e}{^i_\mu}\tensor{e}{^j_\nu}\tensor{\eta}{_{ij}}\equiv\tensor{g}{_{\mu\nu}}$ with inverse~$\tensor{e}{_i^\mu}\tensor{e}{_j^\nu}\tensor{\eta}{^{ij}}\equiv\tensor{g}{^{\mu\nu}}$, with identities~$\tensor{e}{^i_\mu}\tensor{e}{_i^\nu}\equiv\tensor*{\delta}{_\mu^\nu}$ and~$\tensor{e}{^i_\mu}\tensor{e}{_j^\mu}\equiv\tensor*{\delta}{_j^i}$ as kinematic restrictions. There is also a spin connection~$\tensor{\omega}{^{ij}_\mu}\equiv\tensor{\omega}{^{[ij]}_\mu}$, so that the PGT curvature and PGT torsion are respectively
\begin{subequations}
	\begin{align}
		 \tensor{\mathscr{R}}{^{kl}_{ij}}&\equiv 2\tensor{e}{_i^\mu}\tensor{e}{_j^\nu}\big(\PD{_{[\mu|}}\tensor{\omega}{^{kl}_{|\nu]}}+\tensor{\omega}{^k_{m[\mu|}}\tensor{\omega}{^{ml}_{|\nu]}}\big),\label{PGTCurvature}
	\\
		\tensor{\mathscr{T}}{^k_{ij}}&\equiv 2\tensor{e}{_i^\mu}\tensor{e}{_j^\nu}\big(\PD{_{[\mu|}}\tensor{e}{^k_{|\nu]}}+\tensor{\omega}{^k_{m[\mu|}}\tensor{e}{^m_{|\nu]}}\big).\label{PGTTorsion}
	\end{align}
\end{subequations}
Note that~\cref{PGTCurvature,PGTTorsion} have different index conventions to the metric-affine counterparts which will be defined in~\cref{FDef,TDef}. By analogy to Yang--Mills theories, the general parity-preserving action up to quadratic order in the field strengths in~\cref{PGTTorsion,PGTCurvature} is
\begin{align}
	&S_{\text{PGT}}=\int\mathrm{d}^4x e\bigg[
		-\frac{1}{2}\alpha_0\mathscr{R}
	+\frac{1}{6}\left(2\alpha_1+3\alpha_2+\alpha_3\right)\tensor{\mathscr{R}}{_{ijkl}}\tensor{\mathscr{R}}{^{ijkl}}
	\nonumber\\
	&
	+\frac{2}{3}\left(\alpha_1-\alpha_3\right)\tensor{\mathscr{R}}{_{ijkl}}\tensor{\mathscr{R}}{^{ikjl}}
	-\left(\alpha_1+\alpha_2-\alpha_4-\alpha_5\right)\tensor{\mathscr{R}}{_{ij}}\tensor{\mathscr{R}}{^{ij}}
	\nonumber\\
	&
	+\frac{1}{6}\left(2\alpha_1-3\alpha_2+\alpha_3\right)\tensor{\mathscr{R}}{_{ijkl}}\tensor{\mathscr{R}}{^{klij}}
	\nonumber\\
	&
	-\left(\alpha_1-\alpha_2+\alpha_4-\alpha_5\right)\tensor{\mathscr{R}}{_{ij}}\tensor{\mathscr{R}}{^{ji}}
	+\frac{1}{6}\left(2\alpha_1-3\alpha_4+\alpha_6\right)\tensor{\mathscr{R}}{}^2
	\nonumber\\
	&
	+\frac{1}{3}\left(2\beta_1+\beta_3\right)\tensor{\mathscr{T}}{_{ijk}}\tensor{\mathscr{T}}{^{ijk}}
	\nonumber\\
	&
	+\frac{2}{3}\left(\beta_1-\beta_3\right)\tensor{\mathscr{T}}{_{ijk}}\tensor{\mathscr{T}}{^{jik}}
	+\frac{2}{3}\left(\beta_1-\beta_2\right)\tensor{\mathscr{T}}{_{i}}\tensor{\mathscr{T}}{^{j}}
	+L_{\text{M}}
	\bigg],\label{PGTVersion}
\end{align}
where~$\tensor{\mathscr{R}}{_{ij}}\equiv\tensor{\mathscr{R}}{^l_{ilj}}$ and~$\tensor{\mathscr{R}}{}\equiv\tensor{\mathscr{R}}{^i_i}$ and~$\tensor{\mathscr{T}}{_i}\equiv\tensor{\mathscr{T}}{^j_{ij}}$, with the measure~$e\equiv\det\big(\tensor{e}{^i_\mu}\big)\equiv\sqrt{-g}$, where~$g\equiv\det\big(\tensor{g}{_{\mu\nu}}\big)$. The matter Lagrangian~$L_{\text{M}}$ will provide all the sources. In the weak-field regime, we take~$\tensor{\omega}{^{ij}_\mu}$ to be inherently perturbative, with 24 d.o.f, conjugate to which is the perturbative matter spin current~$\tensor{\sigma}{_{ij}^\mu}$~\cite{Rigouzzo:2023sbb,Rigouzzo:2022yan,Karananas:2021zkl}. We define the symbol \lstinline!SpinConnection! accordingly:
\begin{lstlisting}
In[#]:= DefField[SpinConnection[-a, -b, -c], Antisymmetric[{-a, -b}], PrintAs -> "\[Omega]", PrintSourceAs -> "\[Sigma]"]; 
\end{lstlisting}
The output is shown in~\cref{FieldKinematicsSpinConnection}. We perturb the tetrad around the `Kronecker' choice of Minkowski vacuum~\cite{Barker:2023bmr,Blixt:2022rpl,Blixt:2023qbg}
\begin{equation}\label{DefTetradPerturbation}
	\tensor{e}{_i^\mu}\equiv\tensor*{\delta}{_i^\mu}+\tensor{f}{_i^\mu},
	\quad
	\tensor{e}{^i_\mu}\equiv\tensor*{\delta}{^i_\mu}-\tensor{f}{_\mu^i}+\mathscr{O}\big(f^2\big).
\end{equation}
To lowest order in the quadratic action, notice in~\cref{DefTetradPerturbation} how the Kronecker vacuum allows the Greek and Roman indices to be interchangeable: this is good, because \PSALTer{} only knows about one set of spacetime indices on the Minkowski background, and these are strictly associated with Greek indices which represent Cartesian coordinates. By comparing~\cref{DefMetricPerturbation,DefTetradPerturbation} we conclude that~$\tensor*{h}{_{\mu\nu}}\equiv2\tensor{f}{_{(\mu\nu)}}+\mathscr{O}\big(f^2\big)$, and that in general there are 16 d.o.f in the asymmetric~$\tensor{f}{_i^\mu}$. Conjugate to~$\tensor{f}{_i^\mu}$ is the translational source (asymmetric stress-energy tensor)~$\tensor{\tau}{^i_\mu}$. We define the perturbation \lstinline!TetradPerturbation! as:
\begin{lstlisting}
In[#]:= DefField[TetradPerturbation[-a, -b], PrintAs -> "\[ScriptF]", PrintSourceAs -> "\[Tau]"]; 
\end{lstlisting}
The output is shown in~\cref{FieldKinematicsTetradPerturbation}. We also need to define the new coupling coefficients which appear in~\cref{PGTVersion} accordingly:
\begin{lstlisting}
In[#]:= DefConstantSymbol[Alp0, PrintAs -> "\!\(\ * SubscriptBox[\(\[Alpha]\), \(0\)]\)"]; DefConstantSymbol[Alp1, PrintAs -> "\!\(\ * SubscriptBox[\(\[Alpha]\), \(1\)]\)"]; DefConstantSymbol[Alp2, PrintAs -> "\!\(\ * SubscriptBox[\(\[Alpha]\), \(2\)]\)"]; DefConstantSymbol[Alp3, PrintAs -> "\!\(\ * SubscriptBox[\(\[Alpha]\), \(3\)]\)"]; DefConstantSymbol[Alp4, PrintAs -> "\!\(\ * SubscriptBox[\(\[Alpha]\), \(4\)]\)"]; DefConstantSymbol[Alp5, PrintAs -> "\!\(\ * SubscriptBox[\(\[Alpha]\), \(5\)]\)"]; DefConstantSymbol[Alp6, PrintAs -> "\!\(\ * SubscriptBox[\(\[Alpha]\), \(6\)]\)"]; DefConstantSymbol[Bet1, PrintAs -> "\!\(\ * SubscriptBox[\(\[Beta]\), \(1\)]\)"]; DefConstantSymbol[Bet2, PrintAs -> "\!\(\ * SubscriptBox[\(\[Beta]\), \(2\)]\)"]; DefConstantSymbol[Bet3, PrintAs -> "\!\(\ * SubscriptBox[\(\[Beta]\), \(3\)]\)"]; 
\end{lstlisting}
Note that the way these couplings are arranged in~\cref{PGTVersion} looks somewhat convoluted: the reason is that each coupling (with the exception of~$\alpha_0$) actually parameterises the square of a different~$\soonethree$ irrep derived from~\cref{PGTCurvature,PGTTorsion}.

\begin{figure*}[t!]
	\includegraphics[width=\linewidth]{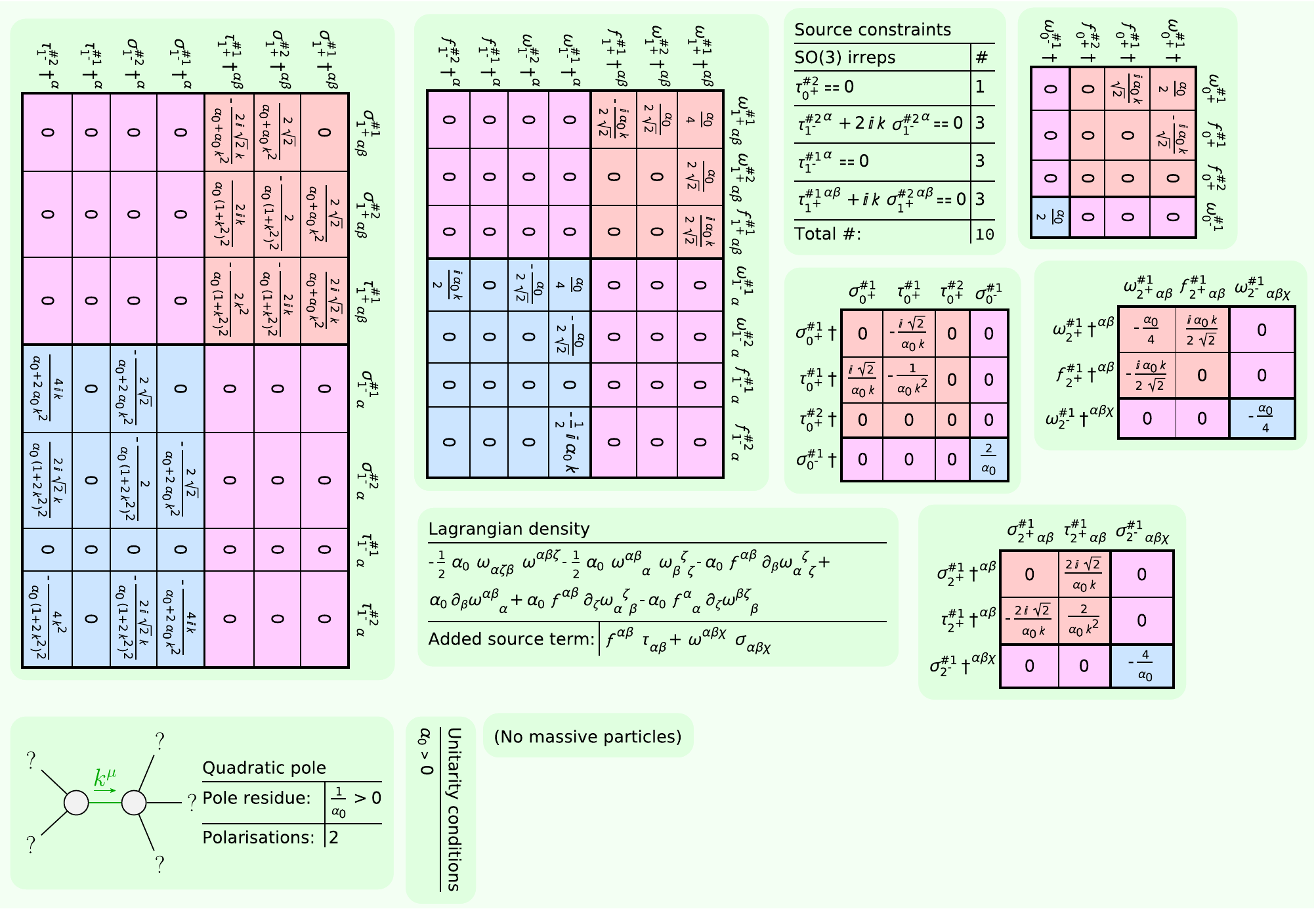}
	\caption{The particle spectrum of Einstein--Cartan (EC) theory, reached by setting all coupling coefficients except for~$\alpha_0$ to zero in~\cref{PGTVersion}. The spectrum itself is consistent with~\cref{ParticleSpectrographFierzPauliTheory}, because EC theory and GR differ only by non-propagating torsion which is sourced by a contact interaction with the matter spin tensor. Note that the presence of ten gauge generators: these signal a local symmetry of the eponymous Poincar\'e group. Four of the generators correspond to translations (inherited from the diffeomorphism symmetry in~\cref{ParticleSpectrographFierzPauliTheory}), and six are new: these correspond to the Lorentz group, and imply the conservation of the matter spin tensor. The no-ghost condition on the graviton~$\alpha_0>0$ is consistent with the convention that~$\alpha_0=\MPl{}^2$. All quantities are defined in~\cref{FieldKinematicsTetradPerturbation,FieldKinematicsSpinConnection}.}
\label{ParticleSpectrographEinsteinCartanTheory}
\end{figure*}

\paragraph*{Einstein--Cartan theory} The simplest PGT to analyse is Einstein--Cartan (EC) theory~\cite{Cartan:1923zea}, whereby the only non-vanishing coupling is~$\alpha_0$. At quadratic order, the action is:
\begin{lstlisting}
In[#]:= ParticleSpectrum[-1/2 * (Alp0 * SpinConnection[-a, -z, -b] * SpinConnection[a, b, z]) - (Alp0 * SpinConnection[a, b, -a] * SpinConnection[-b, z, -z])/2 - Alp0 * TetradPerturbation[a, b] * CD[-b][SpinConnection[-a, z, -z]] + Alp0 * CD[-b][SpinConnection[a, b, -a]] + Alp0 * TetradPerturbation[a, b] * CD[-z][SpinConnection[-a, z, -b]] - Alp0 * TetradPerturbation[a, -a] * CD[-z][SpinConnection[b, z, -b]], TheoryName -> "EinsteinCartanTheory", Method -> "Hard", MaxLaurentDepth -> 3]; 
\end{lstlisting}
The output is shown in~\cref{ParticleSpectrographEinsteinCartanTheory}. The spectrum of EC theory is basically the same as that of GR, though the gauge symmetries and source constraints are more extensive due to the local Poincar\'e invariance.

\begin{figure*}[t!]
	\includegraphics[width=\linewidth]{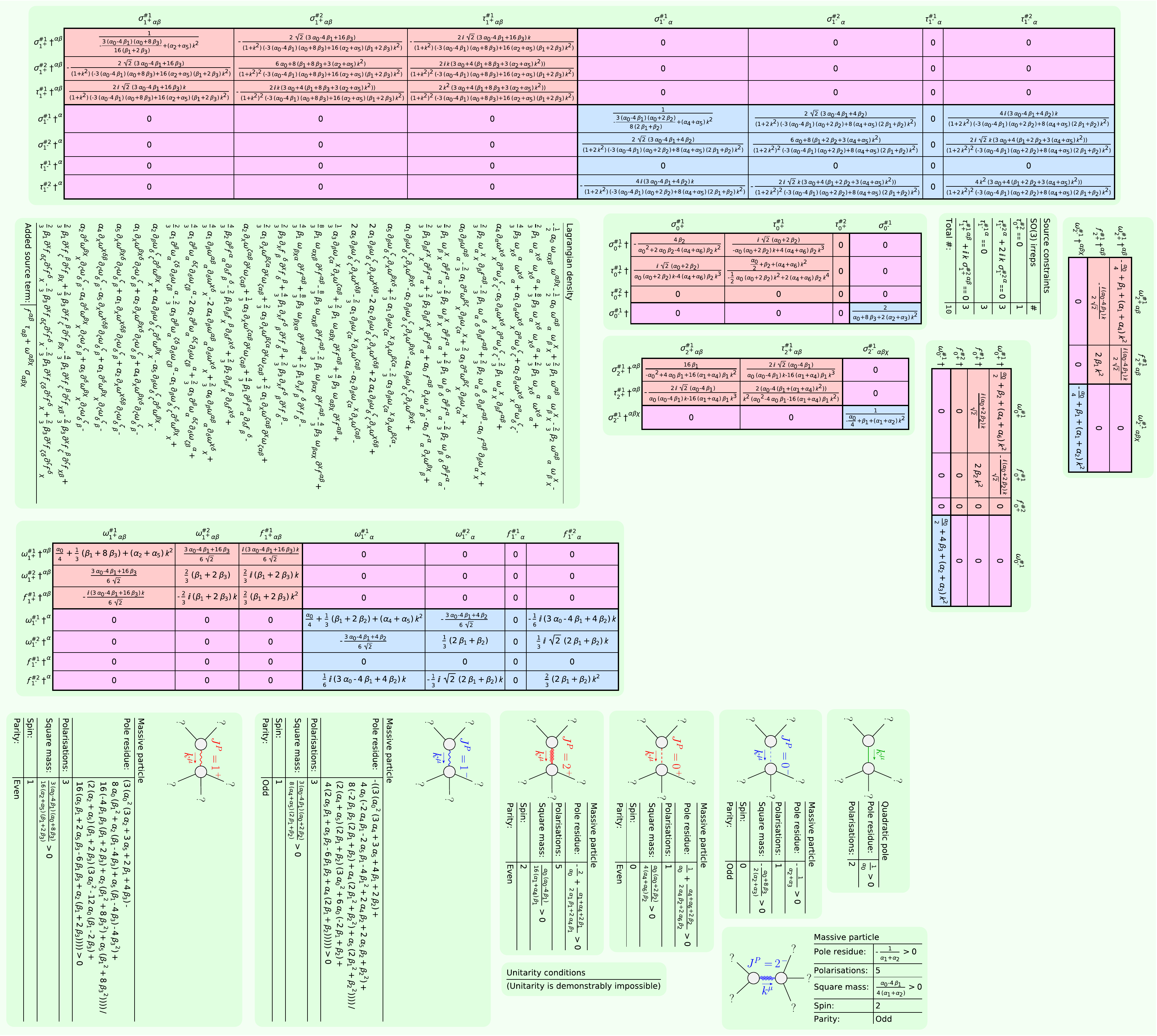}
	\caption{The particle spectrum of the most general parity-preserving PGT. The fields~$\tensor{f}{^i_\mu}$ and~$\tensor{\omega}{^{ij}_\mu}$ contain respectively~$16$ and~$24$ d.o.f. The Poincar\'e symmetry eliminates~$2\times 10$ d.o.f, and two are accounted for by the graviton polarisations. The remaining~$18$ d.o.f are partitioned amongst the six massive species shown above. As is well known, only for special cases of the PGT action in~\cref{PGTVersion} do the masses and pole residues of these species allow for unitarity: the general case shown here is sick. All quantities are defined in~\cref{FieldKinematicsTetradPerturbation,FieldKinematicsSpinConnection}.}
\label{ParticleSpectrographGeneralPGT}
\end{figure*}

\paragraph*{General PGT} It is possible to also study the completely general case, where no assumptions are made about the coupling coefficients. The quadratic part of~\cref{PGTVersion} is inevitably a very long expression:
\begin{lstlisting}
In[#]:= ParticleSpectrum[-1/2 * (Alp0 * SpinConnection[-a, -c, -b] * SpinConnection[a, b, c]) - (Alp0 * SpinConnection[a, b, -a] * SpinConnection[-b, c, -c])/2 + (2 * Bet1 * SpinConnection[a, b, -a] * SpinConnection[-b, c, -c])/3 - (2 * Bet2 * SpinConnection[a, b, -a] * SpinConnection[-b, c, -c])/3 - (2 * (*omitted 6930 characters for brevity*) -z]] * CD[z][TetradPerturbation[d, -c]])/3 - (2 * Bet1 * CD[c][TetradPerturbation[-z, -d]] * CD[z][TetradPerturbation[d, -c]])/3 + (2 * Bet3 * CD[c][TetradPerturbation[-z, -d]] * CD[z][TetradPerturbation[d, -c]])/3, TheoryName -> "GeneralPGT", Method -> "Hard", MaxLaurentDepth -> 3];
\end{lstlisting}
The output is shown in~\cref{ParticleSpectrographGeneralPGT}. The many massive states can be compared to the mass spectrum found in~\cite{Hayashi:1979wj,Hayashi:1980qp}. Note that \PSALTer{} is not designed to provide the cumbersome quadratic action in this input: the user is expected to do this. There are already very sophisticated tools in \xAct{} (most notably \xPert{}) which make the linearisation of arbitrarily complicated tensorial field theories a trivial task. The fact that \xPert{} is not a dependency of \PSALTer{} is intended to be in line with the UNIX philosophy: ``\textit{Write programs that do one thing and do it well}''~\cite{Mcilroy:1978}.

\begin{figure*}[t!]
	\includegraphics[width=\linewidth]{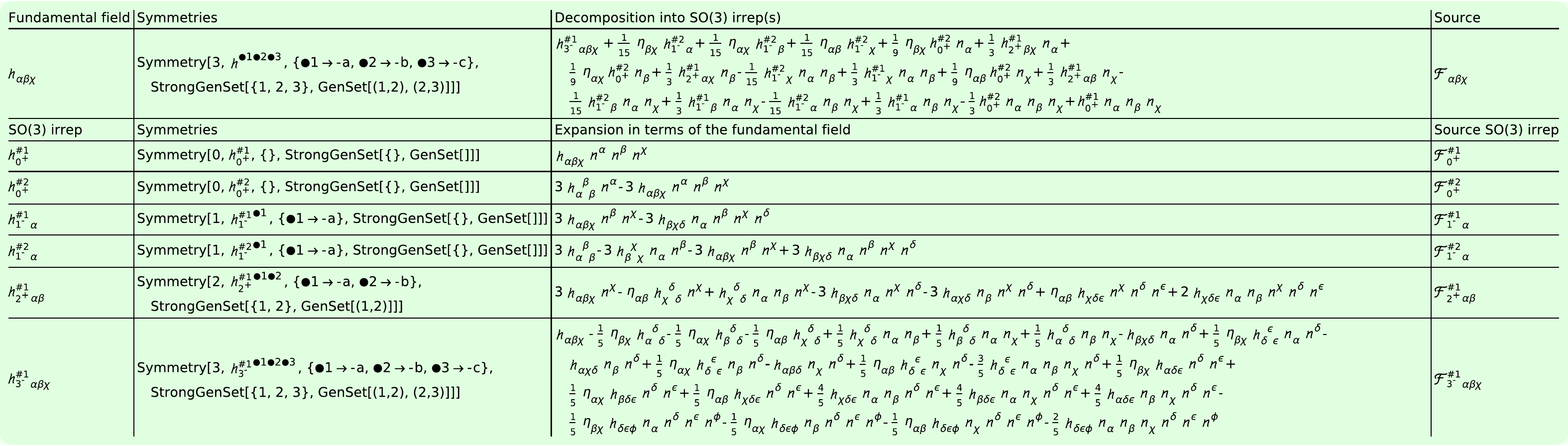}
	\caption{The declaration of a totally symmetric, rank-three tensor field \lstinline!HigherSpinField!. Compared to the pair-antisymmetric case of \lstinline!SpinConnection! in~\cref{FieldKinematicsSpinConnection}, there are fewer~$\sothree{}$ irreps, but one of the new additions is a comparatively complicated~$3^-$ irrep. These definitions are used in~\cref{ParticleSpectrographSinghHagenTheory}.}
\label{FieldKinematicsHigherSpinField}
\end{figure*}

\subsection{Higher-spin bosons}\label{HigherSpinBosons}
\begin{figure}[t!]
	\includegraphics[width=\linewidth]{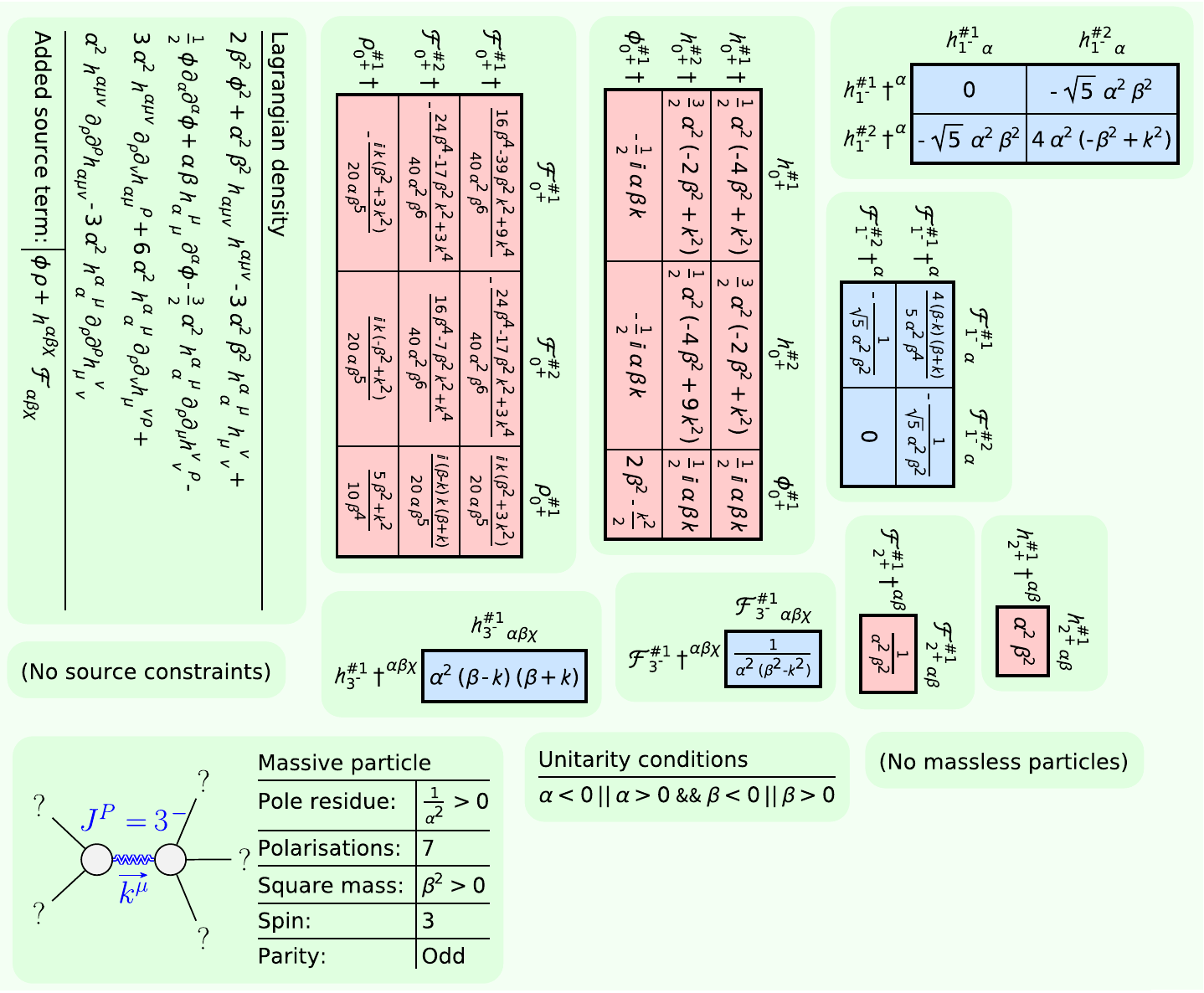}
	\caption{The particle spectrum of Singh--Hagen theory in~\cref{SinghHagenTheory}. All quantities are defined in~\cref{FieldKinematicsScalarField,FieldKinematicsHigherSpinField}.}
\label{ParticleSpectrographSinghHagenTheory}
\end{figure}

\paragraph*{Kinematics} So far in~\cref{TensorTheory,PoincareGaugeTheory} we have considered models which propagate spin-two species, embedded within tensors of second (parity-even) and third (parity-odd) rank. Third-rank tensors are needed to represent spin-three particles. As a minimal example, we consider a rank-three field~$\HigherSpinField{_{\mu\nu\sigma}}\equiv\HigherSpinField{_{(\mu\nu\sigma)}}$ which is totally symmetric in all its indices. The conjugate source will be~$\HigherSpinSource{^{\mu\nu\sigma}}\equiv\HigherSpinSource{^{(\mu\nu\sigma)}}$. Accordingly we define \lstinline!HigherSpinField! as:
\begin{lstlisting}
In[#]:= DefField[HigherSpinField[-a, -b, -c], Symmetric[{-a, -b, -c}], PrintAs -> "\[ScriptH]", PrintSourceAs -> "\[ScriptCapitalF]"]; 
\end{lstlisting}
The output is shown in~\cref{FieldKinematicsHigherSpinField}.
\paragraph*{Singh--Hagen theory} The natural model to study will be Singh--Hagen theory~\cite{Singh:1974qz}, and we will adhere to the conventions set out in~\cite{Mendonca:2019gco}. As we observed in~\cref{Introduction}, freedom from ghosts and tachyons is a basic requirement of most physical models. Higher-spin particles come from higher-rank fields, and higher-rank fields tend to carry more spurious~$\sothree$ irreps: the risk of accidentally importing ghosts can therefore be expected to increase with the spin of the model. To remove spurious d.o.f from the Singh--Hagen model, an additional scalar field must be introduced. We will make use of the field~$\phi$ (i.e. \lstinline!ScalarField!) introduced already in~\cref{ScalarTheory}. The linear Singh--Hagen theory is then
\begin{align}
	S_{\text{F}}
	=
	\int
	&
	\mathrm{d}^4x
	\Big[
	2\beta^2\phi^2
	+\alpha^2\beta^2\HigherSpinField{_{\mu\nu\sigma}}\HigherSpinField{^{\mu\nu\sigma}}
	-3\alpha^2\beta^2\HigherSpinField{_\mu}\HigherSpinField{^\mu}
	+\frac{1}{2}\phi\Box\phi
	\nonumber\\&
	+\alpha\beta\HigherSpinField{_\mu}\PD{^\mu}\phi
	-\frac{3}{2}\alpha^2\HigherSpinField{_\mu}\PD{^\mu}\PD{_\nu}\HigherSpinField{^\nu}
	-3\alpha^2\HigherSpinField{_{\mu\nu\sigma}}\PD{^\sigma}\PD{_\lambda}\HigherSpinField{^{\mu\nu\lambda}}
	\nonumber\\&
	+6\alpha^2\HigherSpinField{_{\mu}}\PD{_\nu}\PD{_\sigma}\HigherSpinField{^{\mu\nu\sigma}}
	+\alpha^2\HigherSpinField{_{\mu\nu\sigma}}\Box\HigherSpinField{^{\mu\nu\sigma}}
	-3\alpha^2\HigherSpinField{_{\mu}}\Box\HigherSpinField{^{\mu}}
	\nonumber\\&
	+\phi\rho
	+\HigherSpinField{_{\mu\nu\sigma}}\HigherSpinSource{^{\mu\nu\sigma}}
	\Big],\label{SinghHagenTheory}
\end{align}
where~$\HigherSpinField{_\mu}\equiv\HigherSpinField{^\nu_{\mu\nu}}$, and~$\beta$ has mass dimension one and~$\alpha$ is dimensionless. Note from~\cref{SinghHagenTheory} that the parameterisation suggested in~\cite{Mendonca:2019gco} is naturally lower- and higher-order in the coupling coefficients: it will be difficult to pass this model to \lstinline!ParticleSpectrum! in a way that is linear in \lstinline!Coupling1! and \lstinline!Coupling2!. When this happens, it is not a disaster, but it means we must select \lstinline!Method->"Easy"! to avoid errors. Whilst the SPO analysis of the Singh--Hagen model is complicated enough that it deserves the dedicated treatment in~\cite{Mendonca:2019gco}, it is still well within the capabilities of \lstinline!Method->"Easy"!. The cases where we strictly need to use \lstinline!Method->"Hard"! are those where there are not only many~$\sothree$ irreps present in the fields, but also many coupling coefficients\footnote{As a rule, \lstinline!Method->"Hard"! is safer, so non-linear coupling coefficient parameterisations should be avoided wherever possible.}. We proceed with the analysis as follows:
\begin{lstlisting}
In[#]:= ParticleSpectrum[Coupling1^2 * Coupling2^2 * HigherSpinField[-a, -m, -n] * HigherSpinField[a, m, n] - 3 * Coupling1^2 * Coupling2^2 * HigherSpinField[a, -a, m] * HigherSpinField[-m, n, -n] + 2 * Coupling2^2 * ScalarField[]^2 + (ScalarField[] * CD[-a][CD[a][ScalarField[]]])/2 + Coupling1 * Coupling2 * HigherSpinField[-a, m, -m] * CD[a][ScalarField[]] - (3 * Coupling1^2 * HigherSpinField[a, -a, m] * CD[-r][CD[-m][HigherSpinField[n, -n, r]]])/2 - 3 * Coupling1^2 * HigherSpinField[a, m, n] * CD[-r][CD[-n][HigherSpinField[-a, -m, r]]] + 6 * Coupling1^2 * HigherSpinField[a, -a, m] * CD[-r][CD[-n][HigherSpinField[-m, n, r]]] + Coupling1^2 * HigherSpinField[a, m, n] * CD[-r][CD[r][HigherSpinField[-a, -m, -n]]] - 3 * Coupling1^2 * HigherSpinField[a, -a, m] * CD[-r][CD[r][HigherSpinField[-m, n, -n]]], TheoryName -> "SinghHagenTheory", Method -> "Easy", MaxLaurentDepth -> 3]; 
\end{lstlisting}
The output is shown in~\cref{ParticleSpectrographSinghHagenTheory}. As expected, the massive~$3^-$ mode is propagating.

\begin{figure*}[t!]
	\includegraphics[width=\linewidth]{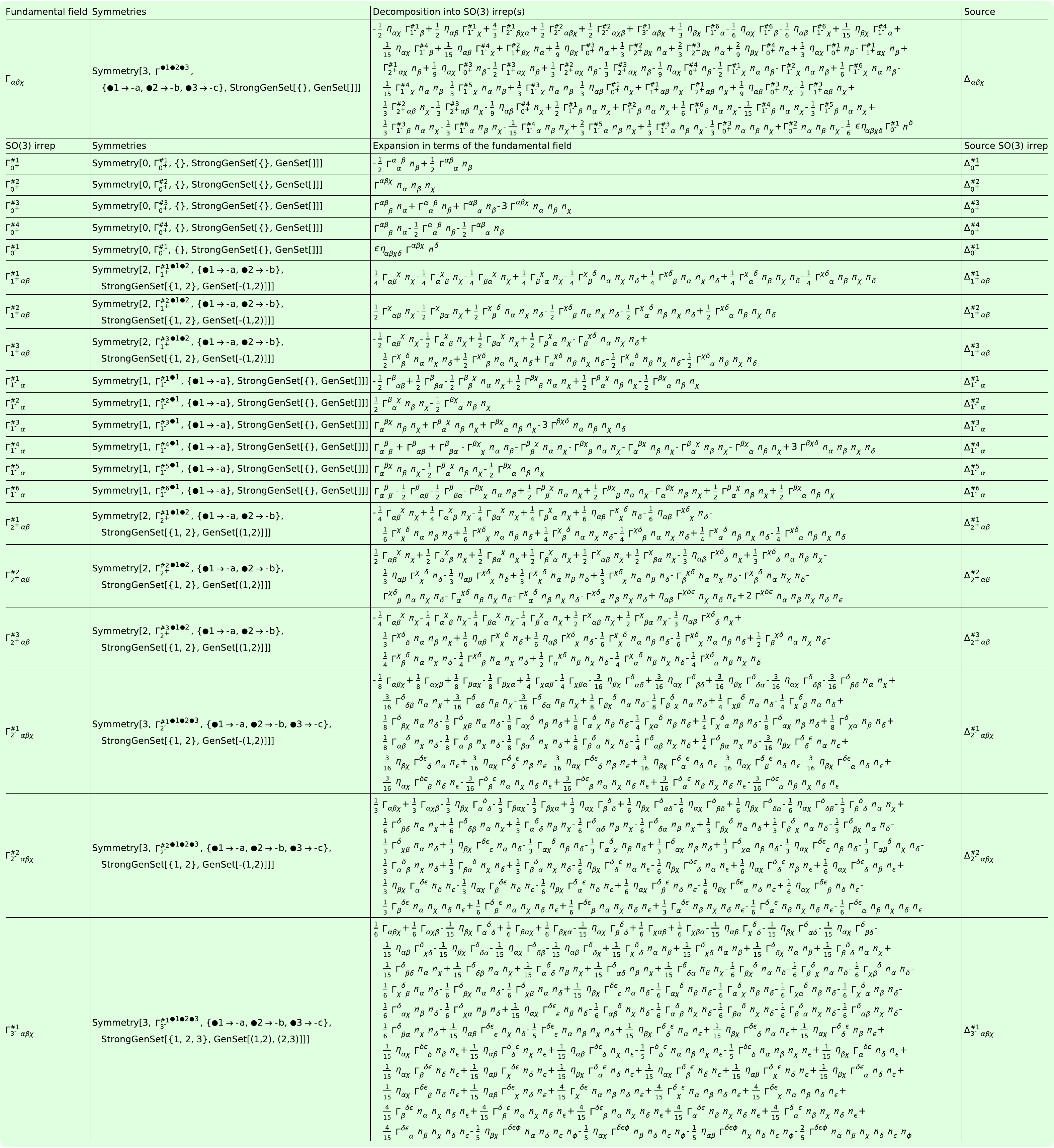}
	\caption{The declaration of an asymmetric, rank-three tensor field \lstinline!Connection!. Compared to the pair-antisymmetric case of \lstinline!SpinConnection! in~\cref{FieldKinematicsSpinConnection}, there are many more~$\sothree{}$ irreps, including a comparatively complicated~$3^-$ irrep. These definitions are used in~\cref{ParticleSpectrographMetricAffineEinsteinHilbertTheory,ParticleSpectrographProjectiveTheory}.}
\label{FieldKinematicsConnection}
\end{figure*}

\subsection{Metric-affine gravity}
\paragraph*{Kinematics} In the PGT formulation of gravity in~\cref{PGTVersion}, the connection field is allowed to become fully independent. However, the spin connection~$\tensor{\omega}{^{ij}_\mu}$ introduces fewer than the maximum possible number of extra d.o.f from a rank-three field, because it is antisymmetric. Meanwhile, the use of asymmetric tetrads~$\tensor{e}{_i^\mu}$ in PGT introduces six extra d.o.f which never propagate because of the Poincar\'e symmetry. These features both change in the metric-affine gravity (MAG) formulation, which introduces non-metricity as well as torsion to the geometry of the spacetime. The kinematics of MAG involve the metric~$\tensor{g}{_{\mu\nu}}$ defined already in~\cref{TensorTheory}, and an independent but completely asymmetric connection~$\MAGA{_\mu^\nu_\rho}$. We take care to use precisely the same conventions for MAG as~\cite{Percacci:2020ddy,Marzo:2021iok,Barker:2024dhb}. For the curvature, torsion and non-metric tensors we therefore have
\begin{subequations}
\begin{align}
	\MAGF{_{\mu\nu}^\rho_\sigma}&\equiv 2\left(\tensor{\partial}{_{[\mu}}\MAGA{_{\nu]}^\rho_\alpha}+\MAGA{_{[\mu|}^\rho_\alpha}\MAGA{_{|\nu]}^\alpha_\sigma}\right),\label{FDef}\\
	\MAGT{_\mu^\alpha_\nu}&\equiv 2\MAGA{_{[\mu|}^\alpha_{|\nu]}},\label{TDef}\\
	\MAGQ{_{\lambda\mu\nu}}&\equiv-\tensor{\partial}{_\lambda}\FieldG{_{\mu\nu}}+\MAGA{_\lambda^\alpha_\mu}\FieldG{_{\alpha\nu}}+\MAGA{_\lambda^\alpha_\nu}\FieldG{_{\mu\alpha}}.\label{QDef}
\end{align}
\end{subequations}
If the additional kinematic restriction~$\MAGQ{_{\lambda\mu\nu}}\equiv 0$ were to be imposed, the MAG torsion and curvature in~\cref{TDef,FDef} would be precisely analogous to the PGT counterparts in~\cref{PGTTorsion,PGTCurvature} respectively, through the relations~$\MAGT{_{\mu}^\alpha_{\nu}}\equiv\tensor{e}{^i_\mu}\tensor{e}{_k^\alpha}\tensor{e}{^j_\nu}\tensor{\mathscr{T}}{^k_{ij}}$ and~$\MAGF{_{\mu\nu}^\rho_\sigma}\equiv\tensor{e}{^i_\mu}\tensor{e}{^j_\nu}\tensor{e}{_k^\rho}\tensor{e}{_l_\sigma}\tensor{\mathscr{R}}{^{k}_{lij}}$, where we pay attention to the different ordering of the indices according to the two sets of conventions. Next,~\cref{FDef,TDef,QDef} can be further contracted to give 
\begin{align}
	\MAGT{_\mu}&\equiv\MAGT{_\alpha^\alpha_\mu},\quad
		\MAGQ{_\mu}\equiv\MAGQ{_{\mu\alpha}^\alpha},\quad
		\MAGQt{_\mu}\equiv\MAGQ{_\alpha^\alpha_\mu},\quad
		\MAGF{}\equiv\MAGF{_{\mu\nu}^{\mu\nu}},\nonumber\\
		\MAGF{_{\mu\nu}}&\equiv\MAGF{_{\mu\nu\alpha}^\alpha}\quad
		\MAGFa{_{\mu\nu}}\equiv\MAGF{_{\alpha\mu\nu}^\alpha},\quad
		\MAGFb{_{\mu\nu}}\equiv\MAGF{_{\alpha\mu}^\alpha_{\nu}}.\label{Contractions}
\end{align}
Some of these quantities have extra nomenclature, so that~$\MAGF{_{\mu\nu}}$ is the \emph{homothetic} curvature~\cite{BeltranJimenez:2016wxw,Iosifidis:2018jwu}, while~$\MAGFb{_{\mu\nu}}$ and~$\MAGFa{_{\mu\nu}}$ are variously \emph{pseudo}-Ricci tensors~\cite{Percacci:2020ddy}, or~$\MAGFa{_{\mu\nu}}$ is the \emph{co}-Ricci~\cite{BeltranJimenez:2016wxw}, and~$\MAGQ{_\mu}$ is the \emph{Weyl} vector~\cite{Iosifidis:2018zwo,BeltranJimenez:2014iie,Iosifidis:2018diy,Helpin:2019vrv,OrejuelaGarcia:2020viw,Ghilencea:2020piz,BeltranJimenez:2020sih,Xu:2020yeg,Yang:2021fjy,Quiros:2021eju,Quiros:2022uns,Yang:2022icz,Burikham:2023bil,Haghani:2023nrm} for which the homothetic curvature is also the Maxwell strength~$\MAGF{_{\mu\nu}}\equiv\tensor{\partial}{_{[\mu}}\MAGQ{_{\nu]}}$, while~$\MAGQt{_\mu}$ does not have a name~\cite{Iosifidis:2018jwu}, and~$\MAGT{_\mu}$ is the \emph{torsion contraction}~\cite{Lin:2018awc,Lin:2019ugq}, and~$\MAGF{}$ is still just the Ricci scalar. By combining~\crefrange{FDef}{Contractions} the general parity-preserving action with 28 coupling coefficients may be constructed 
\begin{align}
	&S_{\text{MAG}}=-\frac{1}{2}\int\mathrm{d}^4x\sqrt{-g}\bigg[
	-a_0\MAGF{}
	+\MAGF{^{\mu\nu\rho\sigma}}\Big(
	c_1\MAGF{_{\mu\nu\rho\sigma}}
	\nonumber\\
	&
	+c_2\MAGF{_{\mu\nu\sigma\rho}}
	+c_3\MAGF{_{\rho\sigma\mu\nu}}
	+c_4\MAGF{_{\mu\rho\nu\sigma}}
	+c_5\MAGF{_{\mu\sigma\nu\rho}}
	+c_6\MAGF{_{\mu\sigma\rho\nu}}
	\Big)
	\nonumber\\
	&
		+\MAGFb{^{\mu\nu}}
		\Big(
			c_7\MAGFb{_{\mu\nu}}
			+c_8\MAGFb{_{\nu\mu}}
		\Big)
		+\MAGFa{^{\mu\nu}}
		\Big(
			c_9\MAGFa{_{\mu\nu}}
	\nonumber\\
	&
			+c_{10}\MAGFa{_{\nu\mu}}
		\Big)
		+\MAGFa{^{\mu\nu}}
		\Big(
			c_{11}\MAGFb{_{\mu\nu}}
			+c_{12}\MAGFb{_{\nu\mu}}
		\Big)
	\nonumber\\
	&
		+\MAGF{^{\mu\nu}}
			\Big(
			c_{13}\MAGF{_{\mu\nu}}
			+c_{14}\MAGFb{_{\mu\nu}}
			+c_{15}\MAGFa{_{\mu\nu}}
			\Big)
		+c_{16}\MAGF{}^2
	\nonumber\\
	&
	+\MAGT{^{\mu\rho\nu}}
		\Big(
			a_1\MAGT{_{\mu\rho\nu}}
			+a_2\MAGT{_{\mu\nu\rho}}
		\Big)
		+a_3\MAGT{_\mu}\MAGT{^\mu}
	+\MAGQ{^{\rho\mu\nu}}
		\Big(
			a_4\MAGQ{_{\rho\mu\nu}}
	\nonumber\\
	&
			+a_5\MAGQ{_{\nu\mu\rho}}
		\Big)
		+a_6\MAGQ{_\mu}\MAGQ{^\mu}
		+a_7\MAGQt{_\mu}\MAGQt{^\mu}
		+a_8\MAGQ{_\mu}\MAGQt{^\mu}
	\nonumber\\
	&
		+a_9\MAGQ{_{\mu\rho\nu}}\MAGT{^{\mu\rho\nu}}
		+\MAGT{^\mu}
		\Big(
		a_{10}\MAGQ{_\mu}
		+a_{11}\MAGQt{_\mu}
		\Big)
		-2L_{\text{M}}
	\bigg].\label{MetricAffineGravityAction}
\end{align}
In the weak-field regime,~$\MAGA{_\mu^\alpha_\nu}$ is already taken to be perturbative, just like the spin connection. The conjugate source is known as the \emph{hypermomentum}~$\tensor{\Delta}{^\mu_\alpha^\nu}$. Accordingly we define the field \lstinline!Connection!:
\begin{lstlisting}
In[#]:= DefField[Connection[-a, -b, -c], PrintAs -> "\[CapitalGamma]", PrintSourceAs -> "\[CapitalDelta]"]; 
\end{lstlisting}
The output is shown in~\cref{FieldKinematicsConnection}. For the metric perturbation, we recycle our conventions for~$\tensor*{h}{_{\mu\nu}}$ in~\cref{DefMetricPerturbation} and the previously-defined field \lstinline!MetricPerturbation!. The couplings defined in~\cref{MetricAffineGravityAction} must also be defined:
\begin{lstlisting}
In[#]:= DefConstantSymbol[A0, PrintAs -> "\!\(\ * SubscriptBox[\(\[ScriptA]\), \(0\)]\)"]; DefConstantSymbol[A1, PrintAs -> "\!\(\ * SubscriptBox[\(\[ScriptA]\), \(1\)]\)"]; DefConstantSymbol[A2, PrintAs -> "\!\(\ * SubscriptBox[\(\[ScriptA]\), \(2\)]\)"]; DefConstantSymbol[A3, PrintAs -> "\!\(\ * SubscriptBox[\(\[ScriptA]\), \(3\)]\)"]; (*omitted 1490 characters for brevity*) PrintAs -> "\!\(\ * SubscriptBox[\(\[ScriptC]\), \(13\)]\)"]; DefConstantSymbol[C14, PrintAs -> "\!\(\ * SubscriptBox[\(\[ScriptC]\), \(14\)]\)"]; DefConstantSymbol[C15, PrintAs -> "\!\(\ * SubscriptBox[\(\[ScriptC]\), \(15\)]\)"]; DefConstantSymbol[C16, PrintAs -> "\!\(\ * SubscriptBox[\(\[ScriptC]\), \(16\)]\)"];
\end{lstlisting}
Note that these couplings are used in~\cref{MetricAffineGravityAction} in a more straightforward way than the couplings in~\cref{PGTVersion}.

\begin{figure*}[t!]
	\includegraphics[width=\linewidth]{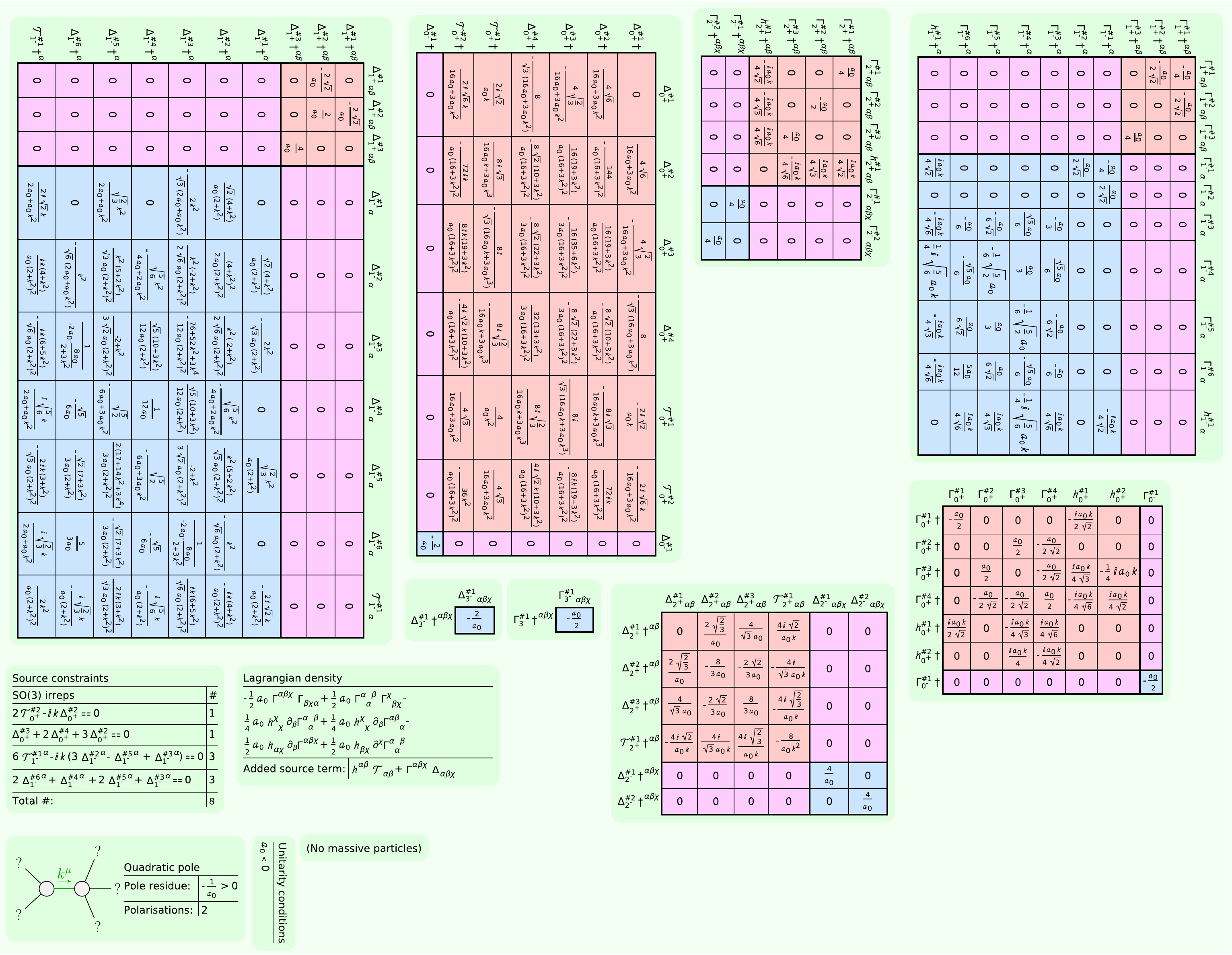}
	\caption{The particle spectrum of the Einstein--Hilbert action in metric-affine gravity, obtained by setting all the couplings in~\cref{MetricAffineGravityAction} to zero except for~$a_0$. The results are consistent with~\cref{ParticleSpectrographEinsteinCartanTheory,ParticleSpectrographFierzPauliTheory}. Unlike in the case of~\cref{ParticleSpectrographEinsteinCartanTheory}, there are only eight symmetries. Four symmetry generators are associated with the diffeomorphism gauge symmetry and the conservation of the matter stress-energy tensor. There are also four extra symmetries which follow from projective invariance. All quantities are defined in~\cref{FieldKinematicsMetricPerturbation,FieldKinematicsConnection}.}
\label{ParticleSpectrographMetricAffineEinsteinHilbertTheory}
\end{figure*}

\paragraph*{Einstein--Hilbert theory} The basic theory to consider is again that with an Einstein--Hilbert action. The linearisation of this action near Minkowski spacetime yields:
\begin{lstlisting}
In[#]:= ParticleSpectrum[-1/2 * (A0 * Connection[a, b, c] * Connection[-b, -c, -a]) + (A0 * Connection[a, -a, b] * Connection[c, -b, -c])/2 - (A0 * MetricPerturbation[c, -c] * CD[-b][Connection[a, -a, b]])/4 + (A0 * MetricPerturbation[c, -c] * CD[-b][Connection[a, b, -a]])/4 - (A0 * MetricPerturbation[-a, -c] * CD[-b][Connection[a, b, c]])/2 + (A0 * MetricPerturbation[-b, -c] * CD[c][Connection[a, -a, b]])/2, TheoryName -> "MetricAffineEinsteinHilbertTheory", Method -> "Hard", MaxLaurentDepth -> 3]; 
\end{lstlisting}
The output is shown in~\cref{ParticleSpectrographMetricAffineEinsteinHilbertTheory}. Despite the addition of all the extra kinematic ingredients in~\cref{FieldKinematicsConnection}, the spectrum is identical to that in~\cref{ParticleSpectrographFierzPauliTheory}. As with the general PGT in~\cref{ParticleSpectrographGeneralPGT}, it is possible to study more general versions of MAG. Some initial analyses were already made using \PSALTer{} in~\cite{Barker:2024ydb}, and the matter is subject to ongoing investigation (see e.g.~\cite{Percacci:2020ddy,Mikura:2024mji}). 

\begin{figure*}[t!]
	\includegraphics[width=\linewidth]{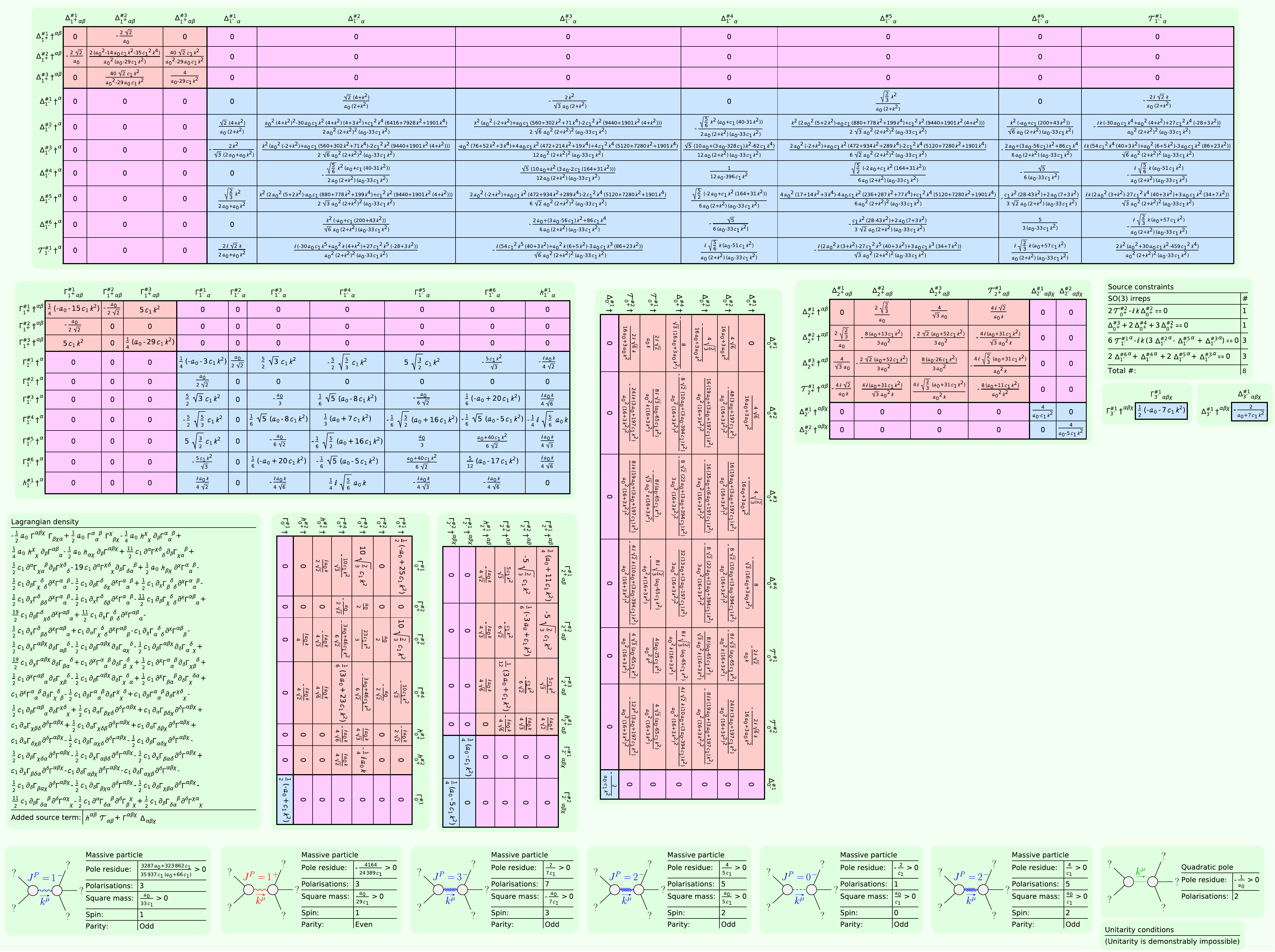}
	\caption{The particle spectrum of one-parameter projective-invariant metric-affine gravity from~\cite{Barker:2024dhb}, as set out in~\cref{ProjTheory}, to be compared with~\cref{ParticleSpectrographMetricAffineEinsteinHilbertTheory}. Whilst the projective invariance is preserved, the particle spectrum is clearly sick. All quantities are defined in~\cref{FieldKinematicsMetricPerturbation,FieldKinematicsConnection}.}
\label{ParticleSpectrographProjectiveTheory}
\end{figure*}

\paragraph*{Projective theory} As an alternative to the Einstein--Hilbert action, we can study the one-parameter projective-invariant MAG which was considered in~\cite{Barker:2024dhb}
\begin{align}
	&S_{\text{Projective}} = \int\mathrm{d}^4x\sqrt{-g}\Bigg[\frac{a_0}{2}\MAGF{}
	+c_1\bigg(\MAGF{^{\rho\sigma\mu\nu}}\Big[
	2\MAGF{_{(\rho\sigma)\mu\nu}}
	\nonumber\\
	&
	+\MAGF{_{\mu\nu\rho\sigma}}
	+\MAGF{_{\nu\sigma\mu\rho}}
	+2\MAGF{_{(\nu\rho)\mu\sigma}}
	\Big]
	-2\MAGFb{^{(\mu\nu)}}
	\MAGFb{_{(\mu\nu)}}
	\nonumber\\
	&
	+\MAGFa{^{\mu\nu}}\Big[
	\MAGFa{_{\mu\nu}}
	-19\MAGFa{_{\nu\mu}}
	\Big]
	+\MAGF{^{\mu\nu}}\Big[
	\MAGF{_{\mu\nu}}
	-\MAGFb{_{\mu\nu}}
	\nonumber\\
	&
	-11\MAGFa{_{\mu\nu}}
	\Big]
	+\MAGFa{^{\mu\nu}}
	\MAGFb{_{\mu\nu}}
	+\MAGF{}^2\bigg)
	+L_{\text{M}}
	\Bigg] \;.
\label{ProjTheory}
\end{align}
In this model, we extend the Einstein--Hilbert term by highly specific quadratic curvature invariants which are associated with the preservation of projective invariance. The linearisation of this action near Minkowski spacetime yields:
\begin{lstlisting}
In[#]:= ParticleSpectrum[-1/2 * (A0 * Connection[a, b, c] * Connection[-b, -c, -a]) + (A0 * Connection[a, -a, b] * Connection[c, -b, -c])/2 - (A0 * MetricPerturbation[c, -c] * CD[-b][Connection[a, -a, b]])/4 + (A0 * MetricPerturbation[c, -c] * CD[-b][Connection[a, b, -a]])/4 - (A0 * MetricPerturbation[-a, (*omitted 3200 characters for brevity*) - (11 * C1 * CD[-b][Connection[-d, -a, b]] * CD[d][Connection[a, c, -c]])/2 - (C1 * CD[a][Connection[-d, -a, b]] * CD[d][Connection[-b, c, -c]])/2 + (C1 * CD[-b][Connection[-d, -a, b]] * CD[d][Connection[c, a, -c]])/2, TheoryName -> "ProjectiveTheory", Method -> "Hard", MaxLaurentDepth -> 1];
\end{lstlisting}
The output is shown in~\cref{ParticleSpectrographProjectiveTheory}. As observerd in~\cite{Barker:2024dhb}, the model is clearly sick: the failure of projective symmetry highlights the severe difficulty in proposing self-consistent models which differ from GR. This concludes the worked examples with code. Further examples can be found in~\cref{FurtherExamples}.


\begin{table}
\caption{\label{IndicesTable} Indices and labels used throughout~\cref{TheoreticalDevelopment}. No summation is assumed over repeated non-spacetime indices.}
\begin{center}
\begin{tabularx}{\linewidth}{c|c|X}
\hline\hline
	Indices & Values & Meaning\\
\hline
	$\mu,\ \nu$ &~$\{0,1,2,3\}$ & Spacetime indices\\
	$\ovl\mu,\ \ovl\nu$ &~$\{0,1,2,3\}$ & `Parallel' spacetime indices which have been projected with~$\tensor*{\delta}{^\mu_\nu}-\tensor{n}{^\mu}\tensor{n}{_\mu}$\\
	$\perp$ & (none) & `Perpendicular' spacetime index which was contracted with~$\tensor{n}{^\mu}$\\
	$X,\ Y$ & (symbolic) & Distinct Lorentz-covariant tensor fields\\
	$J,\ \Jp{}$ &~$\mathbb{Z}^{\geq}$ & Spin\\
	$P,\ \Pp{}$ &~$\{1,-1\}$ & Parity\\
	$s_{J^P},\ \Sp{}_{J^P}$ &~$\mathbb{Z}^{>}$ & Slots for masses (if any) associated with a given~$J^P$ state\\
	$\mu_X,\ \nu_X$ &~$\{0,1,2,3\}^{\mathbb{Z}^{\geq}}$ & Collections of (perhaps) symmetrised spacetime indices which are associated with the field~$X$\\
	$\ovl\mu_{J^P},\ \ovl\nu_{J^P}$ &~$\{0,1,2,3\}^{\mathbb{Z}^{\geq}}$ & Collections of symmetrised `parallel' spacetime indices which are associated with a given~$J^P$ state\\
	$\States{J}{P}{X}{i},\ \States{J}{P}{X}{j}$ &~$\mathbb{Z}^{>}$ & Multiple independent copies of a given~$J^P$ state which are associated with the field~$X$\\
	$a_{J^P},\ \Ap{}_{J^P}$ &~$\mathbb{Z}^{>}$ & Slots for the null eigenvectors (if any) of the~$J^P$ wave operator block~$\WaveOperatorJP{J}{P}$\\
	$b,\ \Bp{}$ &~$\mathbb{Z}^{>}$ & Slots for the null eigenvectors (if any) of the constraint matrix~$\ConstraintMatrix{}$\\
	$c,\ \Cp{}$ &~$\mathbb{Z}^{>}$ & Slots for the non-null eigenvectors (if any) of the reduced propagator~$\NewRes{\En{}}{\Mo{}}\left(\ReducedPropagator{}\right)$\\
\hline\hline
\end{tabularx}
\end{center}
\end{table}

\section{Theoretical development}\label{TheoreticalDevelopment}
\paragraph*{Further conventions} Having motivated the SPO algorithm in~\cref{Introduction} and illustrated its implementation in~\cref{SymbolicImplementation}, we will now describe it at a technical level (see also~\cite{Aurilia:1969bg,Buoninfante:2016iuf,Lin:2018awc,Lin:2020fuo}). Throughout this section, we will develop the theory with simple examples from the vector model introduced in~\cref{GeneralVectorLagrangian}. We will need several different kinds of indices and labels, which we summarise in~\cref{IndicesTable}. Comparing~\cref{GeneralVectorLagrangian} to~\cref{BasicPositionAction} we only need to worry about one field label~$X={\mathcal{B}}$, and so we make the single field definition~$\FieldDown{{\mathcal{B}}}{\mu}\equiv\B{_{\mu}}$, with the straightforward raising rule~$\FieldUp{{\mathcal{B}}}{\nu}\equiv\B{^{\nu}}\equiv\tensor{\eta}{^{\mu\nu}}\B{_{\mu}}$. The source conjugate to~$\B{_{\mu}}$ is~$\SourceUp{{\mathcal{B}}}{\mu}\equiv-\J{^\mu}$. The wave operator is~$\WaveOperatorTensorUpDown{{\mathcal{B}}}{{\mathcal{B}}}{\mu}{\nu}\equiv-\alpha\tensor*{\delta}{^{\mu}_{\nu}}\Box-\beta\tensor{\partial}{^\mu}\tensor{\partial}{_\nu}$ up to total derivatives.

\subsection{Spin projection}\label{SpinProjection}
\paragraph*{Theoretical development} Projection by spin~$J$ and parity~$P$ becomes useful when a preferred frame is available. The particle four-momentum~$\tensor{k}{^\mu}$ is timelike or null for massive and massless particles, respectively. In the timelike case we may define a unit-timelike vector
\begin{equation}\label{UnitTimelike}
	\N{^\mu}\equiv\tensor{k}{^\mu}/k,\quad k\equiv\sqrt{\tensor{k}{^\mu}\tensor{k}{_\mu}},
\end{equation}
so that~$\N{^\mu}$ provides a preferred frame in which it coincides with the observer's four-velocity
\begin{equation}\label{UnitTimelikeComponents}
	\left[\N{^\mu}\right]^\mathrm{T}=\begin{bmatrix}1 & 0 & 0 & 0\end{bmatrix}.
\end{equation}
Such a frame constitutes a Cauchy slicing in the flat spacetime, and we will introduce a `parallel' and `perpendicular' notation for indices with respect to such slices. A `parallel' index, with the overbar notation, is thus defined as having zero contraction with~$\N{^\mu}$. The \emph{general} tensor fields~$\FieldDown{X}{\mu}$ were introduced in~\cref{BasicPositionAction}. It is possible to decompose these fields into~$J^P$ sectors
\begin{subequations}
\begin{align}
	\FieldDown{X}{\mu}&\equiv\sum_{J,P}\sum_{\States{J}{P}{X}{i}}\FieldDownFullState{J}{P}{X}{i}{\mu},\label{DefineDecomposition}
	\\
	\FieldDownFullState{J}{P}{X}{i}{\mu}&\equiv\SPODownUp{J}{P}{X}{X}{i}{i}{\mu}{\nu}\FieldDown{X}{\nu},\label{DefineProjection}
\end{align}
\end{subequations}
where the labels~$\States{J}{P}{X}{i}$ allow multiple independent states with the same~$J^P$ to be contained within the single field~$X$ (see~\cref{IndicesTable}). The particular \emph{spin-projection operators} (SPOs) in~\cref{DefineProjection} are diagonal in the sense that their arguments happen to be equal. In general, the first and second SPO arguments must share field labels with the first and second collections of Lorentz indices, respectively. The SPOs are constructed only from~$\tensor{\eta}{_{\mu\nu}}$,~$\tensor{\epsilon}{_{\mu\nu\sigma\lambda}}$, the unit-timelike vector~$\N{^\mu}$ defined in~\cref{UnitTimelike}, and the imaginary unit. They satisfy the following identities
\begin{subequations}
	\begin{align}
		&\SPODownUp{J}{P}{X}{Y}{i}{j}{\mu}{\nu}\equiv\SPOUpDown{J}{P}{Y}{X}{j}{i}{\nu}{\mu}^*,\label{Hermicity}
		\\
		&\sum_{J,P}\sum_{\States{J}{P}{X}{i}}\SPODownUp{J}{P}{X}{X}{i}{i}{\mu}{\nu}\equiv\tensor*{\mathbb{I}}{^{\nu_X}_{\mu_X}},\label{Completeness}
		\\	
		&\SPODownUp{J}{P}{X}{X}{i}{j}{\mu}{\nu}\SPODownUp{\Jp{}}{\Pp{}}{X}{X}{k}{l}{\nu}{\sigma}
		\nonumber\\
		&\hspace{70pt}\equiv
		\tensor*{\delta}{_{jk}}
		\tensor*{\delta}{_{J\Jp{}}}
		\tensor*{\delta}{_{P\Pp{}}}
		\SPODownUp{J}{P}{X}{X}{i}{l}{\mu}{\sigma},\label{Orthonormality}
		\\
		&P\FieldDown{X}{\mu}^*
		\SPOUpDown{J}{P}{X}{X}{i}{i}{\mu}{\nu}
		\FieldUp{X}{\nu}\geq 0,\label{Positivity}
	\end{align}
\end{subequations}
where~$\tensor*{\mathbb{I}}{^{\nu_X}_{\mu_X}}$ is the product of Kronecker symbols~$\tensor*{\delta}{^\nu_\mu}$, with one factor for each index associated with~$X$. Respectively,~\cref{Hermicity,Completeness,Orthonormality,Positivity} encode Hermicity, completeness, orthonormality and positive-definiteness. To compute the SPOs, we may proceed as follows. By standard means involving e.g. Young tableaux and trace-free decomposition, the \emph{reduced-index}~$J^P$ representations may be extracted by hand so as to define the reduced-index SPOs
\begin{equation}\label{ReducedIndex}
	\FieldDownState{J}{P}{X}{i}{\mu}
	\equiv
	\ReducedSPODownUp{J}{P}{X}{i}{\mu}{\nu}
	\FieldDown{X}{\nu},
\end{equation}
where~$\ParallelFieldIndices{J}{P}{\mu}$ is a reduced collection of `parallel' indices specific to the~$J^P$ state. The term \emph{reduced} here implies that there may be fewer indices than in~$\FieldIndices{X}{\mu}$ for some or all of the fields~$X$ which contain states with this~$J^P$. For example, a high-rank tensor may contain many scalar states, but scalars require no indices. All the reduced indices are necessarily `parallel', in the sense of~\cref{PerpPara}. The reduced-index SPOs in~\cref{ReducedIndex} are constructed from~$\tensor{\eta}{_{\mu\nu}}$,~$\tensor{\epsilon}{_{\mu\nu\sigma\lambda}}$ and~$\tensor{n}{^\mu}$, but \emph{not} the imaginary unit. They need not be normalised in any sense, and their definitions may vary from author to author. The full-index SPOs typically have a more complicated structure than their reduced-index counterparts, and are given by
\begin{equation}\label{Normalisation}
	\begin{aligned}
		&\SPODownUp{J}{P}{X}{Y}{i}{j}{\mu}{\nu}
	\equiv
	\Normalisation{J}{P}{X}{i}
	\Normalisation{J}{P}{Y}{j}
	\\
		&\hspace{90pt}
		\times
		\ReducedSPOUpDown{J}{P}{X}{i}{\sigma}{\mu}
		\ReducedSPODownUp{J}{P}{Y}{j}{\sigma}{\nu},
	\end{aligned}
\end{equation}
where the~$\Normalisation{J}{P}{X}{i}\in\mathbb{C}$ are fixed (each up to a sign) by the requirements of~\crefrange{Hermicity}{Positivity}. Because this normalisation is well suited to computer algebra, it is no longer appropriate to tabulate the full-index SPOs of any theory when presenting the results of spectral analysis\footnote{Such tables occupy a lot of space in older papers~\cite{VanNieuwenhuizen:1973fi,Neville:1978bk,Neville:1979rb,Sezgin:1981xs,Sezgin:1979zf,Kuhfuss:1986rb,Karananas:2014pxa,Karananas:2016ltn,Percacci:2019hxn,Mendonca:2019gco,Marzo:2021esg,Marzo:2021iok,Mikura:2023ruz,Mikura:2024mji,Aurilia:1969bg,Buoninfante:2016iuf,Lin:2018awc,Lin:2019ugq,Lin:2020phk,Lin:2020fuo}.}.

Because of the properties in~\crefrange{Hermicity}{Positivity}, all the cumbersome indices in~\cref{IndicesTable} can be replaced by a more natural matrix notation, 
\begin{equation}\label{ToMatrices1}
	\WaveOperatorTensorUpDown{X}{Y}{\mu}{\nu}\equiv\sum_{J,P}\sum_{\States{J}{P}{X}{i},\States{J}{P}{Y}{j}}\WaveOperatorJPComponents{J}{P}{X}{Y}{i}{j}\SPOUpDown{J}{P}{X}{Y}{i}{j}{\mu}{\nu},
\end{equation}
where, for each~$J^P$, the matrix~$\WaveOperatorJP{J}{P}$ is indexed by all the~$\States{J}{P}{X}{i}$ and~$\States{J}{P}{Y}{j}$. A block-diagonal matrix form for the whole operator can then be defined as
\begin{equation}\label{BuildWaveOperator}
\WaveOperator{}\equiv\bigoplus_{J,P}\WaveOperatorJP{J}{P}.
\end{equation}
Precisely this block-diagonal form was illustrated already throughout~\cref{SymbolicImplementation} in the many example outputs from \lstinline!ParticleSpectrum!. In that case, blocks associated with each~$J$ (but not the sub-blocks associated with each~$P$) are broken into individual boxes, which are arranged in a space-saving mosaic. The matrices can be obtained by the projection
\begin{equation}\label{ToMatrices2}
	\begin{aligned}
	&\WaveOperatorJPComponents{J}{P}{X}{Y}{i}{j}
	\SPOUpDown{J}{P}{X}{Y}{i}{j}{\mu}{\nu}
	\equiv
	\\
	&\hspace{50pt}
	\SPOUpDown{J}{P}{X}{X}{i}{i}{\mu}{\sigma}
	\WaveOperatorTensorUpDown{X}{Y}{\sigma}{\lambda}
	\SPOUpDown{J}{P}{Y}{Y}{j}{j}{\lambda}{\nu}.
	\end{aligned}
\end{equation}
The same techniques apply to the fields. Using~\cref{Completeness,Normalisation} we define the field vector~$\Field{}$ through its inner product
\begin{align}\label{InnerProduct}
	\FieldConj{}\cdot\Field{}&\equiv
	\sum_X\FieldDown{X}{\mu}^*\FieldUp{X}{\mu}
	\nonumber
	\\	
	&
	\equiv\sum_{J,P,X}\sum_{\States{J}{P}{X}{i}}\Normalisation{J}{P}{X}{i}^2\FieldDownState{J}{P}{X}{i}{\mu}^*\FieldUpState{J}{P}{X}{i}{\mu}.
\end{align}
The final equality of~\cref{InnerProduct} leaves the actual~$\Field{}$ components ambiguous, i.e. it does not define the individual elements which match the dimensions of the~$\WaveOperator{}$ matrix. From~\cref{ToMatrices1,BuildWaveOperator,ToMatrices2} the~$\WaveOperator{}$ matrix has one \emph{scalar}-valued element (linear in the Lagrangian couplings of~\cref{BasicPositionAction}, and rational in~$k$ from~\cref{UnitTimelike}), for each~$\big(\States{J}{P}{X}{i},\States{J}{P}{Y}{j}\big)$ pair. However, each~$\States{J}{P}{X}{i}$ state is contributing a \emph{quadratic form} of tensor components to~\cref{InnerProduct}, rather than the square of a scalar. One way to properly accommodate these hidden multiplicities would be to modify~\cref{BuildWaveOperator}, so that~$\WaveOperator{}$ is built up from the blocks~$\WaveOperatorJP{J}{P}\otimes\IdentityJ{2J+1}$, where~$\IdentityJ{2J+1}$ is the~$(2J+1)\times(2J+1)$ identity, and to separate out the azimuthal spin states in~$\Field{}$ with some polarisation basis vectors. Since this is complicated, we gloss over all such details within this section, and treat each~$\States{J}{P}{X}{i}$ component in~$\Field{}$ as if it were indeed a scalar. In doing so, we will however need to be mindful when going over to the massless limit in~\cref{MasslessSpectrum} that, according to~\cref{InnerProduct}, the squares of these components have a more involved structure. By applying the convolution theorem, and remembering that for real fields~$\FieldDown{Y}{\mu}(x)$ we have~$\FieldDown{X}{\mu}(-k)\equiv\FieldDown{X}{\mu}(k)^*$, the position representation of the action in~\cref{BasicPositionAction} can now be re-written using~\cref{BuildWaveOperator,InnerProduct} as
\begin{align}
	S_{\text{F}}
	&
	=
	\int\mathrm{d}^4x
	\sum_X\FieldDown{X}{\mu}\left(x\right)\Bigg[\sum_Y\WaveOperatorTensorUpDown{X}{Y}{\mu}{\nu}\left(\partial\right)\FieldUp{Y}{\nu}\left(x\right)
	-\SourceUp{X}{\mu}\left(x\right)\Bigg]
	\nonumber
	\\
	&
	=
	\int\mathrm{d}^4k
	\sum_X\FieldDown{X}{\mu}\left(-k\right)\Bigg[\sum_Y\WaveOperatorTensorUpDown{X}{Y}{\mu}{\nu}\left(k\right)\FieldUp{Y}{\nu}\left(k\right)
	-\SourceUp{X}{\mu}\left(k\right)\Bigg]
	\nonumber
	\\
	&=\int\mathrm{d}^4k\ 
	\FieldConj{}\cdot\left[\WaveOperator{}\cdot\Field{}
	-\Source{}\right],\label{OperatorMomentumAction}
\end{align}
where we use the source vector~$\Source{}$ by analogy\footnote{Since~$\SourceUp{X}{\mu}\left(x\right)$ are also real, we could equally write~$\int\mathrm{d}^4k\FieldConj{}\cdot\Source{}\equiv\frac{1}{2}\int\mathrm{d}^4k\big[\FieldConj{}\cdot\Source{}+\SourceConj{}\cdot\Field{}\big]$.} to~$\Field{}$. Assuming a non-higher-derivative model, the wave operator in~\cref{OperatorMomentumAction} may be written in terms of three constant, real and a priori \emph{asymmetric} matrices~$\AMat{}$,~$\BMat{}$ and~$\CMat{}$
\begin{equation}\label{MinkowskiLinearisation}
	\WaveOperator{}\left(k\right)\equiv-\AMat{}k^2-i\BMat{} k+\CMat{},
\end{equation}
respectively of mass dimensions zero, one and two, and depending only linearly on the coupling coefficients of the theory --- recall in~\cref{MinkowskiLinearisation} that~$k$ is the momentum norm, as defined in~\cref{UnitTimelike}, and our Fourier convention replaces~$\tensor{\partial}{_\mu}$ with~$-i\tensor{k}{_\mu}$. Taking the vanishing on-shell variation of~\cref{BasicPositionAction} and following the same steps as in~\cref{OperatorMomentumAction}, we find
\begin{equation}\label{Variations}
	\delta S_{\text{F}}
	=\int\mathrm{d}^4k\ 
	\delta\FieldConj{}\cdot\left[\tilde\WaveOperator{}\cdot\Field{}
	-\Source{}\right]\approx 0,
\end{equation}
where the total derivatives used in~\cref{Variations} to extract the field equations as an overall right-factor under the integral ensure that the \emph{physical} part~$\tilde{\mathsf{O}}\left(k\right)$ of the wave operator~$\mathsf{O}\left(k\right)$ --- i.e. the part appearing in the field equations --- is given by the formula
\begin{equation}\label{MinkowskiOperator}
	\tilde{\mathsf{O}}\left(k\right)\equiv
		-\frac{\left(\mathsf{A}+\mathsf{A}^{\mathrm{T}}\right)k^2}{2}
		-\frac{i\left(\mathsf{B}-\mathsf{B}^{\mathrm{T}}\right) k}{2}
		+\frac{\left(\mathsf{C}+\mathsf{C}^{\mathrm{T}}\right)}{2}.
\end{equation}
We conclude from~\cref{MinkowskiOperator} that the physical part~$\tilde{\mathsf{O}}\equiv\frac{1}{2}\left(\mathsf{O}+\mathsf{O}^\dagger\right)$ is just the \emph{Hermitian} part of the original wave operator. Indeed, using similar total derivatives, the original wave operator in~\cref{MinkowskiLinearisation} can always be made to be Hermitian, so that~$\WaveOperator{}\equiv\WaveOperatorConj{}\equiv\tilde{\WaveOperator{}}$. This argument also extends to higher-derivative models. Going forwards, we will always assume Hermicity, since it simplifies the computations. The matrices displayed in \lstinline!ParticleSpectrum! outputs are always Hermitian.

\paragraph*{Examples} To illustrate the `parallel' and `perpenducular' indices, the vector field~$\B{_\mu}$ introduced in~\cref{GeneralVectorLagrangian} has components
\begin{equation}\label{PerpPara}
	\B{_{\ovl{\mu}}}\equiv\big(\tensor*{\delta}{^\nu_\mu}-\N{^{\nu}}\N{_\mu}\big)\B{_\nu}, \quad \B{_\perp}\equiv\N{^\mu}\B{_\mu}.
\end{equation}
The reduced-index~$J^P$ parts of the field defined in~\cref{ReducedIndex} are already obtained by~\cref{PerpPara}, specifically 
\begin{equation}\label{ReducedIndexDefinitions}
	\FieldDownState{0}{+}{{\mathcal{B}}}{1}{\cdot}\equiv\B{_\perp}, \quad 
	\FieldDownState{1}{-}{{\mathcal{B}}}{1}{\mu}\equiv\B{_{\ovl{\mu}}}.
\end{equation}
For each of the~$J^P$ sectors, there is only one state, and so we need not go beyond the label~$\States{J}{P}{{\mathcal{B}}}{i}\equiv\States{J}{P}{{\mathcal{B}}}{1}$ in either case. We are free to choose the normalisations in~\cref{ReducedIndexDefinitions}, and they imply the reduced-index SPOs
\begin{equation}\label{ReducedIndexSPODefinitions}
	\ReducedSPODownUp{0}{+}{{\mathcal{B}}}{1}{\cdot}{\nu}\equiv\N{^\nu}, \quad 
	\ReducedSPODownUp{1}{-}{{\mathcal{B}}}{1}{\mu}{\nu}\equiv\tensor*{\delta}{^\nu_\mu}-\N{^{\nu}}\N{_\mu}.
\end{equation}
From~\cref{ReducedIndexSPODefinitions,Normalisation} we obtain the diagonal, full-index SPOs~$\SPODownUp{0}{+}{{\mathcal{B}}}{{\mathcal{B}}}{1}{1}{\mu}{\nu}\equiv\N{_\mu}\N{^\nu}$ and~$\SPODownUp{1}{-}{{\mathcal{B}}}{{\mathcal{B}}}{1}{1}{\mu}{\nu}\equiv\tensor*{\delta}{^\nu_\mu}-\N{^{\nu}}\N{_\mu}$, with normalisation solutions~$\Normalisation{0}{+}{{\mathcal{B}}}{1}\equiv\pm 1$ and~$\Normalisation{1}{-}{{\mathcal{B}}}{1}\pm 1$. To summarise,~\cref{ReducedIndexDefinitions,ReducedIndexSPODefinitions,DefineDecomposition,DefineProjection} just reduce to the obvious statement
\begin{equation}\label{VectorDecomposition}
	\B{_\mu}\equiv\N{_\mu}\B{_\perp}+\B{_{\ovl{\mu}}}.
\end{equation}
These results were shown in~\cref{FieldKinematicsVectorField}. Whilst the notation introduced throughout~\cref{SpinProjection} is evidently overpowered here\footnote{Note also by comparing~\cref{ReducedIndexDefinitions} with~\cref{FieldKinematicsVectorField} that the reduced-index~$J^P$ representations have a less formal notation in the \PSALTer{} output: we write~$\tensor*{{\mathcal{B}}}{_{0^+}^{\#1}}$ for~$\FieldDownState{0}{+}{{\mathcal{B}}}{1}{\cdot}$, and~$\tensor*{{\mathcal{B}}}{_{1^-\mu}^{\#1}}$ for~$\FieldDownState{1}{-}{{\mathcal{B}}}{1}{\mu}$, omitting the `parallel' index notation.}, for theories with multiple fields and degenerate~$J^P$ states it serves to concretely define the matrices~$\WaveOperatorJP{J}{P}$. These matrices are central to the \PSALTer{} implementation in~\cref{SymbolicImplementation}.

\subsection{Massive spectrum}\label{MassiveSpectrum}
\paragraph*{Theoretical development} In~\cref{SpinProjection} we reduced the position-space action in~\cref{BasicPositionAction} to the canonical momentum-space matrix form in~\cref{OperatorMomentumAction}. We now proceed to evaluate the saturated propagator. The wave operator block for a certain~$J^P$ may have a collection of (normalised, but not necessarily orthogonal) null right eigenvectors 
\begin{equation}
\WaveOperatorJP{J}{P}\cdot\NullVector{J}{P}{a}\equiv 0,\quad
\NullVectorConj{J}{P}{a}\cdot\NullVector{J}{P}{a}\equiv 1,
\end{equation}
so that~$\NullVectorConj{J}{P}{a}$ are the null left eigenvectors by the Hermicity of~$\mathsf{O}$. Because of the presence of null eigenvectors, or repeated zero eigenvalues,~$\WaveOperator{}$ is generically singular. The nullity is equal to the number of independent gauge symmetries (or the number of gauge \emph{generators}) in the theory. Variations~$\delta\upzeta^{\mathrm{T}}(-k)\equiv\delta\FieldConj{}=\bigoplus_{J,P}\sum_{\NullVectors{J}{P}{a}}\GaugeVaryingConj{J}{P}{a}\NullVectorConj{J}{P}{a}$ parameterised by some scalars~$\GaugeVarying{J}{P}{a}\in\mathbb{C}$ are purely gauge, and by~\cref{Variations} they imply the on-shell source constraints
\begin{equation}
	\Source{}\approx\Similarity{}\cdot\Source{},\quad
	\Similarity{}\equiv\bigoplus_{J,P}\left[\mathsf{1}-\sum_{\NullVectors{J}{P}{a}}\NullVector{J}{P}{a}\cdot\NullVectorConj{J}{P}{a}\right].
	\label{FundamentalSourceConstraints}
\end{equation}
When source gauge symmetries or source constraints are present, the na\"ive formula~$\Propagator{}\approx\SourceConj{}\cdot\WaveOperator{}^{-1}\cdot\Source{}$ for the saturated propagator fails. This problem can be solved by the addition of gauge-fixing terms to the original formulation in~\cref{BasicPositionAction}. The gauge-fixed wave operator~$\WaveOperator{}+\bigoplus_{J,P}\sum_{\NullVectors{J}{P}{a}}\GaugeFixing{J}{P}{a}\NullVector{J}{P}{a}\cdot\NullVectorConj{J}{P}{a}$ is invertible for non-zero gauge-fixing parameters~$\GaugeFixing{J}{P}{a}\in\mathbb{R}^*$, and taking care to use the physical sources~$\Similarity{}\cdot\Source{}$, the saturated propagator is
\begin{equation}\label{GaugeFixed}
	\Propagator{}\approx \SourceConj{}\cdot\Similarity{}\cdot
	\left[\WaveOperator{}+\bigoplus_{J,P}\sum_{\NullVectors{J}{P}{a}}\GaugeFixing{J}{P}{a}\NullVector{J}{P}{a}\cdot\NullVectorConj{J}{P}{a}\right]^{-1}
	\cdot
	\Similarity{}
	\cdot\Source{}.
\end{equation}
We do not really care about the advantages of various gauge fixings at tree level, so it is convenient to set all the~$\GaugeFixing{J}{P}{a}$ to unity in~\cref{GaugeFixed} and find on-shell
\begin{equation}\label{PropagatorDefinition}
	\Propagator{}\approx \SourceConj{}\cdot\MoorePenrose{}\cdot\Source{},
\end{equation}
where~$\MoorePenrose{}$ is the unique Moore--Penrose~\cite{Moore:1920,Penrose:1955} or Drazin~\cite{Drazin:1958} pseudoinverse of the Hermitian wave operator~$\WaveOperator{}$
\begin{equation}\label{MoorePenrose}
	\MoorePenrose{}\equiv\Similarity{}
	\cdot
	\left[
	\WaveOperator{}
	+\bigoplus_{J,P}\sum_{\NullVectors{J}{P}{a}}\NullVector{J}{P}{a}\cdot\NullVectorConj{J}{P}{a}
	\right]^{-1}
	\cdot
	\Similarity{}.
\end{equation}
Having obtained the definition of the saturated propagator in~\cref{PropagatorDefinition}, we are ready to extract its poles, which correspond to propagating particle states. According to~\cref{MoorePenrose}, all the poles correspond to the zeroes in~$k$ of the~$J^P$ determinants
\begin{equation}\label{Determinant}
	\det\left[\WaveOperatorJP{J}{P}+\sum_{\NullVectors{J}{P}{a}}\NullVector{J}{P}{a}\cdot\NullVectorConj{J}{P}{a}\right]\propto \prod_{\Masses{J}{P}{s}}\left[k^2-\SquareMass{J}{P}{s}\right],
\end{equation}
where the proportionality in~\cref{Determinant} holds up to constants and even powers of~$k$, and where~$\Masses{J}{P}{s}$ label distinct non-vanishing masses (if any), whose no-tachyon conditions are
\begin{equation}\label{MassiveNoTachyon}
	\SquareMass{J}{P}{s}>0 \quad \forall J,P,\quad \forall\Masses{J}{P}{s}.
\end{equation}
Note that a fundamental assumption will be that for the theory in~\cref{BasicPositionAction} we have
\begin{equation}\label{DistinctMasses}
	\Mass{J}{P}{s}\neq\Mass{\Jp{}}{\Pp{}}{\Sp{}} \quad \forall J,\Jp{},P,\Pp{},\quad \forall\Masses{J}{P}{s},\Masses{\Jp{}}{\Pp{}}{\Sp{}}.
\end{equation}
The possibility of even-power~$k$ prefactors in~\cref{Determinant} suggests that there may be null poles associated with~$\Propagator{}$ in~\cref{PropagatorDefinition}. The question of whether these poles really persist on the lightcone, and whether they correspond to massless states, will need to be more carefully addressed in~\cref{MasslessSpectrum}. The massive states are meanwhile easier to analyse. If~\cref{DistinctMasses} holds, then it can be shown~\cite{Lin:2018awc} that the no-ghost condition is
\begin{equation}\label{MassiveNoGhost}
	\NewRes{k^2}{\SquareMass{J}{P}{s}}\left(P\Tr\MoorePenroseJP{J}{P}\right)>0\quad \forall J,P,\quad \forall\Masses{J}{P}{s}.
\end{equation}
Collectively,~\cref{MassiveNoTachyon,MassiveNoGhost} are the unitarity conditions on the massive sector of the theory in~\cref{BasicPositionAction}, where the additional assumption of~\cref{DistinctMasses} has been introduced.

\paragraph*{Examples} Following~\cref{OperatorMomentumAction}, in momentum space~\cref{GeneralVectorLagrangian} becomes
\begin{align}\label{GeneralVectorLagrangianDecomposed}
	S_{\text{F}}=\int\mathrm{d}^4k\ 
	\Big[&\left[\left(\alpha+\beta\right)k^2+\gamma\right]\BConj{_\perp}\B{^\perp}
	\nonumber\\
	&\hspace{-20pt}\ \ \ 
	+\left[\alpha k^2+\gamma\right]\BConj{_{\ovl{\mu}}}\B{^{\ovl{\mu}}}
	+\BConj{_\perp}\J{^\perp}
	+\BConj{_{\ovl{\mu}}}\J{^{\ovl{\mu}}}\Big],
\end{align}
where~$\B{_\perp}$ and~$\B{_{\ovl{\mu}}}$ are respectively the~$J^P=0^+$ and~$J^P=1^-$ parts of the field, and similarly for the conjugate source. Let us first assume that the couplings in~\cref{GeneralVectorLagrangian} avoid the following finely-tuned configurations
\begin{equation}\label{GeneralMassiveAssumptions}
	\alpha+\beta\neq 0, \quad \alpha\neq 0, \quad \beta\neq 0,\quad \gamma\neq 0.
\end{equation}
We will motivate the inequalities in~\cref{GeneralMassiveAssumptions} by showing that they correspond to special cases of the theory. As a first step we note that they allow the saturated propagator to be defined using~$\Propagator{}\approx\SourceConj{}\cdot\WaveOperator{}^{-1}\cdot\Source{}$ without any singularities
\begin{equation}\label{GeneralMassiveVectorPropagator}
	\Propagator{}\approx\frac{\JConj{_\perp}\J{^\perp}}{\left(\alpha+\beta\right)k^2+\gamma}
	+\frac{\JConj{_{\ovl{\mu}}}\J{^{\ovl{\mu}}}}{\alpha k^2+\gamma}.
\end{equation}
It is then clear from~\cref{GeneralMassiveVectorPropagator,GeneralMassiveAssumptions} that there are two distinct massive poles
\begin{equation}\label{GeneralMasses}
	\SquareMass{0}{+}{1}\approx-\frac{\gamma}{\alpha+\beta},
	\quad
	\SquareMass{1}{-}{1}\approx-\frac{\gamma}{\alpha}.
\end{equation}
The positivity of the square masses in~\cref{GeneralMasses} --- the \emph{no-tachyon} conditions in~\cref{MassiveNoTachyon} --- are easy to ensure via further inequalities on the couplings. Except for special cases,~\cref{DistinctMasses} will also be satisfied automatically. However, the positivity of the residues of~\cref{GeneralMassiveVectorPropagator} --- the \emph{no-ghost} conditions in~\cref{MassiveNoGhost} --- involve
\begin{equation}\label{NoGhost}
	\NewRes{k^2}{\SquareMass{0}{+}{1}}\left(\Propagator{}\right)\approx\frac{\JConj{_\perp}\J{^\perp}}{\left(\alpha+\beta\right)},
	\quad
	\NewRes{k^2}{\SquareMass{1}{-}{1}}\left(\Propagator{}\right)\approx\frac{\JConj{_{\ovl{\mu}}}\J{^{\ovl{\mu}}}}{\alpha}.
\end{equation}
By moving into the rest frame of either particle we can use~\cref{UnitTimelikeComponents} to argue that~$\J{^{\perp}}\equiv\J{_{\perp}}$, but~$\J{^{\ovl{\mu}}}\equiv-\J{_{\ovl{\mu}}}$ (it is this change in sign which motivates the~$P$ factor in~\cref{MassiveNoGhost}), and by assembling all the conditions from~\cref{GeneralMasses,NoGhost} it is then straightforward to show that the model in~\cref{GeneralMassiveAssumptions} must contain a ghost. This model was considered in~\cref{ParticleSpectrographSickMaxwellTheory}. We now consider a special case by relaxing~\cref{GeneralMassiveAssumptions} to
\begin{equation}\label{LongitudinalMassiveAssumptions}
	\alpha=0, \quad \beta\neq 0,\quad \gamma\neq 0.
\end{equation}
With~\cref{LongitudinalMassiveAssumptions} we find that~\cref{GeneralMassiveVectorPropagator} becomes 
\begin{equation}\label{LongitudinalMassiveVectorPropagator}
	\Propagator{}\approx\frac{\JConj{_\perp}\J{^\perp}}{\beta k^2+\gamma}
	+\frac{\JConj{_{\ovl{\mu}}}\J{^{\ovl{\mu}}}}{\gamma}.
\end{equation}
Once again, there are no singularities or source constraints involved simply because~$\gamma\neq 0$. However this time the whole of the~$1^-$ sector is non-propagating. The~$0^+$ mass in~\cref{GeneralMasses} survives, and it is easy to see that~\cref{LongitudinalMassiveAssumptions} is made fully self-consistent by the conditions
\begin{equation}
	\alpha=0, \quad \beta>0, \quad \gamma<0.
\end{equation}
This model was considered in~\cref{ParticleSpectrographLongitudinalMassive}. As another alternative to the general case, we next relax~\cref{GeneralMassiveAssumptions} to
\begin{equation}\label{TransverseMassiveAssumptions}
	\beta=-\alpha\neq 0,\quad \gamma\neq 0.
\end{equation}
With~\cref{TransverseMassiveAssumptions} the saturated propagator becomes
\begin{equation}\label{TransverseMassiveVectorPropagator}
	\Propagator{}\approx\frac{\JConj{_\perp}\J{^\perp}}{\gamma}
	+\frac{\JConj{_{\ovl{\mu}}}\J{^{\ovl{\mu}}}}{\alpha k^2+\gamma},
\end{equation}
and by comparing~\cref{TransverseMassiveVectorPropagator} with~\cref{LongitudinalMassiveVectorPropagator} we see that the result is complementary to the longitudinal case. The fully consistent version of~\cref{TransverseMassiveAssumptions} is
\begin{equation}\label{TransverseMassiveAssumptionsConsistent}
	\beta=-\alpha> 0,\quad \gamma< 0.
\end{equation}
This case was Proca theory, as studied already in~\cref{ParticleSpectrographProcaTheory}. Finally, it is worth commenting on the case
\begin{equation}\label{DegenerateAssumptions}
	\alpha\neq 0,\quad \beta=0, \quad \gamma\neq 0.
\end{equation}
Using~\cref{DegenerateAssumptions} we find from~\cref{GeneralMasses} that~$\SquareMass{0}{+}{1}\approx\SquareMass{1}{-}{1}$. In this case, two unrelated~$J^P$ sectors share a mass, and the assumption in~\cref{DistinctMasses} no longer holds. This is not a physically mysterious scenario, and extensions to the algorithm could in principle be developed. For the case in hand, it is clear that the no-ghost condition of the model~\cref{DegenerateAssumptions} is violated because the residue of the propagator 
\begin{equation}\label{DegeneratePropagator}
	\Propagator{}\approx\frac{\JConj{_\mu}\J{^\mu}}{\alpha k^2+\gamma},
\end{equation}
is not positive-definite in any frame. So far, we were concerned only with the massive poles in~\cref{GeneralMasses}. The residues of such poles are very easy to inspect because the propagator retains its covariant~$J^P$-decomposed form. However, theories can also contain \emph{massless} poles at~$k=0$. By comparing with~\cref{UnitTimelike,UnitTimelikeComponents}, we see that the SPO formulation will break down as we approach the null cone. 

\subsection{Massless spectrum}\label{MasslessSpectrum}
\paragraph*{Theoretical development} In the massless case, it is no longer possible to define~\cref{UnitTimelike,UnitTimelikeComponents}. Prior to taking the~$\NewRes{k}{0}\big(\Propagator{}\big)$ limit, the SPO decomposition may still be used to obtain~$\Propagator{}$ according to~\cref{PropagatorDefinition}. In this limit, however, the individual~$J^P$ states~$\SourceUpFullState{J}{P}{X}{i}{\mu}$ of the source --- defined analogously to the field states~$\FieldDownFullState{J}{P}{X}{i}{\mu}$ in~\cref{DefineProjection} --- will tend to disintegrate. This makes it difficult to study the pole residues covariantly, as we did throughout~\cref{MassiveSpectrum}. For this reason, it is useful to move to an entirely component-based approach, with a preferred choice of frame
\begin{equation}\label{MomentumComponents}
	\left[\tensor{k}{^\mu}\right]^\mathrm{T}=\begin{bmatrix}\En{} & 0 & 0 & \Mo{}\end{bmatrix}, \quad k^2=\En{}^2-\Mo{}^2,
\end{equation}
in which~$\En{}$ and~$\Mo{}$ are respectively the particle energy and momentum. We use~$2p\NewRes{\En{}}{\Mo{}}\big(\Propagator{}\big)$ to study the~$\NewRes{k}{0}\big(\Propagator{}\big)$ limit. Not all components of the sources should contribute to the saturated propagator. When massive poles are studied covariantly, the pseudoinverse~$\MoorePenrose{}$ annihilates the null space of~$\WaveOperator{}$, and so the parts of~$\Source{}$ proportional to this null space are safely eliminated in~\cref{PropagatorDefinition} to leave only the constrained, on-shell sources in~$\Propagator{}$. For massless poles, where tensor components are used instead, only the on-shell components of the sources need to be included: we will calculate these in advance. The null space of~$\WaveOperator{}$ implies a total of~$\ConstraintTotal{}$ independent constraint equation components from a collection of (generally fewer than~$\ConstraintTotal{}$) covariant constraint equations~$\NullVectorConj{J}{P}{a}\cdot\Source{}\approx 0$. These equations define the shell of the source currents, and we can express them in a matrix-component form
\begin{widetext}
\begin{align}		
	\begin{bmatrix}
		\sum_{\States{\SpinMin{}}{\ParityMin{}}{{\FieldMin{}}}{i}}
		\NullVectorComponents{\SpinMin{}}{\ParityMin{}}{\VectorMin{}}{{\FieldMin{}}}{i}
		\ReducedSPOUpDown{\SpinMin{}}{\ParityMin{}}{\FieldMin{}}{i}{{0\dots 0}}{{0\dots0}} 
		& 
		\dots
		& 
		\sum_{\States{\SpinMin{}}{\ParityMin{}}{{\FieldMax{}}}{i}}
		\NullVectorComponents{\SpinMin{}}{\ParityMin{}}{\VectorMin{}}{{\FieldMax{}}}{i}
		\ReducedSPOUpDown{\SpinMin{}}{\ParityMin{}}{\FieldMax{}}{i}{{0\dots0}}{{3\dots3}}
		\\
		\vdots & \ddots & \vdots
		\\
		\sum_{\States{\SpinMax{}}{\ParityMax{}}{{\FieldMin{}}}{i}}
		\NullVectorComponents{\SpinMax{}}{\ParityMax{}}{\VectorMax{}}{{\FieldMin{}}}{i}
		\ReducedSPOUpDown{\SpinMax{}}{\ParityMax{}}{\FieldMin{}}{i}{{3\dots3}}{{0\dots0}}
		&
		\dots
		&
		\sum_{\States{\SpinMax{}}{\ParityMax{}}{{\FieldMax{}}}{i}}
		\NullVectorComponents{\SpinMax{}}{\ParityMax{}}{\VectorMax{}}{{\FieldMax{}}}{i}
		\ReducedSPOUpDown{\SpinMax{}}{\ParityMax{}}{{\FieldMax{}}}{i}{{3\dots3}}{{3\dots3}} 
	\end{bmatrix}
	 & 
	\begin{bmatrix}
		\SourceUp{\FieldMin{}}{{0\dots 0}}
		\vphantom{%
		\sum_{\States{\SpinMax{}}{\ParityMax{}}{{\FieldMax{}}}{i}}
		\NullVectorComponents{\SpinMax{}}{\ParityMax{}}{\VectorMax{}}{{\FieldMax{}}}{i}
		\ReducedSPOUpDown{\SpinMax{}}{\ParityMax{}}{{\FieldMax{}}}{i}{{3\dots3}}{{3\dots3}} 
		}
		\\
		\vdots\\
		\vphantom{%
		\sum_{\States{\SpinMax{}}{\ParityMax{}}{{\FieldMax{}}}{i}}
		\NullVectorComponents{\SpinMax{}}{\ParityMax{}}{\VectorMax{}}{{\FieldMax{}}}{i}
		\ReducedSPOUpDown{\SpinMax{}}{\ParityMax{}}{{\FieldMax{}}}{i}{{3\dots3}}{{3\dots3}} 
		}
		\SourceUp{\FieldMax{}}{{3\dots 3}}
	\end{bmatrix}
	\approx\ZeroVector{}.
	\label{MatrixEquation}
\end{align}
\end{widetext}
where~$\FieldMin{}$ and~$\FieldMax{}$ are respectively the first and last fields in the whole series of independent tensor fields, and likewise~$\SpinMin{}$ and~$\SpinMax{}$ and~$\ParityMin{}$ and~$\ParityMax{}$ and~$\VectorMin{}$ and~$\VectorMax$ are the first and last spins, parities and null vector labels which are found to be present in the given model. We can write~\cref{MatrixEquation} as~$\ConstraintMatrix{}\cdot\SourceVector{}\approx 0$ where~$\ConstraintMatrix{}$ is a complex~$\ConstraintTotal{}\times\ComponentTotal{}$ matrix and the component vector~$\SourceVector{}$ is a convenient way to store a total of~$\ComponentTotal{}$ off-shell components of the sources. The on-shell sources can then be reparameterised in terms of a reduced number of variables~$\K{b}\in\mathbb{C}$ 
\begin{equation}\label{DefinitionOfX}
	\SourceVector{}\approx\sum_b\K{b}\big(\En{}-\Mo{}\big)^\MinimumPower{b}\ConstraintNullVector{b},
	\quad \ConstraintMatrix{}\cdot\ConstraintNullVector{b}\equiv 0,
	\quad \ConstraintNullVectorConj{b}\cdot\ConstraintNullVector{b}\equiv 1,
\end{equation}
where the~$\ConstraintNullVector{b}\in\mathbb{C}^\ComponentTotal{}$ are the~$\ComponentTotal{}-\ConstraintTotal{}$ right null vectors\footnote{There is no expectation that~$\ConstraintNullVectorConj{b}\cdot\ConstraintNullVector{\Bp{}}\equiv \tensor{\delta}{_{b\Bp{}}}$.} of~$\ConstraintMatrix{}$, and~$\MinimumPower{b}\in\mathbb{Z}^{\geq}$ is defined as the minimum power such that
\begin{equation}\label{NoSpuriousPoles}
	\NewRes{\En{}}{\Mo{}}\left(\big(\En{}-\Mo{}\big)^\MinimumPower{b}\ConstraintNullVector{b}\right)\equiv 0.
\end{equation}
Without~\cref{NoSpuriousPoles} it would be possible for spurious massless poles to enter into~$\Propagator{}$ through the reparameterisation. We can then identify the~$\K{b}$ with the elements of a column vector~$\ReducedSource{}\in\mathbb{C}^{\ComponentTotal{}-\ConstraintTotal{}}$ so that~\cref{PropagatorDefinition} can be re-expressed as 
\begin{equation}\label{ReducedPropagatorDefinition}
	\Propagator{}\approx \SourceConj{}\cdot\MoorePenrose{}\cdot\Source{}\approx\ReducedSourceConj{}\cdot\ReducedPropagator{}\cdot\ReducedSource{},
\end{equation}
for some Hermitian~$(\ComponentTotal{}-\ConstraintTotal{})\times(\ComponentTotal{}-\ConstraintTotal{})$ matrix~$\ReducedPropagator{}$ which depends on the momentum components~$\En{}$ and~$\Mo{}$ and the coupling coefficients in~\cref{BasicPositionAction}. In turn, there may be some \emph{non}-null eigenvectors~$\ReducedEigenVector{c}\in\mathbb{C}^{\ComponentTotal{}-\ConstraintTotal{}}$ with eigenvalues~$\ReducedEigenValue{c}\in\mathbb{R}^*$ such that
\begin{equation}\label{DefinitionOfEigen}
	\NewRes{\En{}}{\Mo{}}\left(\ReducedPropagator{}\right)\cdot\ReducedEigenVector{c}\equiv\ReducedEigenValue{c}\ReducedEigenVector{c}.
\end{equation}
Each such eigenvector signals the presence of a propagating massless d.o.f. The situation where~$\ReducedEigenValue{c}=\ReducedEigenValue{\Cp{}}$ for one or more pairs~$c$ and~$\Cp{}$ indicates that a single massless particle has multiple polarisations. The total no-ghost condition for the massless sector is
\begin{equation}\label{TotalNoGhost}
	\ReducedEigenValue{c}>0, \quad \forall c.
\end{equation}
It should be noted from~\cref{DefinitionOfX,ReducedPropagatorDefinition,DefinitionOfEigen} that the~$\ReducedEigenValue{c}$ in general are rational functions not only in the coupling coefficients, but also of the square of the null momentum component~$\Mo{}^2$>0. This extra variable can sometimes obfuscate the no-ghost conditions in~\cref{TotalNoGhost}, and in such cases it may be helpful to seek further momentum-dependent reparameterisations of~$\ReducedSource{}$ which simplify the eigenvalues.

\paragraph*{Examples} It can be shown that there are no massless poles for the theories considered in~\cref{GeneralMassiveAssumptions,LongitudinalMassiveAssumptions,TransverseMassiveAssumptions,DegenerateAssumptions}. We will therefore consider extra critical cases of the theory in which such poles are present. We first assume 
\begin{equation}\label{GeneralVectorAssumptions}
\alpha+\beta\neq 0, \quad \alpha\neq 0,\quad \gamma=0,
\end{equation}
so that according to the principles above, the saturated propagator implied by~\cref{GeneralVectorLagrangianDecomposed} is
\begin{equation}\label{GeneralVectorPropagator}
	\Propagator{}\approx\frac{\JConj{_\perp}\J{^\perp}}{\left(\alpha+\beta\right)k^2}
	+\frac{\JConj{_{\ovl{\mu}}}\J{^{\ovl{\mu}}}}{\alpha k^2}.
\end{equation}
It follows from~\cref{GeneralVectorAssumptions} that there are no gauge symmetries in the model, hence no source constraints. In this case~$\ComponentTotal=4$, but~$\ConstraintTotal{}=0$ and so~$\ConstraintMatrix{}$ vanishes. In a notational departure from~\cref{MomentumComponents,MatrixEquation} it is equivalent to `pad'~$\ConstraintMatrix{}$ with rows of zeros so as to always give it dimensions~$\ComponentTotal{}\times\ComponentTotal{}$
\begin{align}	
	\begin{bmatrix}
		0 & 0 & 0 & 0\\
		0 & 0 & 0 & 0\\
		0 & 0 & 0 & 0\\
		0 & 0 & 0 & 0 
	\end{bmatrix}
	&
	\begin{bmatrix}
		\J{^0}\\
		\J{^1}\\
		\J{^2}\\
		\J{^3}
	\end{bmatrix}
	\approx\ZeroVector{}.
\end{align}
In this case the null space of~$\ConstraintMatrix{}$ spans all components of the source~$J{^{\mu}}$, and we may as well allocate the source coefficients~$\K{i}$ in~\cref{DefinitionOfX} as follows
\begin{equation}\label{AllocateSourceCoefficients}
	\J{^0}\approx\K{0}, \quad 
	\J{^1}\approx\K{1}, \quad 
	\J{^2}\approx\K{2}, \quad 
	\J{^3}\approx\K{3}.
\end{equation}
By substituting~\cref{MomentumComponents,AllocateSourceCoefficients} into~\cref{GeneralVectorPropagator} and collecting the fractions we have
\begin{align}\label{GeneralVectorPropagatorExpanded}
	\Propagator{}&\approx\Big[
		\left(\alpha+\beta\right)\left(\En{}^2-\Mo{}^2\right)
		\Big(
		\KConj{0}\K{0}
		-\KConj{1}\K{1}
		-\KConj{2}\K{2}
		\nonumber\\
		&\ \ \ \ \ \ \ \ 
		-\KConj{3}\K{3}
		\Big)
		-\beta\left(\En{}\KConj{0}-\Mo{}\KConj{3}\right)
		\left(\En{}\K{0}-\Mo{}\K{3}\right)
		\Big]
		\nonumber\\
		&\ \ \ 
		\times\Big[
			\left(\En{}^2-\Mo{}^2\right)^2
			\alpha\left(\alpha+\beta\right)\Big]^{-1}.
\end{align}
Taking the null residue of~\cref{GeneralVectorPropagatorExpanded} we find
\begin{align}\label{GeneralVectorResidue}
	2\Mo{}\NewRes{\En{}}{\Mo{}}\left(\Propagator{}\right)&\approx
	\frac{2\alpha+\beta}{2\alpha\left(\alpha+\beta\right)}
	\left(\KConj{0}\K{0}-\KConj{3}\K{3}\right)
	\nonumber\\
	&\ \ \ 
	-\frac{1}{\alpha}
	\left(\KConj{1}\K{1}+\KConj{2}\K{2}\right).
\end{align}
The second term in~\cref{GeneralVectorResidue} can be made to be a positive-definite quadratic form if~$\alpha<0$, so~$\NewRes{\En{}}{\Mo{}}\left(\ReducedPropagator{}\right)$ contains the~$2\times 2$ identity, and from this we read off the healthy massless propagation of two transverse polarisations. The same can \emph{never} be said of the first term in~\cref{GeneralVectorResidue} so long as the conditions in~\cref{GeneralVectorAssumptions} hold: two additional modes are propagating, and one of them is strictly a ghost. This model was considered in~\cref{ParticleSpectrographSickMaxwellTheory}. There are precisely two nontrivial ways to relax~\cref{GeneralVectorAssumptions}, and each of these will lead to a different critical case of the vector theory. We start with the (perhaps less familiar) case
\begin{equation}\label{LongitudinalVectorAssumptions}
 \alpha= 0, \quad \beta\neq 0,\quad \gamma=0,
\end{equation}
for which the saturated propagator in~\cref{GeneralVectorPropagator} is replaced by
\begin{equation}\label{LongitudinalVectorPropagator}
	\Propagator{}\approx\frac{\JConj{_\perp}\J{^\perp}}{\beta k^2},
	\quad
	\J{^{\ovl{\mu}}}\equiv\J{^{\mu}}-\N{^{\mu}}\J{^{\perp}}\equiv\J{^{\mu}}-\frac{\tensor{k}{^{\mu}}\tensor{k}{_\nu}\J{^\nu}}{k^2}\approx 0.
\end{equation}
The second equality in~\cref{LongitudinalVectorPropagator} is a source constraint: it can be understood as a physical consequence of the second term in~\cref{GeneralVectorPropagator} diverging as~\cref{LongitudinalVectorAssumptions} is imposed. From the form in which the source constraint is written, it is easy to see that the three spatial components of the source current can all be eliminated in terms of~$J{^0}$. In component form and using~\cref{MomentumComponents,MatrixEquation} the constraint reads 
\begin{align}
	\frac{1}{\En{}^2-\Mo{}^2}
	\begin{bmatrix}
		-\Mo{}^2 & 0 & 0 & \En{}\Mo{}\\
		0 & \En{}^2-\Mo{}^2 & 0 & 0\\
		0 & 0 & \En{}^2-\Mo{}^2 & 0\\
		-\En{}\Mo{} & 0 & 0 & \En{}^2
	\end{bmatrix}
	&
	\begin{bmatrix}
		\J{^0}\\
		\J{^1}\\
		\J{^2}\\
		\J{^3}
	\end{bmatrix}
	\approx\ZeroVector{}.
\end{align}
There is one (minimally rescaled) null vector which we parameterise by~$\K{0}$ in~\cref{DefinitionOfX} to give the physical source
\begin{equation}\label{LongitudinalAllocateSourceCoefficients}
	\J{^0}\approx\En{}\K{0},\quad \J{^1}\approx\J{^2}\approx 0, \quad \J{^3}\approx\Mo{}\K{0},
\end{equation}
where~\cref{LongitudinalAllocateSourceCoefficients} supersedes~\cref{AllocateSourceCoefficients}. Now substituting~\cref{MomentumComponents,LongitudinalAllocateSourceCoefficients} into~\cref{LongitudinalVectorPropagator} as before we have
\begin{equation}\label{LongitudinalResidue}
	\Propagator{}\approx\frac{1}{\beta}\KConj{0}\K{0},\quad 
	2\Mo{}\NewRes{\En{}}{\Mo{}}\left(\Propagator{}\right)\approx 0.
\end{equation}
The conclusion in this case is that the conditions~\cref{LongitudinalVectorAssumptions} lead to an entirely \emph{empty} spectrum. This case was studied in~\cref{ParticleSpectrographLongitudinalMassless}. Finally, we proceed to the well-known case
\begin{equation}\label{TransverseVectorAssumptions}
	\beta=-\alpha\neq 0, \quad \gamma=0,
\end{equation}
which corresponds to the classical electromagnetism of Maxwell. From~\cref{TransverseVectorAssumptions} the saturated propagator in~\cref{GeneralVectorPropagator} becomes
\begin{equation}\label{TransverseVectorPropagator}
	\Propagator{}\approx\frac{\JConj{_{\ovl{\mu}}}\J{^{\ovl{\mu}}}}{\alpha k^2},\quad 
	\N{^\mu}\J{^\perp}\equiv\frac{\tensor{k}{^{\mu}}\tensor{k}{_\nu}\J{^\nu}}{k^2}\approx 0,
\end{equation}
and again~\cref{TransverseVectorPropagator} constitutes a (single) constraint which we can translate with~\cref{MomentumComponents,MatrixEquation} to
\begin{align}	
	\frac{1}{\En{}^2-\Mo{}^2}
	\begin{bmatrix}
		\En{}^2 & 0 & 0 & -\En{}\Mo{}\\
		0 & 0 & 0 & 0\\
		0 & 0 & 0 & 0\\
		\En{}\Mo{} & 0 & 0 & -\Mo{}^2
	\end{bmatrix}
	&
	\begin{bmatrix}
		\J{^0}\\
		\J{^1}\\
		\J{^2}\\
		\J{^3}
	\end{bmatrix}
	\approx\ZeroVector{}.
\end{align}
There are three (minimally rescaled) null vectors in~\cref{DefinitionOfX} parameterised by~$\K{0}$,~$\K{1}$ and~$\K{2}$ according to
\begin{equation}\label{TransverseAllocateSourceCoefficients}
	\J{^0}\approx\Mo{}\K{0}, \quad
	\J{^1}\approx\K{1}, \quad
	\J{^2}\approx\K{2}, \quad
	\J{^3}\approx\En{}\K{0}.
\end{equation}
Now again, we substitute~\cref{TransverseAllocateSourceCoefficients} into the propagator in~\cref{TransverseVectorPropagator} to obtain
\begin{equation}\label{TransverseResidueLong}
	\Propagator{}\approx-\frac{1}{\alpha}\KConj{0}\K{0}
	-\frac{1}{\alpha\left(\En{}^2-\Mo{}^2\right)}\left(\KConj{1}\K{1}+\KConj{2}\K{2}\right).
\end{equation}
The null residue of~\cref{TransverseResidueLong} is
\begin{equation}\label{TransverseResidue}
	2\Mo{}\NewRes{\En{}}{\Mo{}}\left(\Propagator{}\right)\approx-\frac{1}{\alpha}\left(\KConj{1}\K{1}+\KConj{2}\K{2}\right),
\end{equation}
so the~$\alpha<0$ condition is indeed enough to protect the two massless polarisations in this scenario. The final consistent model implied by~\cref{TransverseVectorAssumptions} is
\begin{equation}\label{TransverseVectorAssumptionsConsistent}
	\beta=-\alpha>0, \quad \gamma=0,
\end{equation}
and it so happens that~\cref{TransverseVectorAssumptionsConsistent} is a limit of~\cref{TransverseMassiveAssumptionsConsistent}. The Maxwell theory was considered in~\cref{ParticleSpectrographMaxwellTheory}. Note that by choosing pairs of propagators from~\cref{GeneralVectorResidue,LongitudinalResidue,TransverseResidue}, various common terms can be identified. However, a holistic picture of the physics does not emerge: the various critical cases of the theory must be studied piecewise. This final example concludes the theoretical development of the SPO algorithm.

\section{Conclusions}\label{Conclusions}
\paragraph*{Results of this paper} We presented the Particle Spectrum for Any Tensor Lagrangian (\PSALTer{}) software initiative. \PSALTer{} is a contributing package to the open-source \xAct{} project, and is designed for use with \Mathematica{}. The initial release of \PSALTer{} has functionality for any parity-preserving action of the form~\cref{BasicPositionAction}, where the tensor fields have up to three indices (with any symmetries on those indices). We tested \PSALTer{} against a wide variety of possible actions, most of which were physically motivated and well studied in the literature. Each of these examples can be reproduced on a modern personal computer in a matter of minutes.

\paragraph*{Further work} Longer-term developments of the \PSALTer{} framework may include:
\begin{enumerate}
	\item Compatability of non-linear coupling coefficient parameterisations with \lstinline!Method->"Hard"!.
	\item Extension to parity-violating theories (see e.g.~\cite{Karananas:2014pxa}).
	\item Extension to tensors of rank four or higher.
	\item Extension to multi-term symmetries.
	\item Extension to traceless fields, or fields whose kinematics are defined other than by index symmetries (examples include the Curtright field~\cite{Curtright:1980yk}).
	\item Extension to complex fields.
	\item Recursive surveys of theory-space which exhaustively identify all critical cases of a given multi-parameter model leading to qualitatively distinct spectra (see e.g.~\cite{Lin:2018awc,Lin:2019ugq,Lin:2020phk,Lin:2020fuo}).
	\item Extension to vacua comprising finite values of the background fields.
	\item Perturbative reconstruction of the QFT, including vertices, Feynman rules and Ward identities.
	\item Radiative corrections to the tree-level spectra (see e.g.~\cite{Marzo:2024pyn}).
\end{enumerate}
We rank these anticipated features roughly by increasing order of difficulty.

\paragraph*{A note of caution} As a final comment, it must be emphasised that self-consistency of the linear particle spectrum is a necessary but not sufficient condition for a theory to be healthy. In particular, the \emph{non}-linear particle spectrum may be populated by additional modes, which vanish in the linear analysis: this phenomenon is called \emph{strong coupling}. Strong coupling is always fatal, if not to the proposed model as a whole, then at least to the vacuum around which the linearisation is performed. So far, only the Dirac--Bergmann algorithm~\cite{Dirac:1950pj,Anderson:1951ta,Bergmann:1954tc} is guaranteed to diagnose strong coupling (this is also called non-linear Hamiltonian or canonical analysis). Some recent progress has been made to automate the Dirac--Bergmann algorithm~\cite{Barker:2022kdk}, but the resulting software is not so versatile as \PSALTer{}\footnote{The software, called \HiGGS{}, is restricted to PGTs such as those studied in~\cref{PoincareGaugeTheory}, meanwhile \PSALTer{} is theory-agnostic.}. The Hamiltonian analysis is sensitive only to the classical properties of the model, but quantum inconsistencies can also appear. These may be associated with (e.g.) anomalies, or unitarity bounds. Whilst \PSALTer{} is unable to thoroughly investigate any of these issues, it can provide limited insight in some scenarios. For example, strongly coupled modes are often associated with the non-linear breakdown of so-called \emph{accidental} symmetries~\cite{Percacci:2020ddy}. Helpfully, \PSALTer{} is guaranteed to find all the symmetries of the linearised model, via the source constraints\footnote{To be clear, if any of the symmetries were not identified, then the propagator would involve \lstinline!ComplexInfinity!, resulting in an error.}. Once the linear symmetries are available, they can then be directly sought in the non-linear model: their non-linear absence would be a loud (but not conclusive) signal for strong coupling. Similarly, \PSALTer{} can directly detect local Weyl invariance\footnote{The relevant source constraint is that the trace of the stress-energy tensor must vanish~\cite{Weyl:1918pdp,Weyl:1919fi}.}, which is a loud (but again, not conclusive) signal for the Weyl anomaly~\cite{Capper:1974ic}. Whilst strong coupling is a non-linear phenomenon, there is a priori no guarantee that, when present, it will become manifest at cubic or any particular higher order in the perturbative expansion~\cite{BeltranJimenez:2020lee}. Nonetheless, the inclusion of interaction terms will facilitate the analysis of a wealth of quantum effects. The initial release of \PSALTer{} provides a robust foundation for automating such quantum gravity calculations in the future. The current functionality aims to facilitate responsible model-building in the gravity community.

\begin{acknowledgments}
	This work used the DiRAC Data Intensive service (CSD3 \href{www.csd3.cam.ac.uk}{www.csd3.cam.ac.uk}) at the University of Cambridge, managed by the University of Cambridge University Information Services on behalf of the STFC DiRAC HPC Facility (\href{www.dirac.ac.uk}{www.dirac.ac.uk}). The DiRAC component of CSD3 at Cambridge was funded by BEIS, UKRI and STFC capital funding and STFC operations grants. DiRAC is part of the UKRI Digital Research Infrastructure.

This work also used the Newton server, access to which was provisioned by Will Handley using an ERC grant.

\PSALTer{} was improved by many useful discussions with Jaakko Annala, Stephanie Buttigieg, Will Handley, Mike Hobson, Manuel Hohmann, Damianos Iosifidis, Georgios Karananas, Anthony Lasenby, Yun-Cherng Lin, Oleg Melichev, Yusuke Mikura, Vijay Nenmeli, Roberto Percacci, Syksy Räsänen, Cillian Rew, Zhiyuan Wei, David Yallup, Haoyang Ye, and Sebastian Zell.

WB is grateful for the kind hospitality of Leiden University and the Lorentz Institute, and the support of Girton College, Cambridge.
The work of CM was supported by the Estonian Research Council grants PRG1677, RVTT3, RVTT7, and the CoE program TK202 ``\emph{Fundamental Universe}''.
CR  acknowledges support from a Science and Technology Facilities Council (STFC) Doctoral Training Grant. 
\end{acknowledgments}

\bibliography{Manuscript,NotINSPIRE}

\appendix

\begin{widetext}
\section{Sources and installation}\label{Install}
	In this appendix we briefly summarise the structure of the \PSALTer{} source files. The \PSALTer{} package should be installed once the \xAct{} suite of packages has been installed. Information about \xAct{} can be found at \href{http://www.xact.es/}{\texttt{xact.es}}. The \PSALTer{} package is made available at the public \GitHub{} repository \href{https://github.com/wevbarker/PSALTer}{\texttt{github.com/wevbarker/PSALTer}}, which provides detailed installation instructions for a variety of operating systems, including for the proprietary systems \Windows{} and \Mac{}. As an example, we will consider here a straightforward \Linux{} installation\footnote{Note that the syntax highlighting for \Bash{} in this section differs from that used for the \WolframLanguage{} in~\cref{SymbolicImplementation,TheoreticalDevelopment}.}. The following \Bash{} command will download \PSALTer{} into the home directory:
\begin{lstlisting}[language=Special]
[user@system ~]$\dollar$ git clone https://github.com/wevbarker/PSALTer
\end{lstlisting}
	The package itself is provided as roughly~$\SI{8e3}{}$ source lines of code distributed among plaintext \WolframLanguage{} files with the usual \texttt{.m} or \texttt{.wl} extensions. There is also a smaller collection of graphics files and their sources. There are no binaries, nor does the software need to be compiled. The directory structure can be displayed from the command line:
\begin{lstlisting}[breaklines=true,style=ascii-tree,language=Special]
[user@system ~]$\dollar$ tree PSALTer
PSALTer
├── LICENSE.md
├── README.md
└── xAct
    └── PSALTer
        ├── Documentation
        │   └── English
        │       ├── Documentation.nb
        │       └── Documentation.pdf
        ├── Kernel
        │   └── init.wl
        ├── Logos
        │   ├── ASCIILogo.txt
        │   ├── FieldKinematics.pdf
        │   ├── FieldKinematics.png
        │   ├── GitHubLogo.pdf
        │   ├── GitHubLogo.png
        │   ├── GitHubLogo.sh
        │   ├── GitHubLogo.svgz
        │   ├── GitLabLogo.pdf
        │   ├── GitLabLogo.png
        │   ├── GitLabLogo.sh
        │   ├── GitLabLogo.svgz
        │   ├── ParticleSpectrograph.pdf
        │   └── ParticleSpectrograph.png
        ├── PSALTer.m
        └── Sources
            ├── DefField
            │   ├── AppendToField.m
            │   ├── CombineRules
            │   │   ├── DefAllComponentValues.m
            │   │   ├── DefSummary.m
            │   │   ├── ValidateSO3Irreps
            │   │   │   ├── ValidateInverseField.m
            │   │   │   ├── ValidateInverseMode.m
            │   │   │   ├── ValidateSpatial.m
            │   │   │   ├── ValidateSymmetryField.m
            │   │   │   ├── ValidateSymmetryMode.m
            │   │   │   └── ValidateTraceless.m
            │   │   └── ValidateSO3Irreps.m
            │   ├── CombineRules.m
            │   ├── DefFiducialField.m
            │   ├── DefSO3Irrep
            │   │   ├── DefSymbol.m
            │   │   ├── MakeAutomaticallyNotAntisymmetric.m
            │   │   ├── MakeAutomaticallyTraceless.m
            │   │   └── MakeUniqueQuadratic.m
            │   ├── DefSO3Irrep.m
            │   ├── RegisterFieldRank0.m
            │   ├── RegisterFieldRank1.m
            │   ├── RegisterFieldRank2Antisymmetric.m
            │   ├── RegisterFieldRank2.m
            │   ├── RegisterFieldRank2Symmetric.m
            │   ├── RegisterFieldRank3Antisymmetric12.m
            │   ├── RegisterFieldRank3Antisymmetric13.m
            │   ├── RegisterFieldRank3Antisymmetric23.m
            │   ├── RegisterFieldRank3.m
            │   ├── RegisterFieldRank3Symmetric12.m
            │   ├── RegisterFieldRank3Symmetric13.m
            │   ├── RegisterFieldRank3Symmetric23.m
            │   ├── RegisterFieldRank3TotallyAntisymmetric.m
            │   ├── RegisterFieldRank3TotallySymmetric.m
            │   ├── SummariseField
            │   │   ├── DecompositionTable.m
            │   │   ├── ExpansionTable.m
            │   │   └── FieldMosaic.m
            │   └── SummariseField.m
            ├── DefField.m
            ├── ParticleSpectrum
            │   ├── CombineAssociations
            │   │   ├── CacheContexts.m
            │   │   ├── DefPlaceholderSpins.m
            │   │   ├── GenerateAnsatz
            │   │   │   └── CatalogueInvariant.m
            │   │   ├── GenerateAnsatz.m
            │   │   └── NormaliseRescalings.m
            │   ├── CombineAssociations.m
            │   ├── ConstructLinearAction.m
            │   ├── ConstructMassiveAnalysis
            │   │   ├── MassiveAnalysisOfSector
            │   │   │   ├── IsolatePoles.m
            │   │   │   └── PoleToSquareMass.m
            │   │   ├── MassiveAnalysisOfSector.m
            │   │   ├── MassiveGhost.m
            │   │   └── SimplifyMasses.m
            │   ├── ConstructMassiveAnalysis.m
            │   ├── ConstructMasslessAnalysis
            │   │   ├── ConstructLightcone
            │   │   │   ├── AllIndependentComponents
            │   │   │   │   └── IndependentComponents.m
            │   │   │   ├── AllIndependentComponents.m
            │   │   │   ├── ConstraintComponentToLightcone.m
            │   │   │   ├── MakeConstraintComponentList.m
            │   │   │   ├── MakeFreeSourceVariables
            │   │   │   │   └── DefFreeSourceVariables.m
            │   │   │   ├── MakeFreeSourceVariables.m
            │   │   │   └── RescaleNullVector.m
            │   │   ├── ConstructLightcone.m
            │   │   ├── ConvertLightcone
            │   │   │   ├── ConstrainInLightcone.m
            │   │   │   ├── ExaminePoleOrder
            │   │   │   │   ├── ExtractSecularEquation.m
            │   │   │   │   ├── MasslessAnalysisOfTotal
            │   │   │   │   │   └── ExtractPart.m
            │   │   │   │   ├── MasslessAnalysisOfTotal.m
            │   │   │   │   ├── NullResidue.m
            │   │   │   │   ├── ReparameteriseSources
            │   │   │   │   │   ├── ExtractDenominator.m
            │   │   │   │   │   └── ExtractReparameterisationMatrix.m
            │   │   │   │   └── ReparameteriseSources.m
            │   │   │   ├── ExaminePoleOrder.m
            │   │   │   ├── ExpressInLightcone.m
            │   │   │   ├── FullyCanonicalise.m
            │   │   │   ├── FullyExpandSources.m
            │   │   │   ├── MakeSaturatedMatrix.m
            │   │   │   ├── MatrixFromSymbolic.m
            │   │   │   ├── MatrixToSymbolic.m
            │   │   │   └── Repartition.m
            │   │   └── ConvertLightcone.m
            │   ├── ConstructMasslessAnalysis.m
            │   ├── ConstructSaturatedPropagator
            │   │   ├── ConjectureInverse
            │   │   │   ├── IntermediateRules.m
            │   │   │   ├── MakeSymbolic.m
            │   │   │   ├── ManualPseudoInverse
            │   │   │   │   ├── CarefullyOrthogonalise.m
            │   │   │   │   └── ParameterisedNullVectorQ.m
            │   │   │   ├── ManualPseudoInverse.m
            │   │   │   ├── UnmakeSymbolic
            │   │   │   │   ├── ConsolidateFinalElement.m
            │   │   │   │   ├── ConsolidateUnmakeSymbolic
            │   │   │   │   │   ├── GrabExpression.m
            │   │   │   │   │   └── SimplifyIfSmall.m
            │   │   │   │   ├── ConsolidateUnmakeSymbolic.m
            │   │   │   │   ├── GradualExpandSubTask
            │   │   │   │   │   └── GradualExpand.m
            │   │   │   │   ├── GradualExpandSubTask.m
            │   │   │   │   ├── InitialExpand
            │   │   │   │   │   └── BatchExpanded.m
            │   │   │   │   └── InitialExpand.m
            │   │   │   └── UnmakeSymbolic.m
            │   │   ├── ConjectureInverse.m
            │   │   └── DistributeConjugate.m
            │   ├── ConstructSaturatedPropagator.m
            │   ├── ConstructSourceConstraints
            │   │   ├── ConjectureNullSpace
            │   │   │   ├── CleanNullVector.m
            │   │   │   ├── CommonNullVector
            │   │   │   │   └── IsNullVectorOfSpace.m
            │   │   │   ├── CommonNullVector.m
            │   │   │   ├── CreateList.m
            │   │   │   ├── EnsureLinearInCouplings.m
            │   │   │   └── RemoveReferencesToMomentum.m
            │   │   ├── ConjectureNullSpace.m
            │   │   ├── NonTrivialDot.m
            │   │   └── ToCovariantForm.m
            │   ├── ConstructSourceConstraints.m
            │   ├── ConstructUnitarityConditions.m
            │   ├── ConstructWaveOperator
            │   │   ├── ConstructOperator
            │   │   │   └── GetHermitianPart.m
            │   │   ├── ConstructOperator.m
            │   │   └── FourierLagrangian.m
            │   ├── ConstructWaveOperator.m
            │   ├── SummariseResults
            │   │   ├── CLIPrint
            │   │   │   └── Status.m
            │   │   ├── CLIPrint.m
            │   │   ├── DetailCell.m
            │   │   ├── MakeLabel.m
            │   │   ├── PrintSourceConstraints.m
            │   │   ├── PrintSpectrum
            │   │   │   ├── PrintMassiveSpectrum.m
            │   │   │   ├── PrintMasslessSpectrum.m
            │   │   │   ├── PrintParticle
            │   │   │   │   ├── FeynmanDiagramHexic.pdf
            │   │   │   │   ├── FeynmanDiagramHexic.tex
            │   │   │   │   ├── FeynmanDiagramQuadratic.pdf
            │   │   │   │   ├── FeynmanDiagramQuadratic.tex
            │   │   │   │   ├── FeynmanDiagramQuartic.pdf
            │   │   │   │   ├── FeynmanDiagramQuartic.tex
            │   │   │   │   ├── FeynmanDiagramSpin0ParityEven.pdf
            │   │   │   │   ├── FeynmanDiagramSpin0ParityEven.tex
            │   │   │   │   ├── FeynmanDiagramSpin0ParityOdd.pdf
            │   │   │   │   ├── FeynmanDiagramSpin0ParityOdd.tex
            │   │   │   │   ├── FeynmanDiagramSpin1ParityEven.pdf
            │   │   │   │   ├── FeynmanDiagramSpin1ParityEven.tex
            │   │   │   │   ├── FeynmanDiagramSpin1ParityOdd.pdf
            │   │   │   │   ├── FeynmanDiagramSpin1ParityOdd.tex
            │   │   │   │   ├── FeynmanDiagramSpin2ParityEven.pdf
            │   │   │   │   ├── FeynmanDiagramSpin2ParityEven.tex
            │   │   │   │   ├── FeynmanDiagramSpin2ParityOdd.pdf
            │   │   │   │   ├── FeynmanDiagramSpin2ParityOdd.tex
            │   │   │   │   ├── FeynmanDiagramSpin3ParityEven.pdf
            │   │   │   │   ├── FeynmanDiagramSpin3ParityEven.tex
            │   │   │   │   ├── FeynmanDiagramSpin3ParityOdd.pdf
            │   │   │   │   ├── FeynmanDiagramSpin3ParityOdd.tex
            │   │   │   │   ├── GetDiagram.m
            │   │   │   │   ├── MakeDiagrams.sh
            │   │   │   │   └── Preamble.tex
            │   │   │   ├── PrintParticle.m
            │   │   │   ├── PrintSecularEquation.m
            │   │   │   └── StripFactors.m
            │   │   ├── PrintSpectrum.m
            │   │   ├── PrintUnitarityConditions.m
            │   │   ├── ResultsMosaic
            │   │   │   ├── GraphicsMosaic
            │   │   │   │   └── ShrinkPackRectangles.m
            │   │   │   └── GraphicsMosaic.m
            │   │   ├── ResultsMosaic.m
            │   │   ├── ShowIfSmall.m
            │   │   ├── SplitWignerGrid.m
            │   │   ├── SummariseTheory.m
            │   │   └── WignerGrid.m
            │   ├── SummariseResults.m
            │   ├── UpdateTheoryAssociation.m
            │   ├── ValidateLagrangian.m
            │   ├── ValidateMaxLaurentDepth.m
            │   ├── ValidateMethod.m
            │   └── ValidateTheoryName.m
            ├── ParticleSpectrum.m
            ├── ReloadPackage
            │   ├── Colours.m
            │   ├── DefGeometry.m
            │   ├── Diagnostic.m
            │   ├── MonitorParallel
            │   │   ├── ParallelGrid
            │   │   │   └── RaggedBlock.m
            │   │   ├── ParallelGrid.m
            │   │   └── ReMagnify.m
            │   ├── MonitorParallel.m
            │   ├── NewFramed.m
            │   ├── NewParallelSubmit.m
            │   ├── ToNewCanonical.m
            │   └── Vectorize.m
            └── ReloadPackage.m

47 directories, 189 files
\end{lstlisting}
To perform the installation, the sources need only be copied to the location of the other \xAct{} sources. For a global installation of \xAct{} this may require:
\begin{lstlisting}[language=Special]
[user@system ~]$\dollar$ cd PSALTer/xAct
[user@system xAct]$\dollar$ sudo cp -r PSALTer /usr/share/Mathematica/Applications/xAct/
\end{lstlisting}
For a local installation of \xAct{}, the path may be vary:
\begin{lstlisting}[language=Special]
[user@system xAct]$\dollar$ cp -r PSALTer ~/.Mathematica/Applications/xAct/
\end{lstlisting}
Prior to running the software, it is necessary to install the popular \Inkscape{} package. This should be done using the package manager native to your \Linux{} distribution, for example:
\begin{lstlisting}[language=Special]
[user@system xAct]$\dollar$ sudo pacman -S inkscape
\end{lstlisting}
To get started, there is a documentation notebook, which contains many examples beyond those illustrated in this paper:
\begin{lstlisting}[language=Special]
[user@system xAct]$\dollar$ cd PSALTer/Documentation/English
[user@system English]$\dollar$ mathematica Documentation.nb &
\end{lstlisting}
Before trying the examples in this notebook or in~\cref{SymbolicImplementation}, it will also be necessary to install \RectanglePacking{}~\cite{RectanglePacking:2024}. This is not part of \xAct{}, and it is easiest to install it from within a notebook:
\begin{lstlisting}
In[#]:= PacletInstall["JasonB/RectanglePacking"];
\end{lstlisting}
It may also be helpful to run \PSALTer{} with a stable internet connection, since some of the functions used may need to be imported from the online Wolfram Function Repository --- this process should happen automatically. All the details provided in this appendix may change with future versions of \PSALTer{}. Up-to-date installation instructions will be maintained at \href{https://github.com/wevbarker/PSALTer}{\texttt{github.com/wevbarker/PSALTer}}.
\end{widetext}

\section{Further examples}\label{FurtherExamples}
\begin{figure*}[h]
	\includegraphics[width=\linewidth]{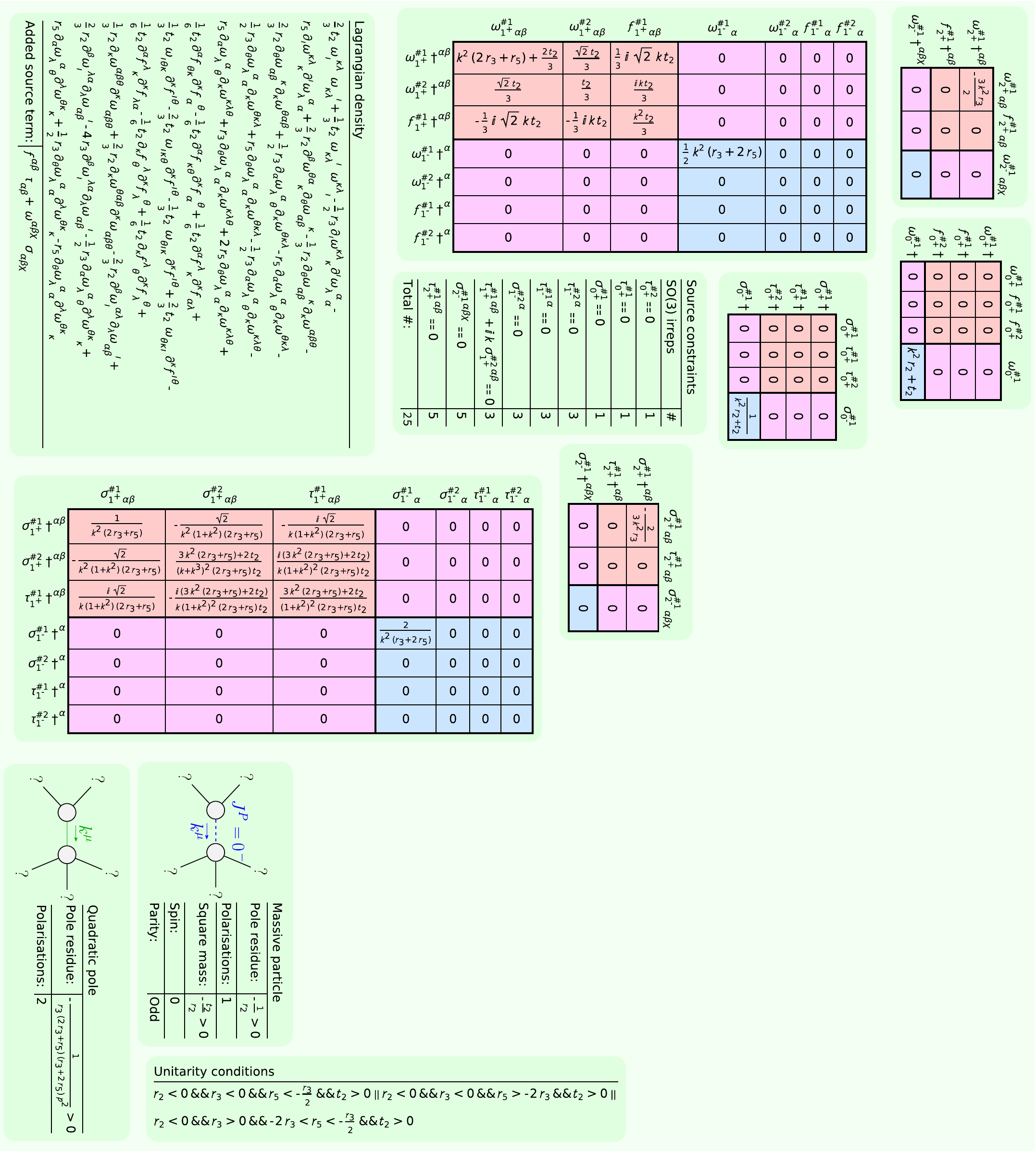}
	\caption{Particle spectrum of Case~$\#1$ from~\cite{Lin:2019ugq}. All quantities are defined in~\cref{FieldKinematicsTetradPerturbation,FieldKinematicsSpinConnection}.}
\label{ParticleSpectrographCase1}
\end{figure*}

\begin{figure*}[h]
	\includegraphics[width=\linewidth]{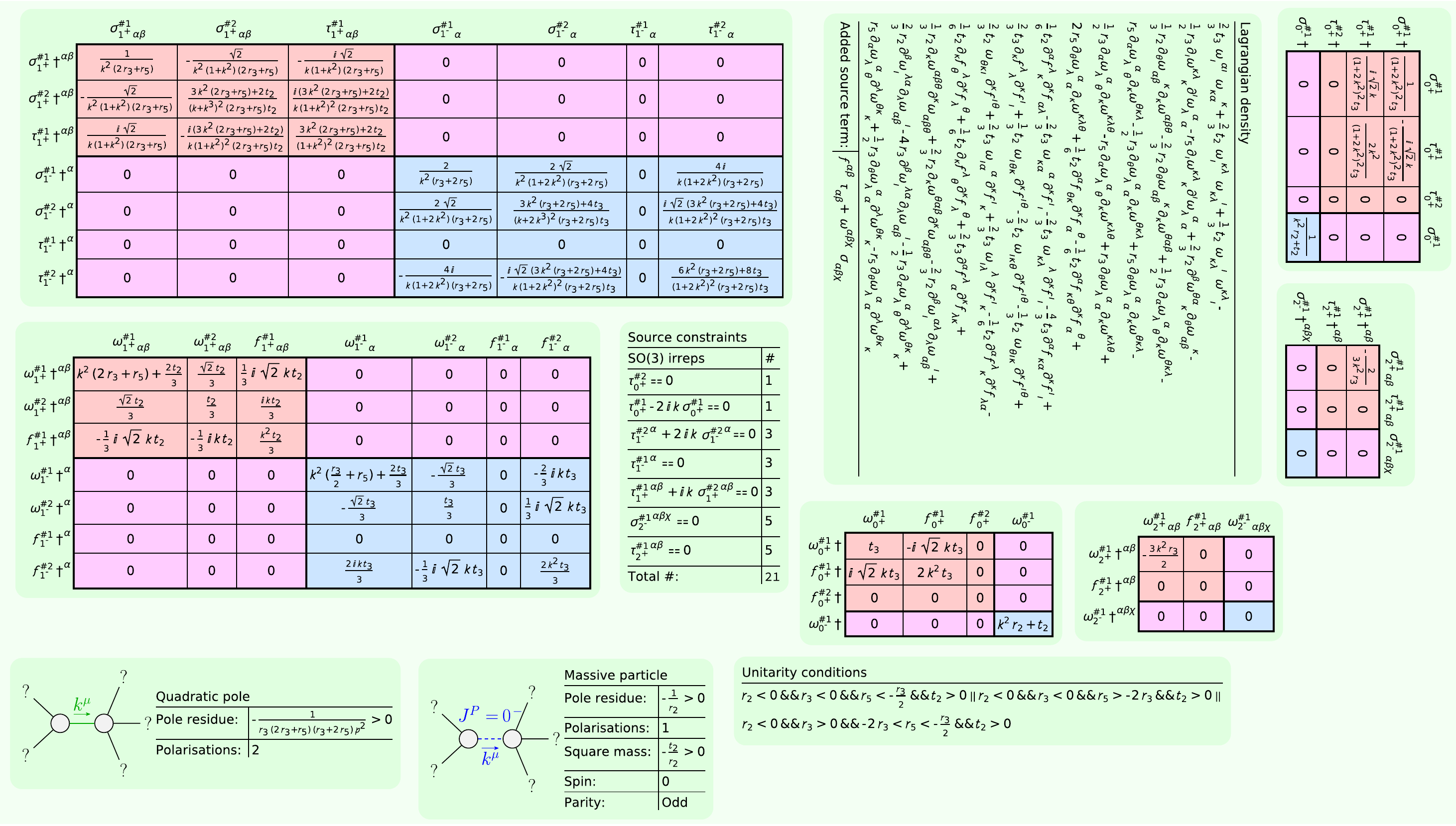}
	\caption{Particle spectrum of Case~$\#2$ from~\cite{Lin:2019ugq}. All quantities are defined in~\cref{FieldKinematicsTetradPerturbation,FieldKinematicsSpinConnection}.}
\label{ParticleSpectrographCase2}
\end{figure*}

\begin{figure*}[h]
	\includegraphics[width=\linewidth]{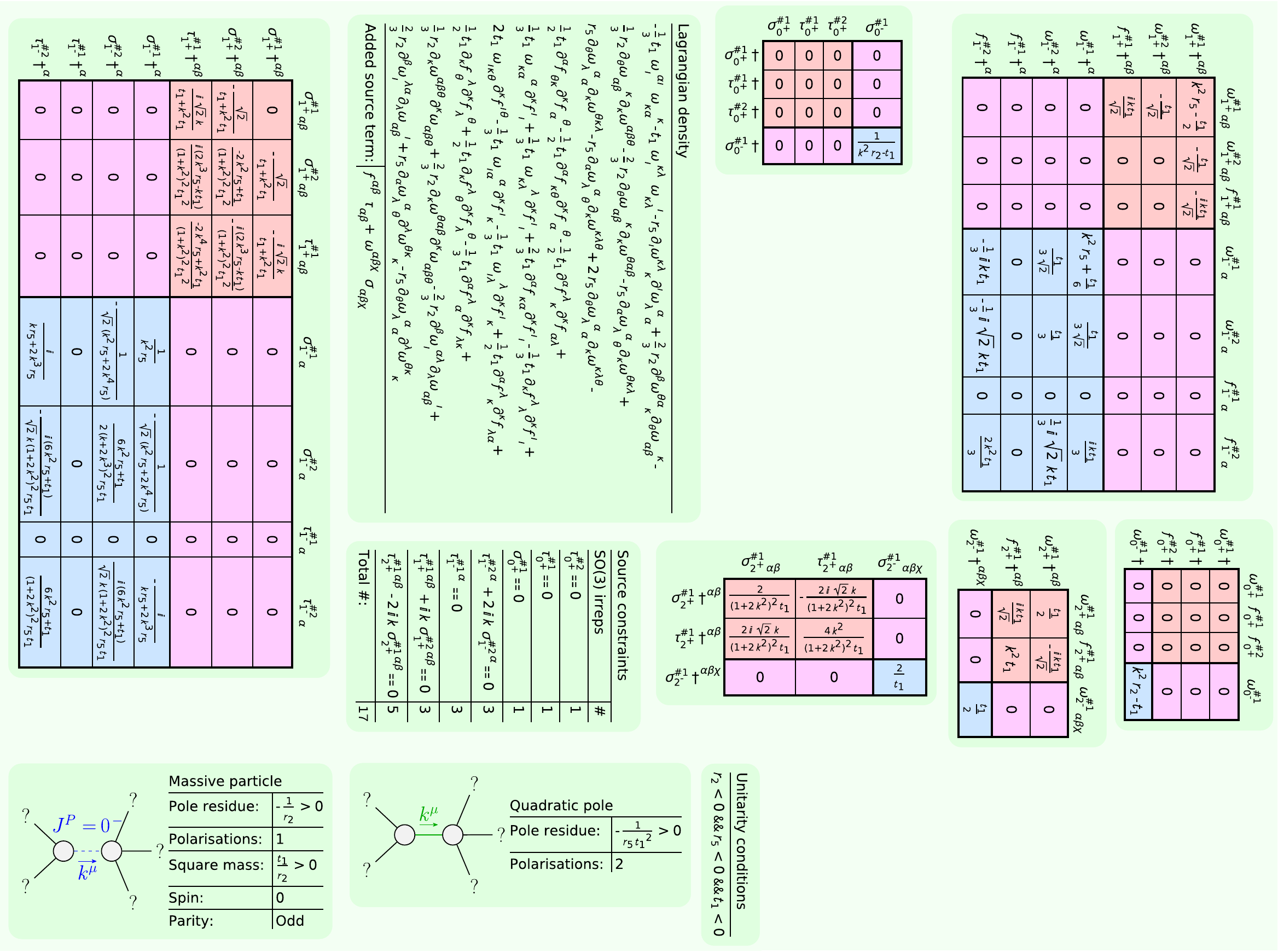}
	\caption{Particle spectrum of Case~$\#3$ from~\cite{Lin:2019ugq}. All quantities are defined in~\cref{FieldKinematicsTetradPerturbation,FieldKinematicsSpinConnection}.}
\label{ParticleSpectrographCase3}
\end{figure*}

\begin{figure*}[h]
	\includegraphics[width=\linewidth]{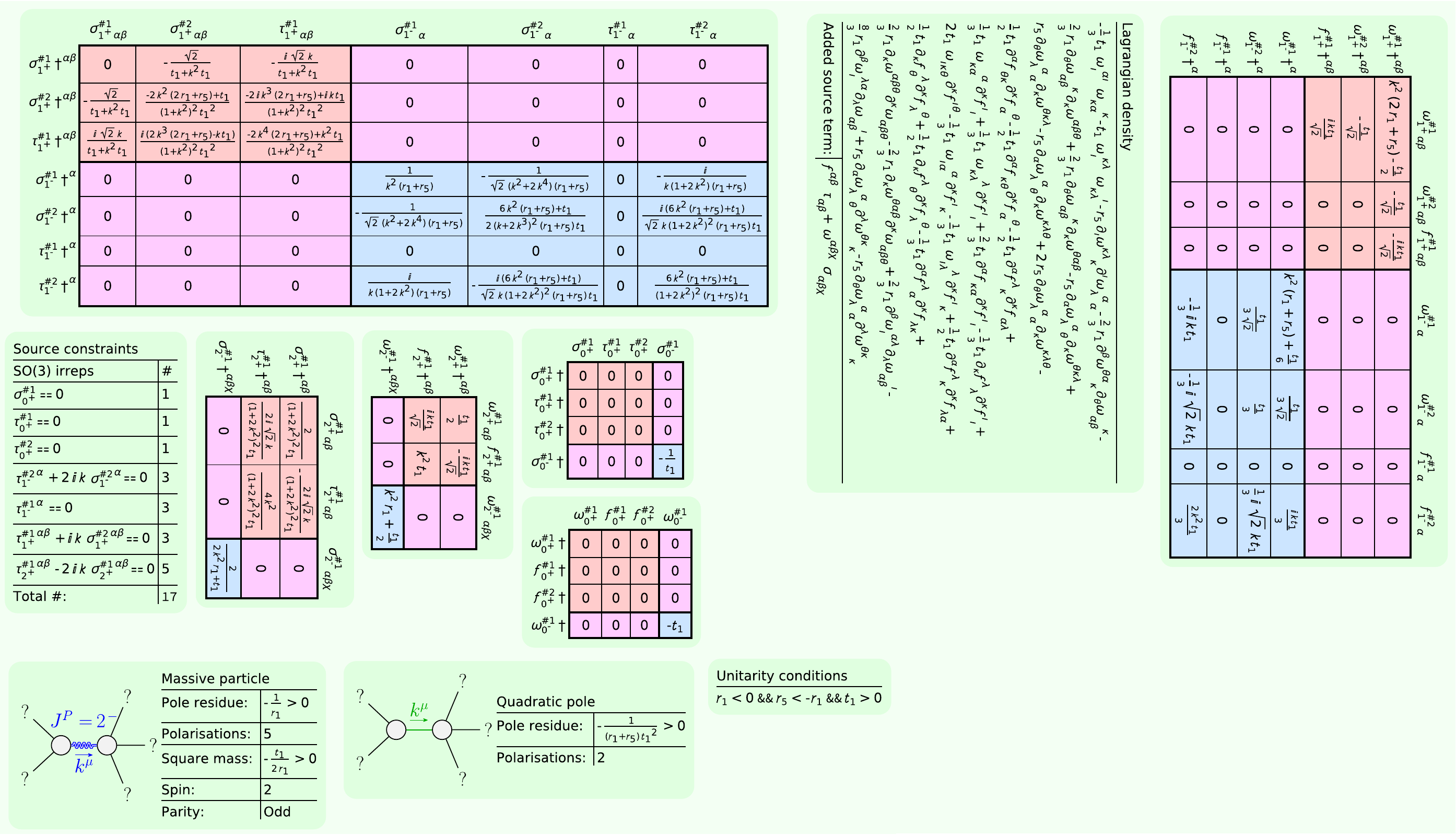}
	\caption{Particle spectrum of Case~$\#4$ from~\cite{Lin:2019ugq}. All quantities are defined in~\cref{FieldKinematicsTetradPerturbation,FieldKinematicsSpinConnection}.}
\label{ParticleSpectrographCase4}
\end{figure*}

\begin{figure*}[h]
	\includegraphics[width=\linewidth]{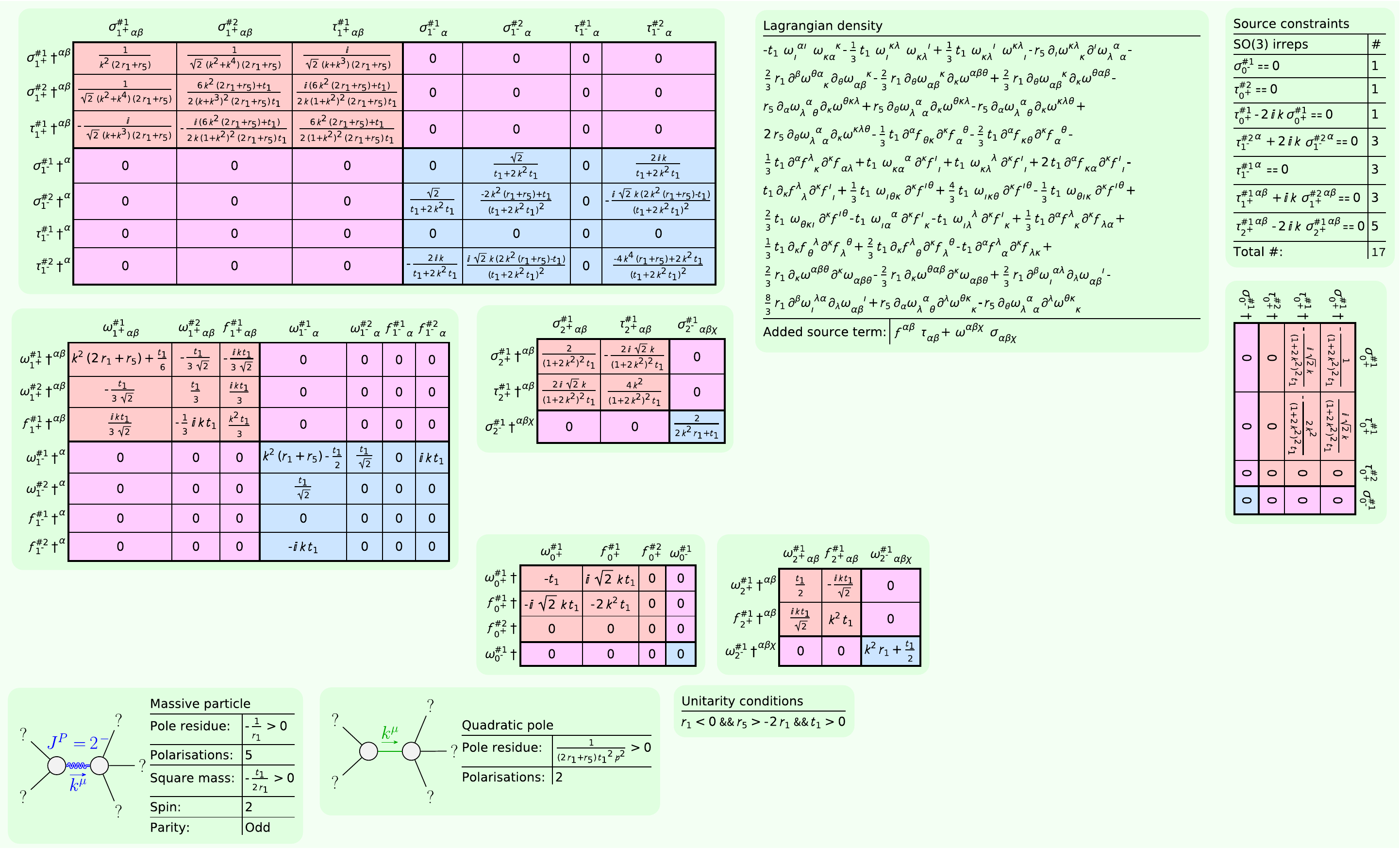}
	\caption{Particle spectrum of Case~$\#5$ from~\cite{Lin:2019ugq}. All quantities are defined in~\cref{FieldKinematicsTetradPerturbation,FieldKinematicsSpinConnection}.}
\label{ParticleSpectrographCase5}
\end{figure*}

\begin{figure*}[h]
	\includegraphics[width=\linewidth]{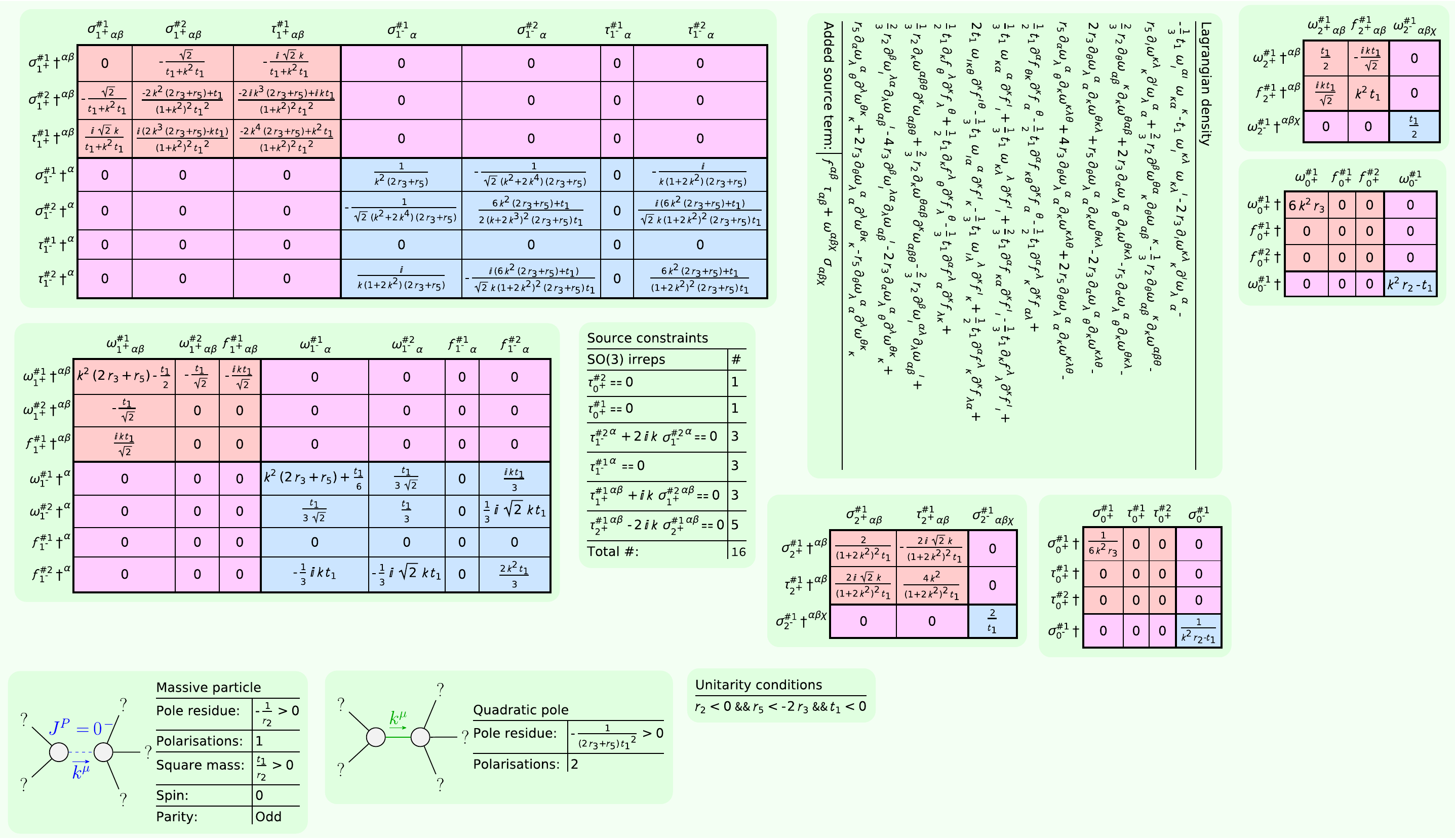}
	\caption{Particle spectrum of Case~$\#6$ from~\cite{Lin:2019ugq}. All quantities are defined in~\cref{FieldKinematicsTetradPerturbation,FieldKinematicsSpinConnection}.}
\label{ParticleSpectrographCase6}
\end{figure*}

\begin{figure*}[h]
	\includegraphics[width=\linewidth]{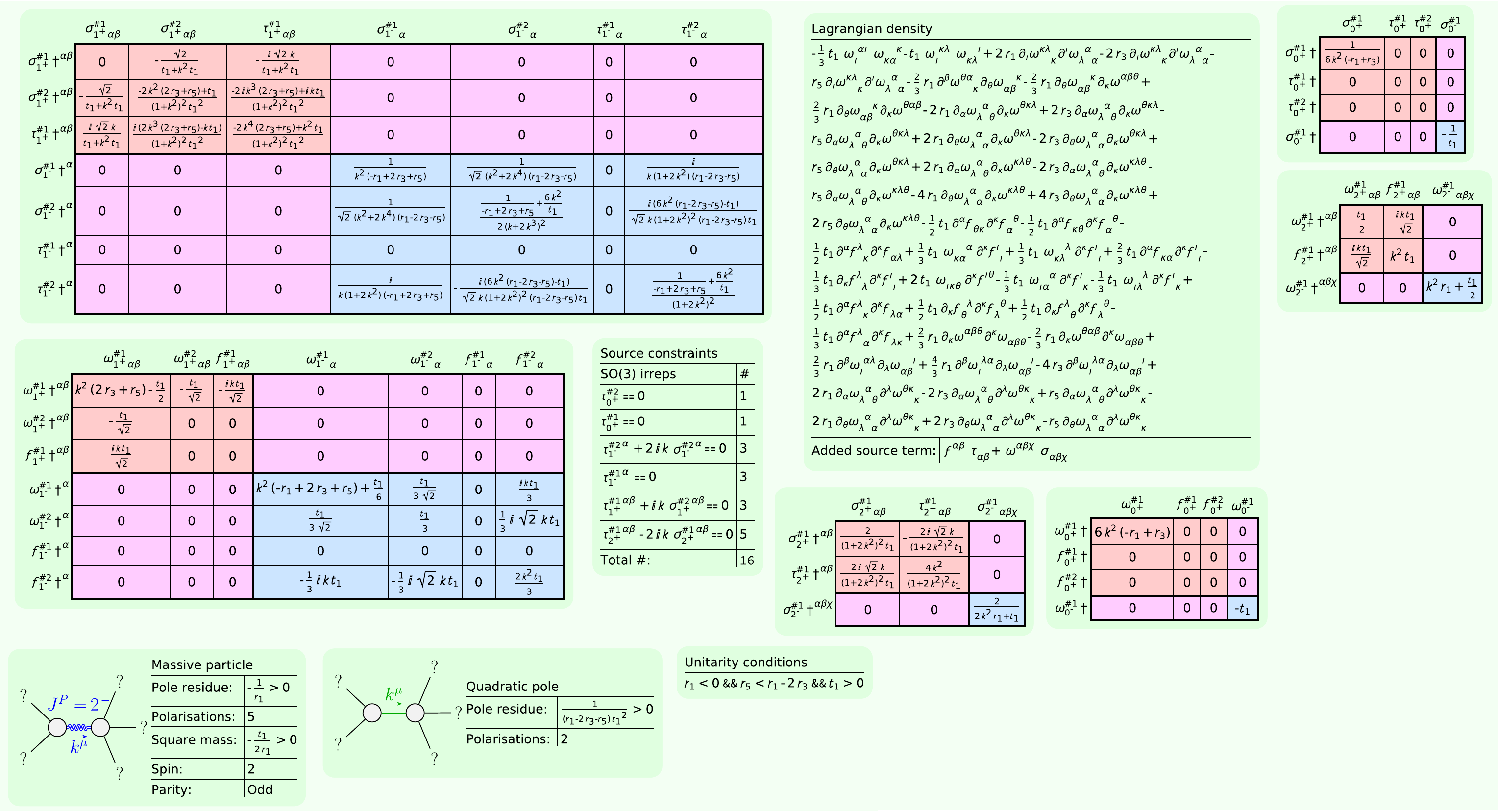}
	\caption{Particle spectrum of Case~$\#7$ from~\cite{Lin:2019ugq}. All quantities are defined in~\cref{FieldKinematicsTetradPerturbation,FieldKinematicsSpinConnection}.}
\label{ParticleSpectrographCase7}
\end{figure*}

\begin{figure*}[h]
	\includegraphics[width=\linewidth]{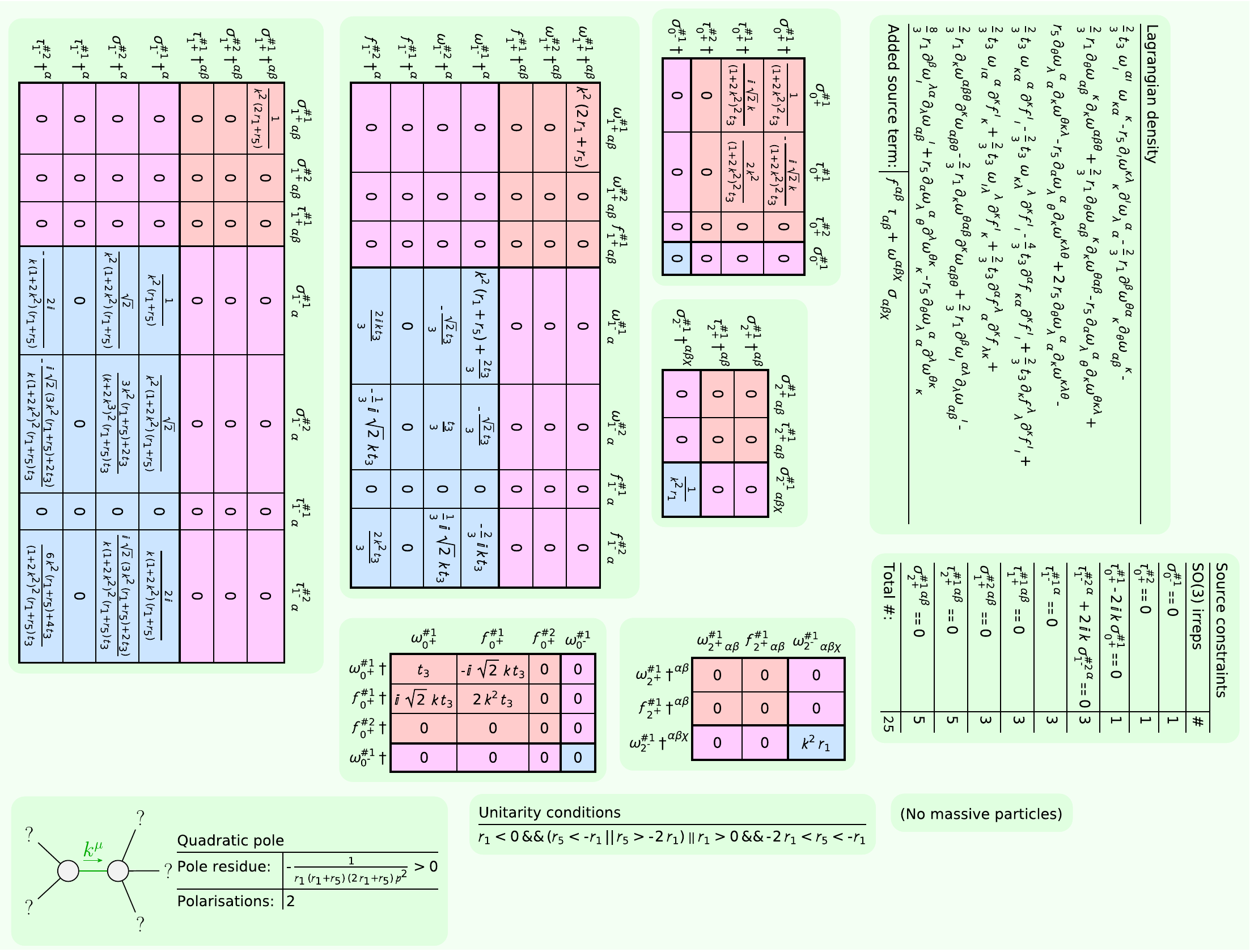}
	\caption{Particle spectrum of Case~$\#8$ from~\cite{Lin:2019ugq}. All quantities are defined in~\cref{FieldKinematicsTetradPerturbation,FieldKinematicsSpinConnection}.}
\label{ParticleSpectrographCase8}
\end{figure*}

\begin{figure*}[h]
	\includegraphics[width=\linewidth]{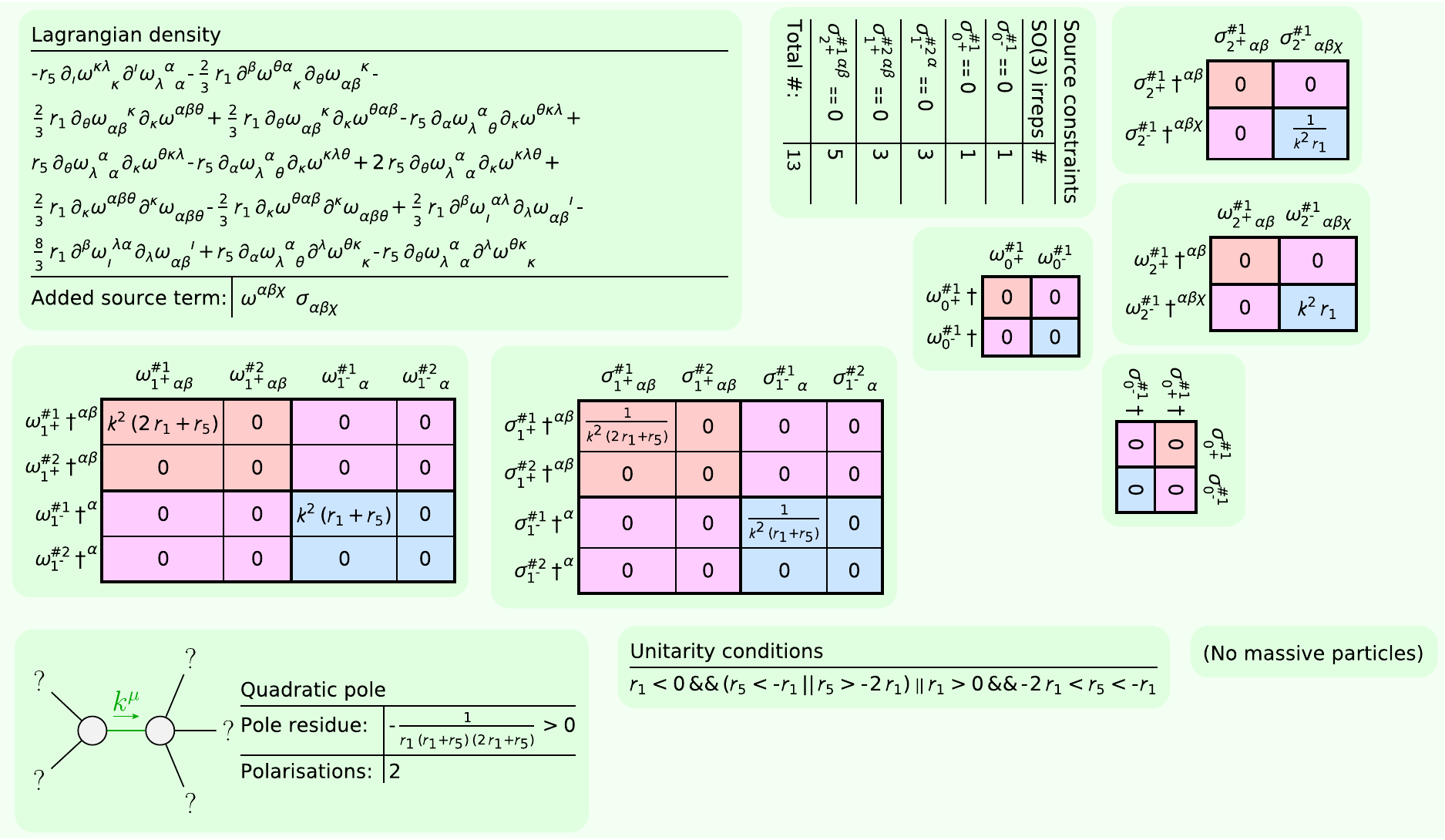}
	\caption{Particle spectrum of Case~$\#9$ from~\cite{Lin:2019ugq}. All quantities are defined in~\cref{FieldKinematicsTetradPerturbation,FieldKinematicsSpinConnection}.}
\label{ParticleSpectrographCase9}
\end{figure*}

\begin{figure*}[h]
	\includegraphics[width=\linewidth]{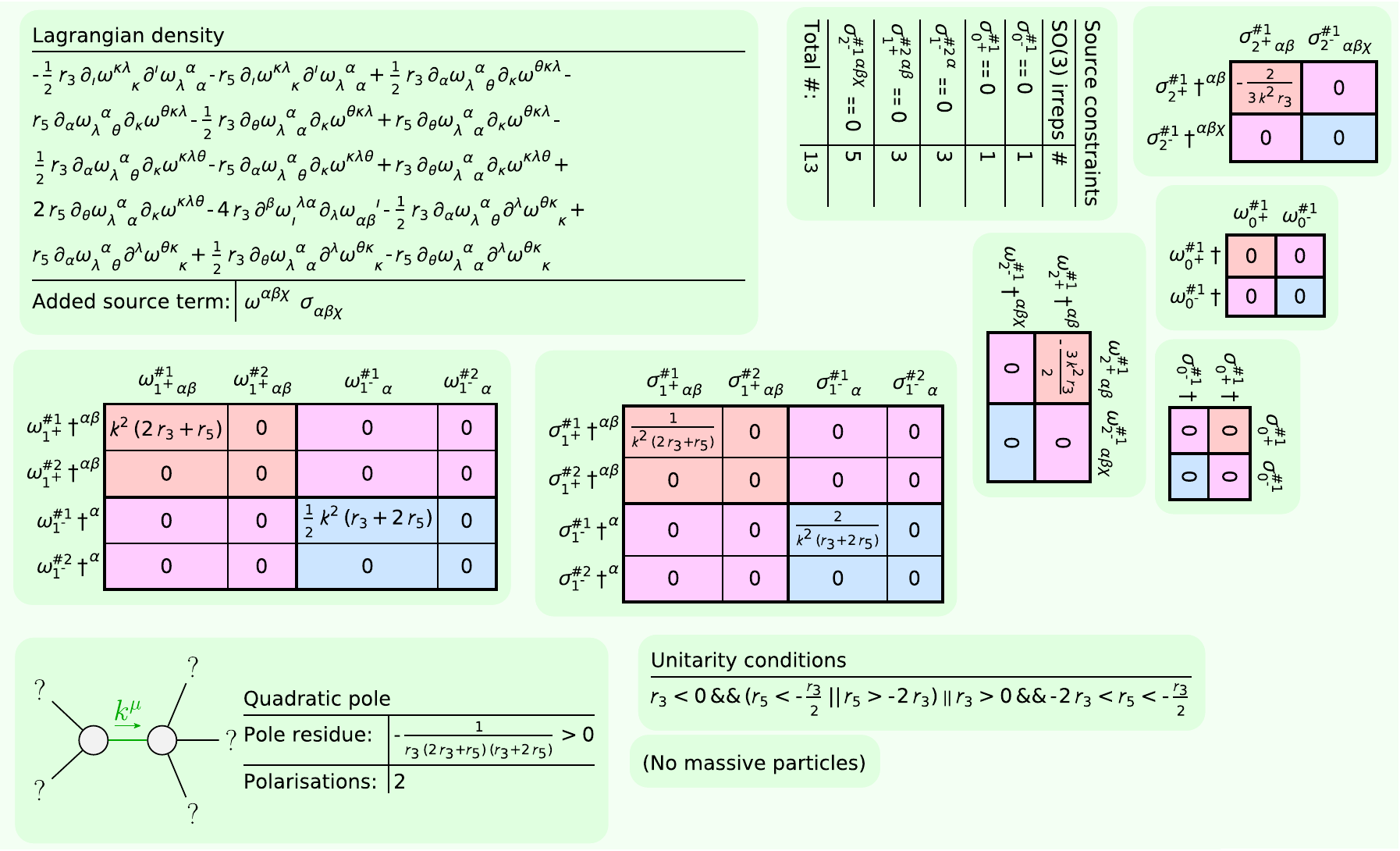}
	\caption{Particle spectrum of Case~$\#10$ from~\cite{Lin:2019ugq}. All quantities are defined in~\cref{FieldKinematicsTetradPerturbation,FieldKinematicsSpinConnection}.}
\label{ParticleSpectrographCase10}
\end{figure*}

\begin{figure*}[h]
	\includegraphics[width=\linewidth]{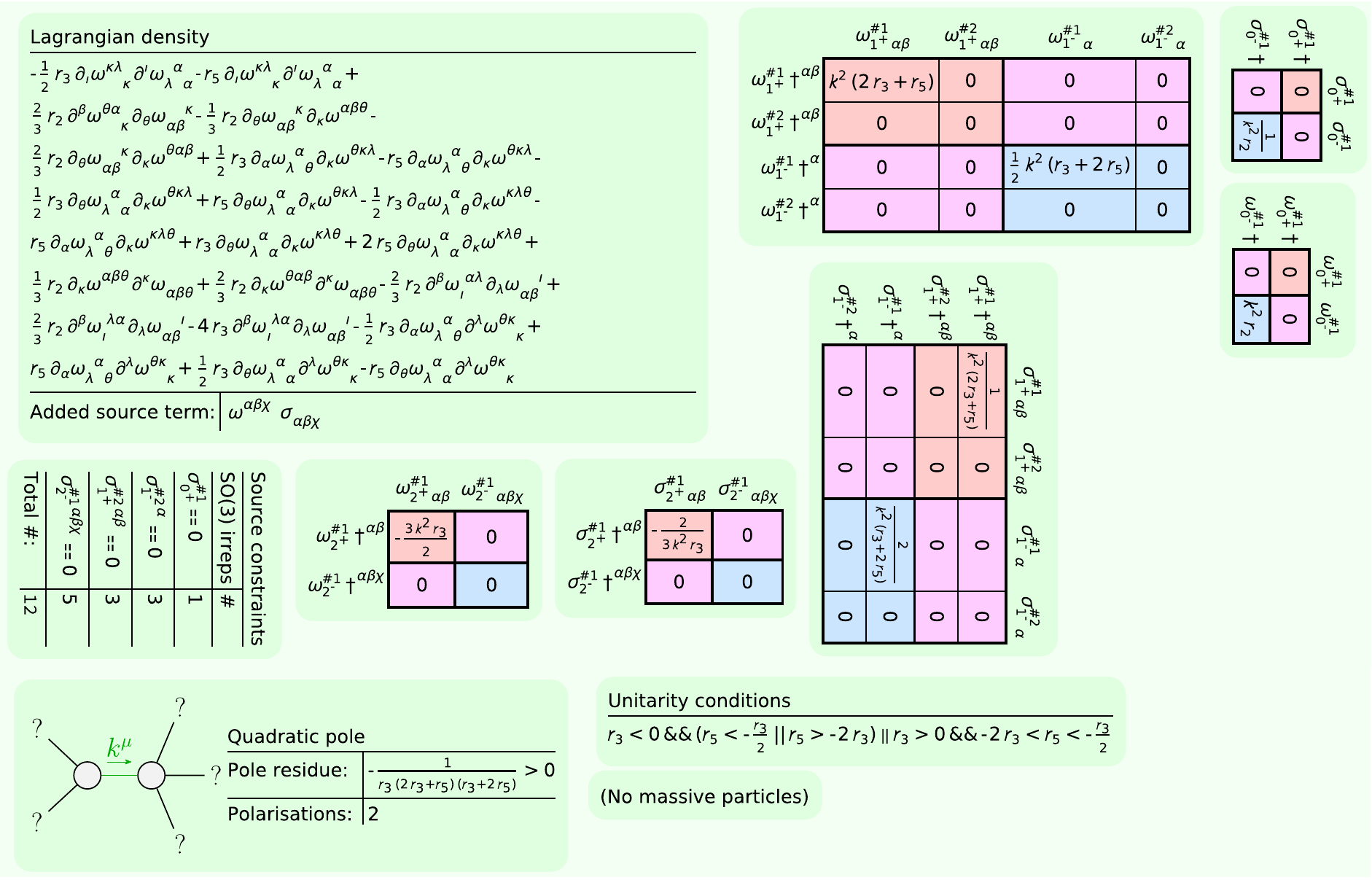}
	\caption{Particle spectrum of Case~$\#11$ from~\cite{Lin:2019ugq}. All quantities are defined in~\cref{FieldKinematicsTetradPerturbation,FieldKinematicsSpinConnection}.}
\label{ParticleSpectrographCase11}
\end{figure*}

\begin{figure*}[h]
	\includegraphics[width=\linewidth]{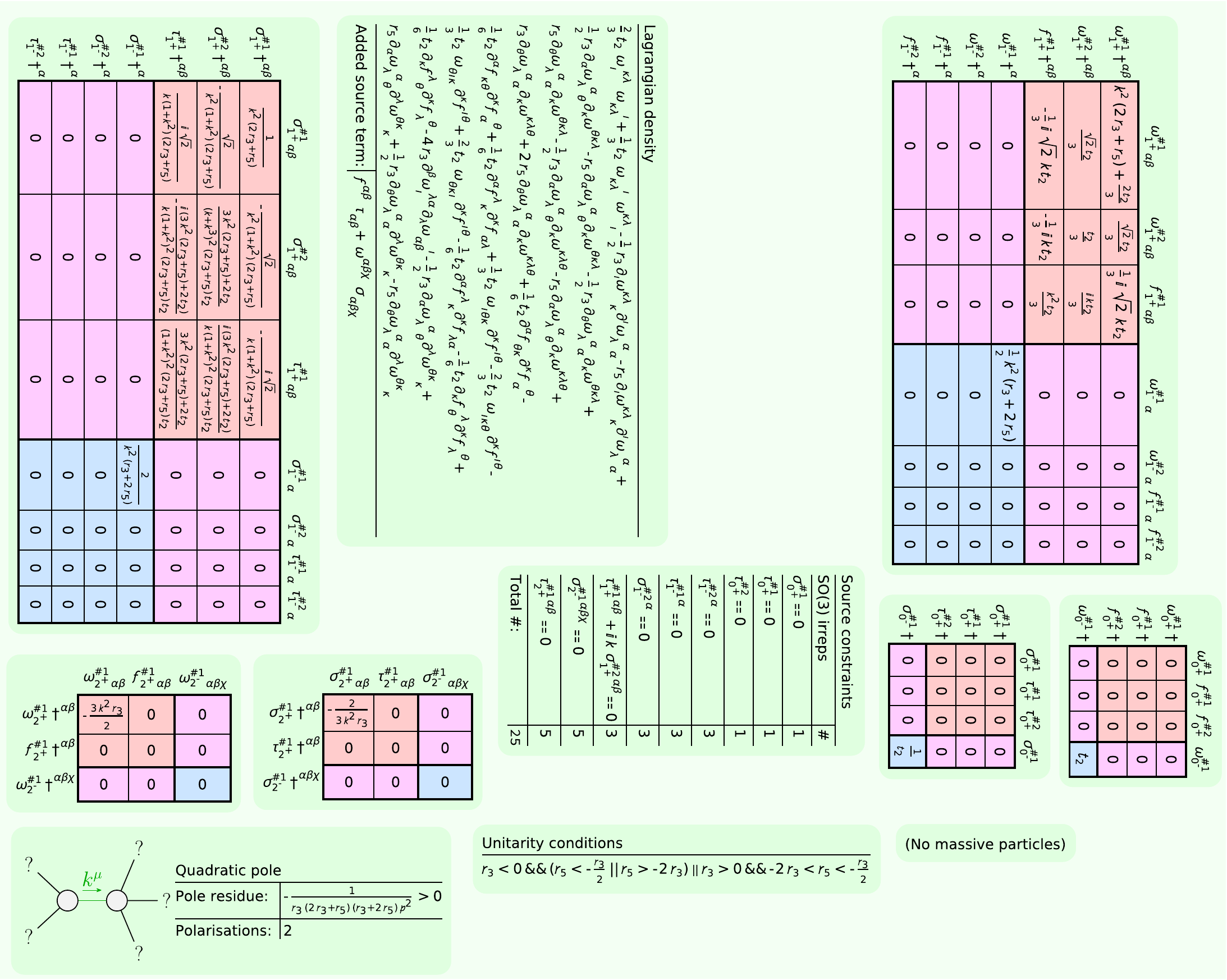}
	\caption{Particle spectrum of Case~$\#12$ from~\cite{Lin:2019ugq}. All quantities are defined in~\cref{FieldKinematicsTetradPerturbation,FieldKinematicsSpinConnection}.}
\label{ParticleSpectrographCase12}
\end{figure*}

\begin{figure*}[h]
	\includegraphics[width=\linewidth]{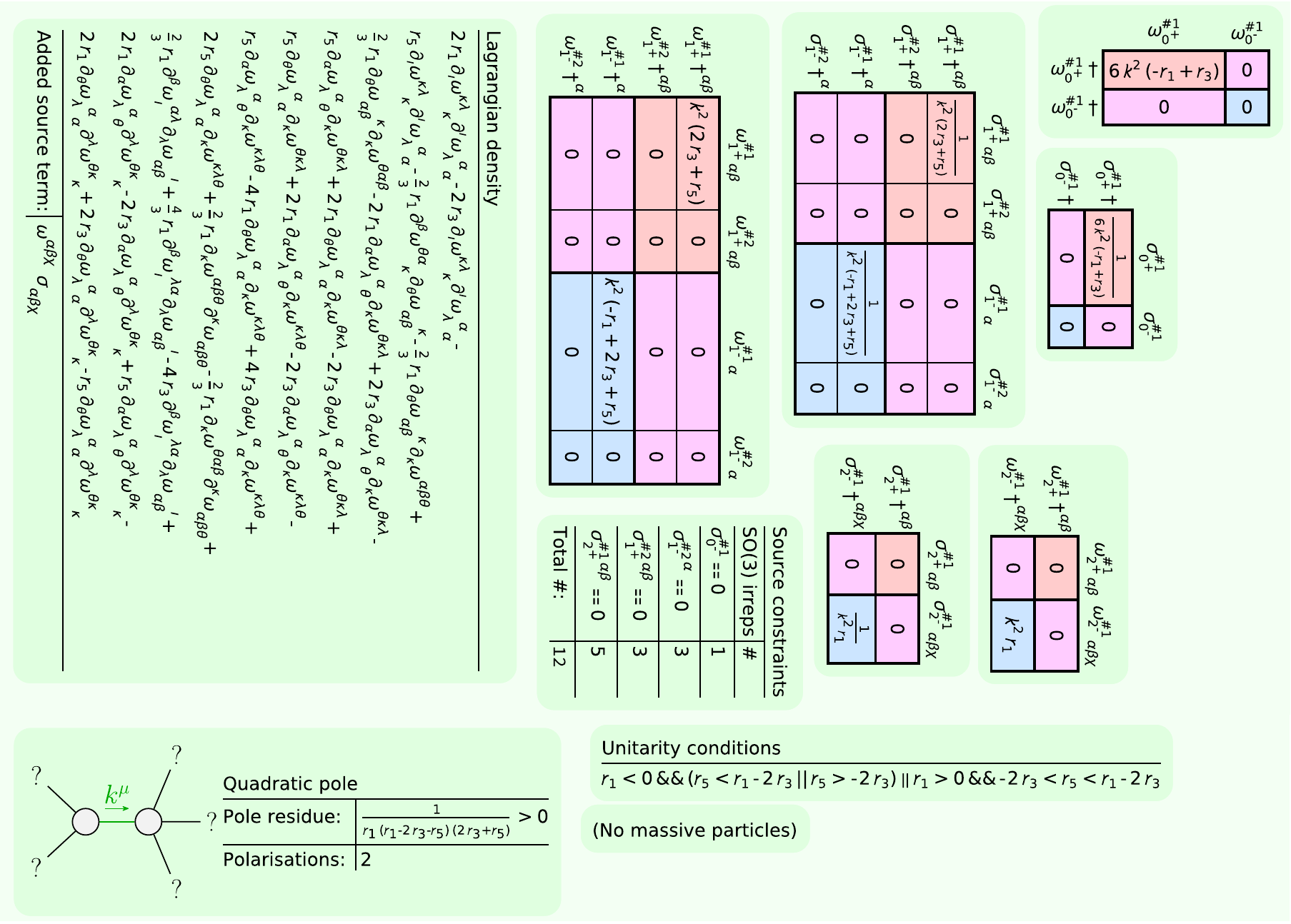}
	\caption{Particle spectrum of Case~$\#13$ from~\cite{Lin:2019ugq}. All quantities are defined in~\cref{FieldKinematicsTetradPerturbation,FieldKinematicsSpinConnection}.}
\label{ParticleSpectrographCase13}
\end{figure*}

\begin{figure*}[h]
	\includegraphics[width=\linewidth]{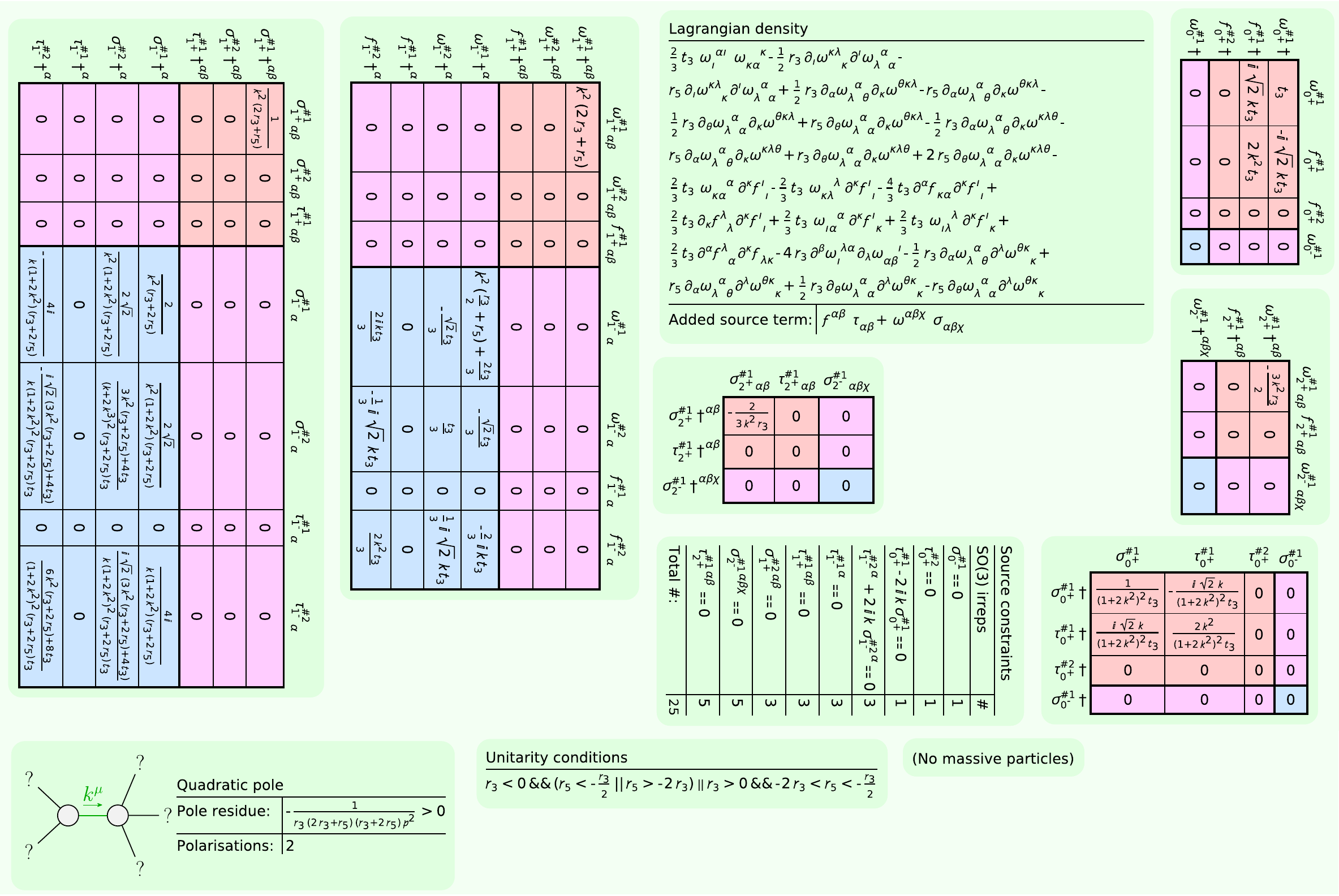}
	\caption{Particle spectrum of Case~$\#14$ from~\cite{Lin:2019ugq}. All quantities are defined in~\cref{FieldKinematicsTetradPerturbation,FieldKinematicsSpinConnection}.}
\label{ParticleSpectrographCase14}
\end{figure*}

\begin{figure*}[h]
	\includegraphics[width=\linewidth]{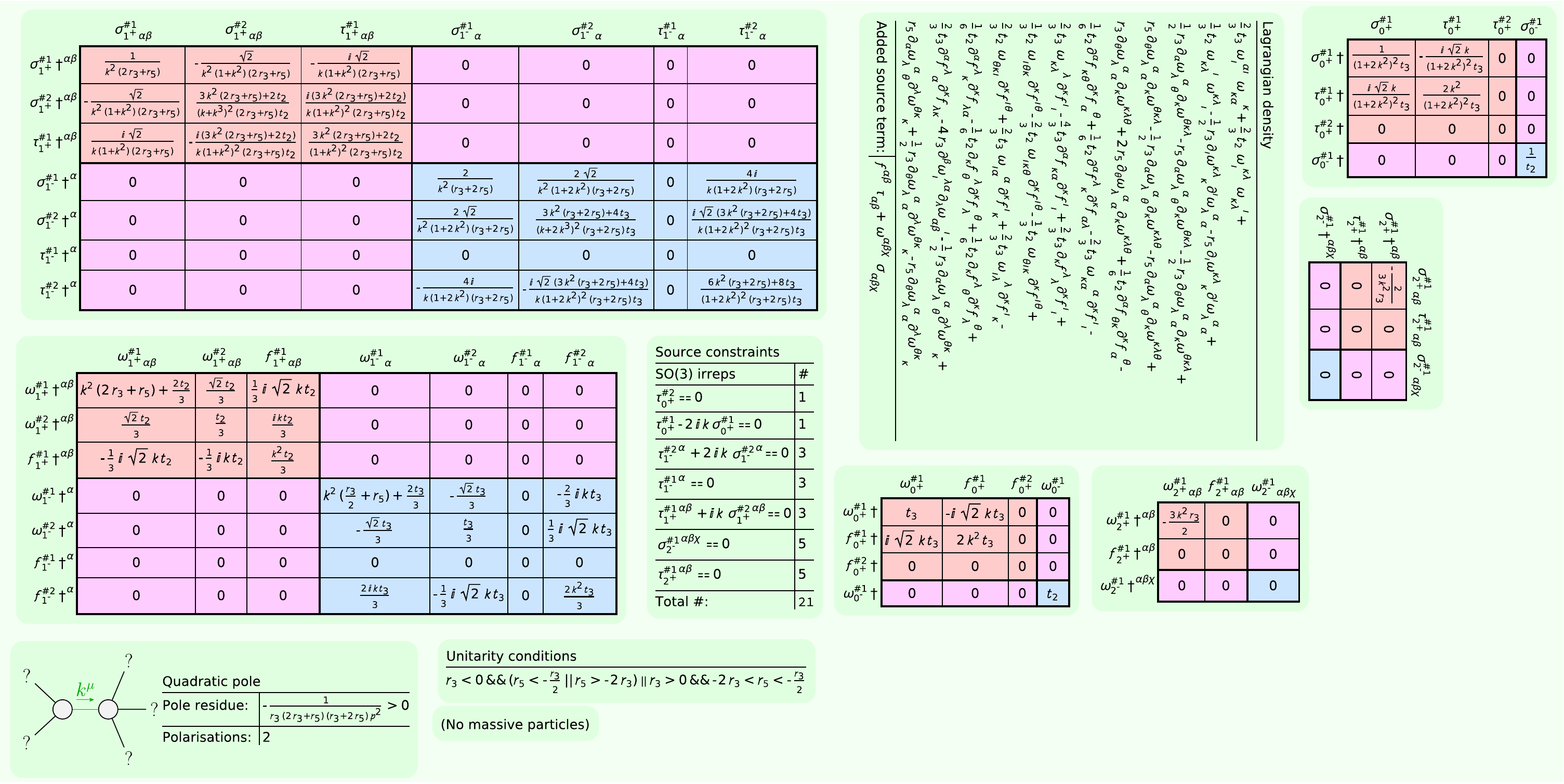}
	\caption{Particle spectrum of Case~$\#15$ from~\cite{Lin:2019ugq}. All quantities are defined in~\cref{FieldKinematicsTetradPerturbation,FieldKinematicsSpinConnection}.}
\label{ParticleSpectrographCase15}
\end{figure*}

\begin{figure*}[h]
	\includegraphics[width=\linewidth]{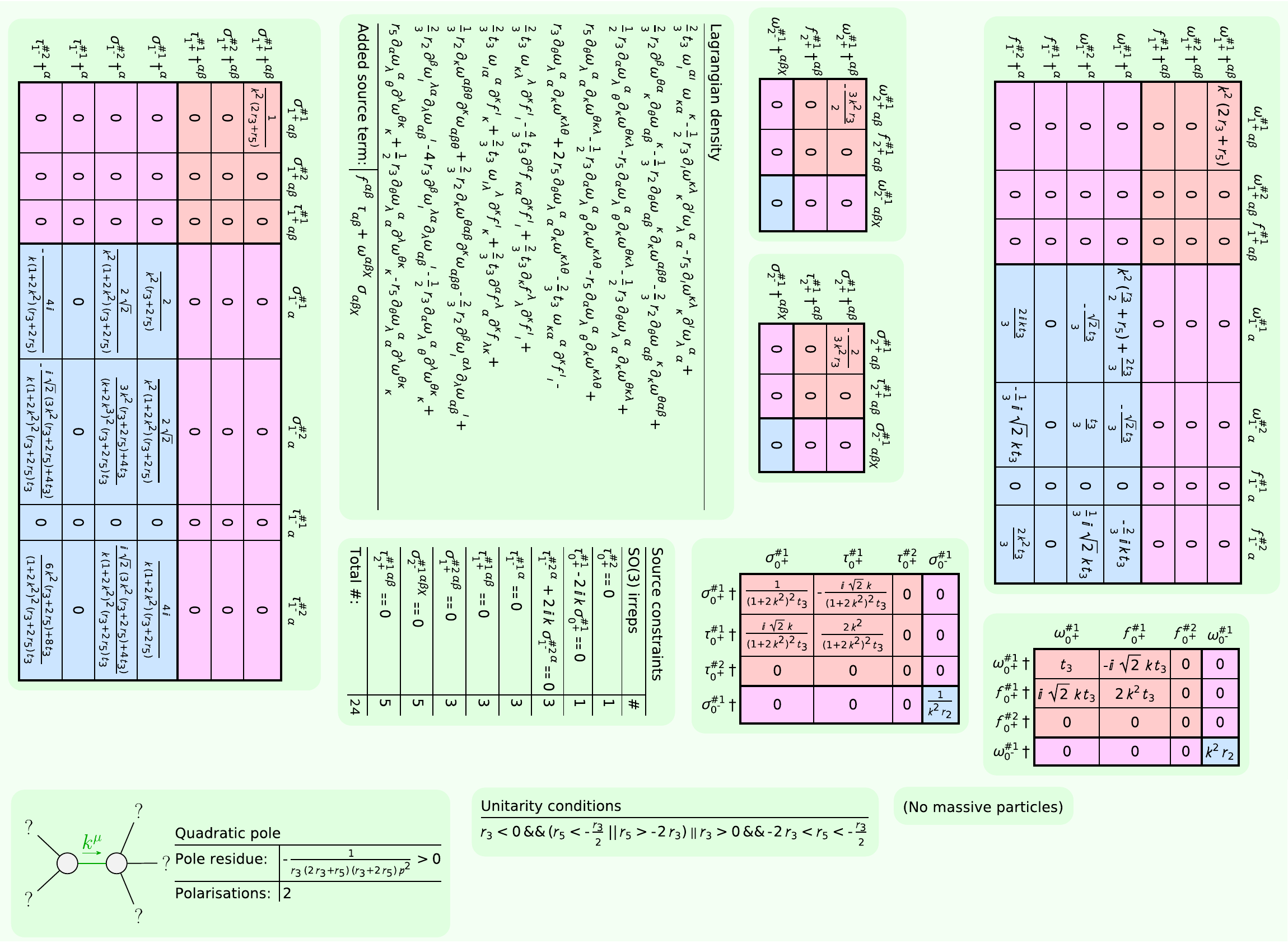}
	\caption{Particle spectrum of Case~$\#16$ from~\cite{Lin:2019ugq}. All quantities are defined in~\cref{FieldKinematicsTetradPerturbation,FieldKinematicsSpinConnection}.}
\label{ParticleSpectrographCase16}
\end{figure*}

\begin{figure*}[h]
	\includegraphics[width=\linewidth]{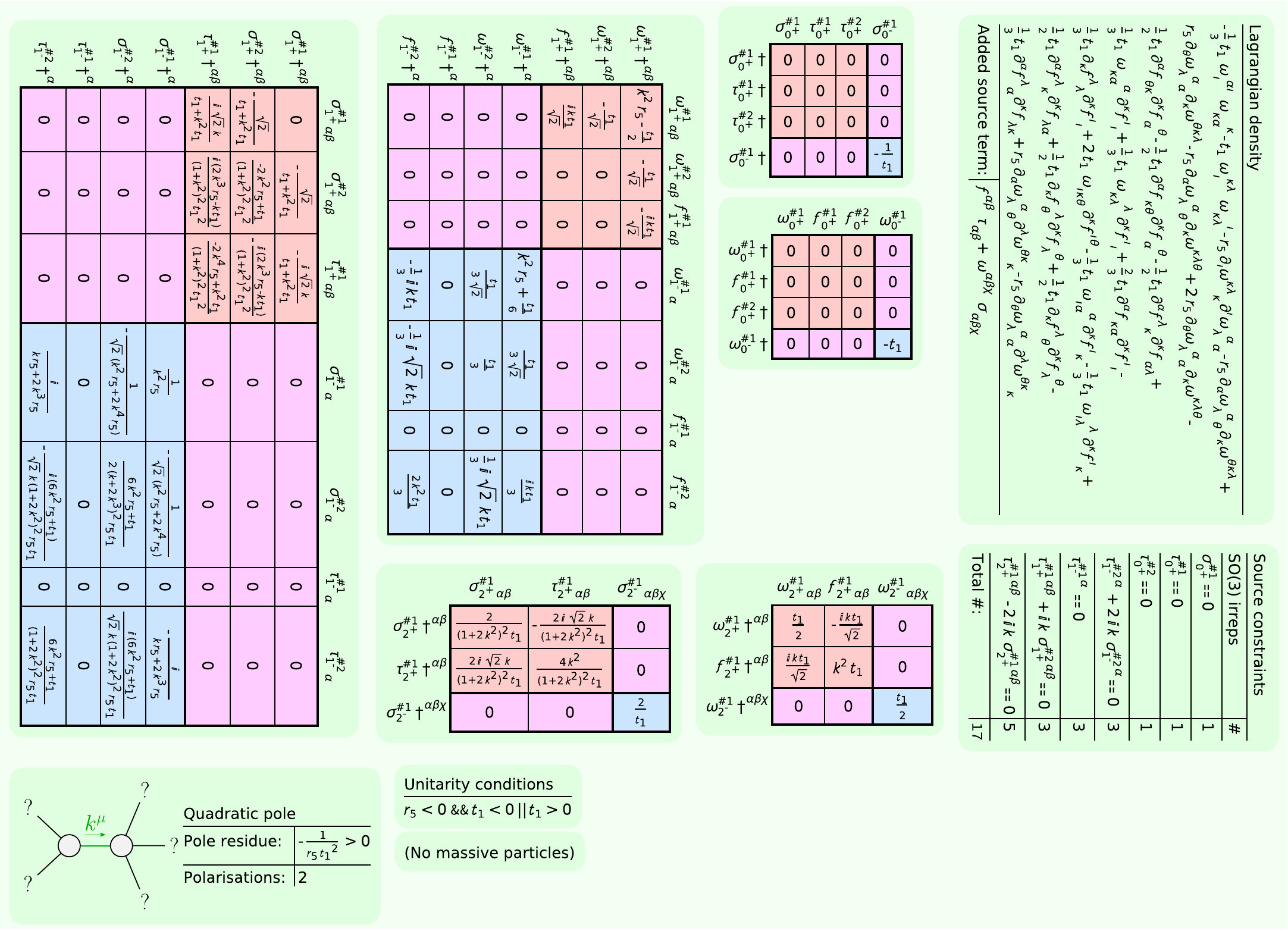}
	\caption{Particle spectrum of Case~$\#17$ from~\cite{Lin:2019ugq}. All quantities are defined in~\cref{FieldKinematicsTetradPerturbation,FieldKinematicsSpinConnection}.}
\label{ParticleSpectrographCase17}
\end{figure*}

\begin{figure*}[h]
	\includegraphics[width=\linewidth]{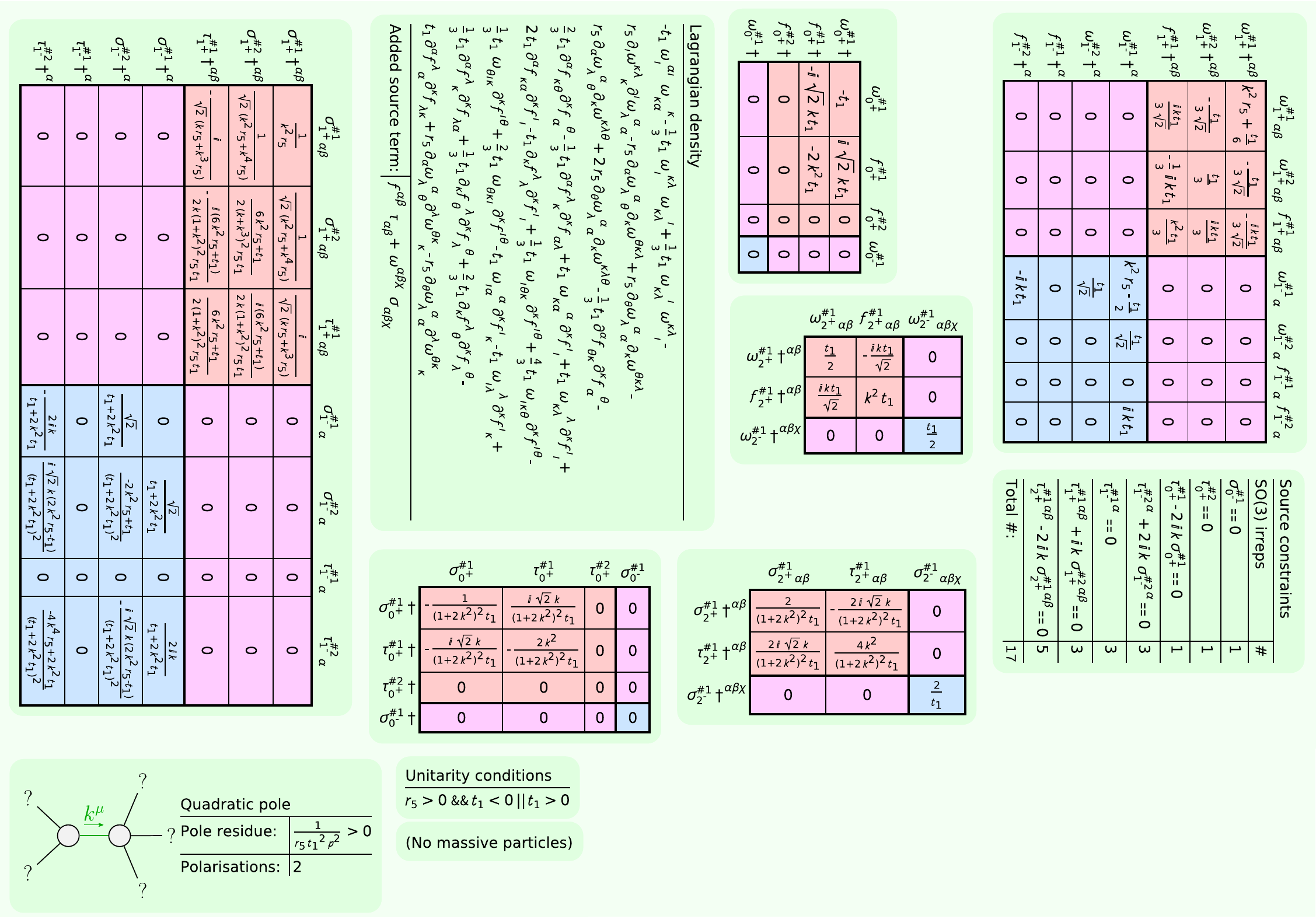}
	\caption{Particle spectrum of Case~$\#18$ from~\cite{Lin:2019ugq}. All quantities are defined in~\cref{FieldKinematicsTetradPerturbation,FieldKinematicsSpinConnection}.}
\label{ParticleSpectrographCase18}
\end{figure*}

\begin{figure*}[h]
	\includegraphics[width=\linewidth]{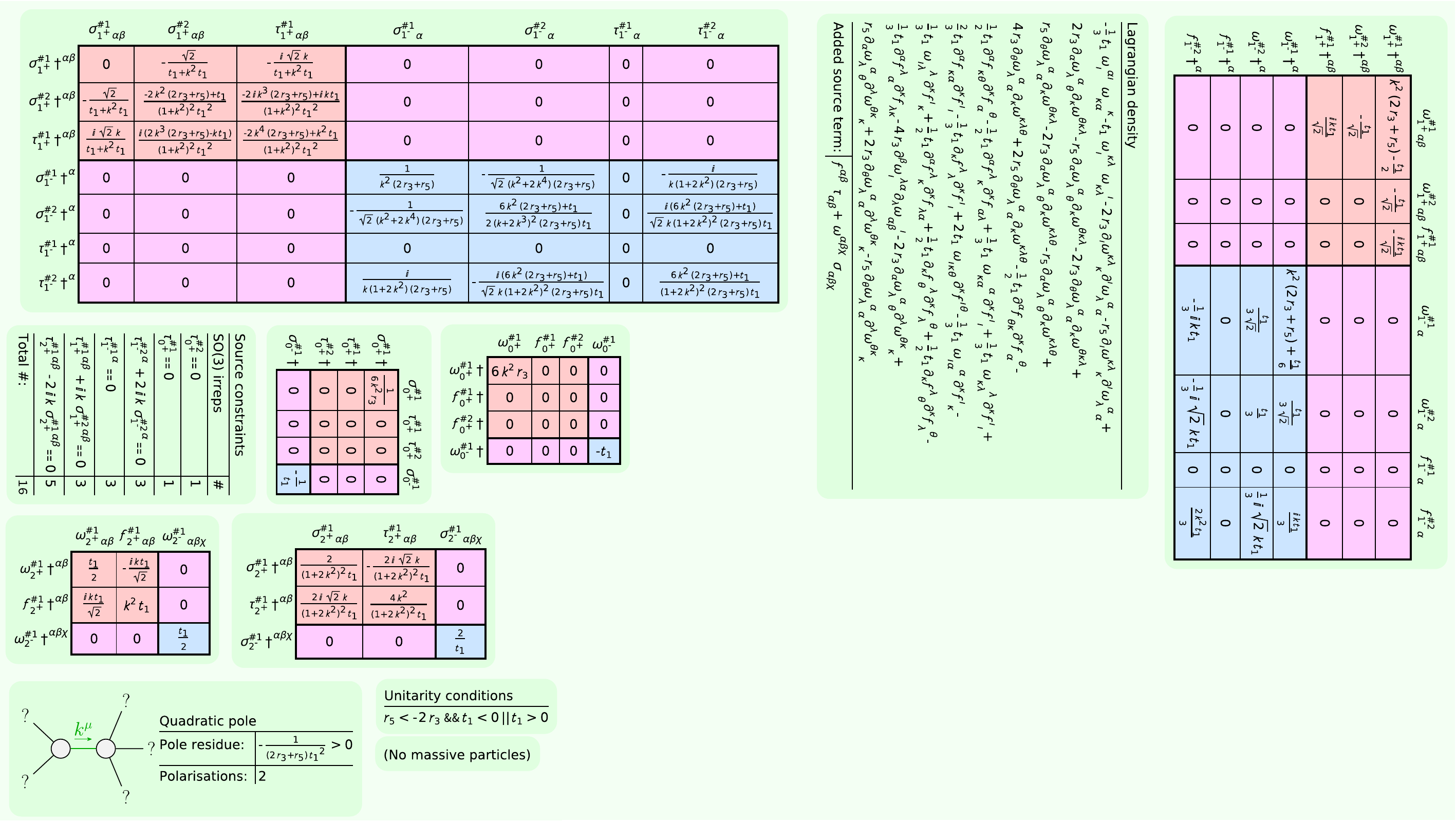}
	\caption{Particle spectrum of Case~$\#19$ from~\cite{Lin:2019ugq}. All quantities are defined in~\cref{FieldKinematicsTetradPerturbation,FieldKinematicsSpinConnection}.}
\label{ParticleSpectrographCase19}
\end{figure*}

In this appendix we provide further particle spectrographs produced by \PSALTer{}. The key reference~\cite{Lin:2019ugq} provided 58 new cases of PGT (see~\cref{PoincareGaugeTheory}) which were unitary and power-counting renormalisable. In~\crefrange{ParticleSpectrographCase1}{ParticleSpectrographCase19} we re-compute the spectra of the first 19 cases: those which contain massless species.


\end{document}